\documentclass[journal]{IEEEtran}
\IEEEoverridecommandlockouts
\usepackage{cite}
\usepackage{amsmath,amssymb,amsfonts}
\usepackage{algorithmic}
\usepackage{graphicx}
\usepackage{textcomp}
\usepackage[dvipsnames]{xcolor}
\usepackage[draft,pdfborder={0 0 0}]{hyperref}
\def\BibTeX{{\rm B\kern-.05em{\sc i\kern-.025em b}\kern-.08em
    T\kern-.1667em\lower.7ex\hbox{E}\kern-.125emX}}

\usepackage[commandnameprefix=ifneeded,commentmarkup=uwave]{changes}
\definechangesauthor[name=Pascal, color=blue]{P}
\definechangesauthor[name=Charles, color=red]{C}

\usepackage{glossaries} % For acronyms
\usepackage{fancyhdr}
\usepackage{lipsum}
\usepackage{tabularx,colortbl}
\usepackage{multirow}
\usepackage{dcolumn}
\usepackage{booktabs}
\usepackage{bm}
\usepackage{tabularx}
\usepackage{algorithm,algorithmic}
\usepackage{multicol}
\usepackage{pgfplots}
\pgfplotsset{compat=1.12}

\usepgfplotslibrary{fillbetween}
\usetikzlibrary{patterns}
\usepackage{comment}
\usepackage{soul}% highliting
\usepackage{dsfont} % for the field symbol used in one of the refs
\usepackage{tikz}
\usetikzlibrary{positioning}
\usetikzlibrary {shapes.geometric}
\tikzset{near start abs/.style={xshift=1cm}}
\usetikzlibrary{positioning,calc}
\usetikzlibrary {arrows.meta}
\usepackage{amsmath,amsfonts,amssymb}
\usetikzlibrary{automata}
\usepackage{threeparttable}
\usepackage{dirtytalk}
\usetikzlibrary{plotmarks, shapes, positioning, arrows, decorations.markings, fit, calc, patterns, circuits.logic.US}
\definechangesauthor[name={Poncha}, color=brown]{jp}
\usetikzlibrary{shapes.misc, positioning}
\usepackage{circuitikz}
\usepackage{caption}
\usepackage{subcaption}
\usepackage{microtype} % Removes white spaces
\usepackage[most]{tcolorbox}
\usepackage{balance} %% Compact refs
\begin{document}
\tikzstyle{mux_r}=[trapezium, trapezium stretches = true, rotate=-90, minimum width=1cm, minimum height=0.5cm, draw]
\tikzstyle{branch}=[fill,shape=circle,minimum size=5pt,inner sep=0pt]
\tikzstyle{ckd_node}=[rectangle, minimum width=0.5cm, minimum height=0.5cm, draw]
\tikzstyle{line:thick}=[line width=0.5mm, arrows = {-Latex[width=5pt, length=5pt]}]
\tikzstyle{line:thin}=[line width=0.1mm, arrows = {-Latex[width=5pt, length=5pt]}]
\tikzset{my node/.style={trapezium, fill=#1!20, draw=#1!75, text=black}}
\tikzstyle{node:ALU}=[muxdemux, muxdemux def={NL=2, NR=1, NB=1, w=2, inset w=1, inset Lh=2, inset Rh=0, square pins=1, Lh=6, Rh=3}]
\tikzstyle{line:thick_dot}=[ decorate, decoration={snake, amplitude=.2mm, segment length=3mm}, line width=0.2mm]
\tikzstyle{line:thick0}=[line width=0.1mm, arrows = {-Latex[width=1pt, length=1pt]}]
\tikzstyle{line:thin0}=[line width=0.1mm, arrows = {-Latex[width=1pt, length=1pt]}]
\tikzstyle{clk}=[isosceles triangle,  minimum width=2.8mm, minimum height=0.8mm]
\tikzstyle{line:thickAr}=[line width=0.5mm, arrows = {Latex[width=5pt, length=5pt]-Latex[width=5pt, length=5pt]}]

%% list of acronyms
\newacronym{alu}{ALU}{arithmetic logic unit}
\newacronym{ascii}{ASCII}{American standard code for information interchange}
\newacronym{asic}{ASIC}{application-specific integrated circuit}
\newacronym{at}{A$\times$T}{area $\times$ time}
\newacronym{bia}{BIA}{binary inversion algorithm}
\newacronym{bip}{BIP}{Bitcoin improvement proposal}
\newacronym{ckd}{CKD}{child key derivation}
\newacronym{cc}{CC}{clock cycle}
\newacronym{clb}{CLB}{configurable logic block}
\newacronym{cu}{CU}{control unit}
\newacronym{cryp}{crypto}{cryptocurrency}
\newacronym{dsp}{DSP}{digital signal processor}
\newacronym{dfi}{DeFi}{decentralized finance}
\newacronym{dpa}{DPA}{differential power analysis}
\newacronym{ec}{EC}{elliptic curve}
\newacronym{ecc}{ECC}{elliptic curve cryptography}
\newacronym{ecdsa}{ECDSA}{elliptic curve digital signature algorithm}
\newacronym{eea}{EEA}{Extended Euclidean algorithm}
\newacronym{eip}{EIP}{Ethereum improvement protocol}
\newacronym{erp}{ERP}{Ethereum request for comment}
\newacronym{epo}{EPO}{energy per operation}
\newacronym{fpga}{FPGA}{field programmable gate array}
\newacronym{fsm}{FSM}{finite state machine}
\newacronym{gf}{GF}{Galois field}
\newacronym{hd}{HD}{hierarchically deterministic}
\newacronym{hmac}{HMAC}{hash-based message authentication code}
\newacronym{hsm}{HSM}{hardware security module}
\newacronym{io}{IO}{input/output}
\newacronym{le}{LE}{logic element}
\newacronym{lsb}{LSB}{least significant bit}
\newacronym{lut}{LUT}{look-up table}
\newacronym{luts}{LUTs}{look up tables}
\newacronym{malu}{MALU}{modular arithmetic logic unit}
\newacronym{mcu}{MCU}{microcontroller unit}
\newacronym{mcus}{MCUs}{microcontroller units}
\newacronym{md}{MD}{Montgomery domain}
\newacronym{msb}{MSB}{most significant bit}
\newacronym{mse}{MSE}{mean square error}
\newacronym{nd}{ND}{non-deterministic}
\newacronym{nist}{NIST}{national institute of standards and technology}
\newacronym{nr}{NR}{not reported}
\newacronym{od}{OD}{original domain}
\newacronym{pa}{ECPA}{elliptic curve point addition}
\newacronym{pbkdf}{PBKDF-2}{password-based key derivation function-2}
\newacronym{pd}{ECPD}{elliptic curve point doubling}
\newacronym{pl}{PL}{programmable logic}
\newacronym{pm}{ECPM}{elliptic curve point multiplication}
\newacronym{prng}{PRNG}{pseudo random number generators}
\newacronym{ps}{PS}{processing system}
\newacronym{pod}{PD}{power density}
\newacronym{qrng}{QRNG}{quantum random number generator}
\newacronym{ram}{RAM}{random access memory}
\newacronym{rng}{RNG}{random number generator}
\newacronym{rtl}{RTL}{register transfer level}
\newacronym{secp}{SECP256K1}{standards for efficient cryptography prime 256 bits Koblitz 1}
\newacronym{sca}{SCA}{side channel analysis}
\newacronym{se}{SE}{secure elements}
\newacronym{sha}{SHA}{secure hash algorithm}
\newacronym{spa}{SPA}{simple power analysis}
\newacronym{trng}{TRNG}{true random number generator}
\newacronym{tpa}{TPA}{throughput per area}
\newacronym{tpw}{TPW}{throughput per Watt}
\newacronym{tps}{TPS}{transactions per second}
\newacronym{usb}{USB}{universal serial bus}

\newacronym{iso}{ISO}{International Standards Organization}
\newacronym{ansi}{ANSI}{American National Standards Institute}
\newacronym{ff}{FF}{flip flop}
\newacronym{fia}{FIA}{fault injection attack}
\newacronym{ema}{EMA}{electromagnetic analysis}
\newacronym{em}{EM}{electromagnetic}
\newacronym{sema}{SEMA}{simple electromagnetic analysis}
\newacronym{dema}{DEMA}{differential electromagnetic analysis}
\newacronym{snr}{SNR}{signal-to-noise ratio}
\newacronym{pin}{PIN}{personal identification number}
\newacronym{xdc}{XDC}{Xilinx design constraint}

\title{EthVault: A Secure and Resource-Conscious FPGA-Based Ethereum Cold Wallet}

\author{Joel Poncha Lemayian, Ghyslain Gagnon, Kaiwen Zhang, and Pascal Giard
\thanks{
    Joel Poncha Lemayian, Ghyslain Gagnon, and Pascal Giard are with the Department of Electrical Engineering, \'{E}cole de technologie sup\'{e}rieure (\'{E}TS), Montr\'{e}al, Canada (e-mail: joel-poncha.lemayian.1@ens.etsmtl.ca,  pascal.giard@etsmtl.ca).
    }
\thanks{
    Kaiwen Zhang is with the Department of Software Engineering and IT, \'{E}cole de technologie sup\'{e}rieure (\'{E}TS), Montr\'{e}al, Canada (e-mail: kaiwen.zhang@etsmtl.ca).
}
}
  
\maketitle
%https://arxiv.org/pdf/1807.05764.pdf
  \begin{abstract} 
    Cryptocurrency blockchain networks safeguard digital assets using cryptographic keys, with wallets playing a critical role in generating, storing, and managing these keys. Wallets, typically categorized as hot and cold, offer varying degrees of security and convenience. However, they are generally software-based applications running on microcontrollers. Consequently, they are vulnerable to malware and side-channel attacks, allowing perpetrators to extract private keys by targeting critical algorithms, such as ECC, which processes private keys to generate public keys and authorize transactions. To address these issues, this work presents EthVault, the first hardware architecture for an Ethereum hierarchically deterministic cold wallet, featuring hardware implementations of key algorithms for secure key generation. Also, an ECC architecture resilient to side-channel and timing attacks is proposed. Moreover, an architecture of the child key derivation function, a fundamental component of cryptocurrency wallets, is proposed. The design minimizes resource usage, meeting market demand for small, portable cryptocurrency wallets. FPGA implementation results validate the feasibility of the proposed approach. The ECC architecture exhibits uniform execution behavior across varying inputs, while the complete design utilizes only 27\%, 7\%, and 6\% of LUTs, registers, and RAM blocks, respectively, on a Xilinx Zynq UltraScale+ FPGA.
\end{abstract}


\glsresetall

\section{Introduction}
\label{sec:intro}

\IEEEPARstart{C}{ryptographic} keys play a vital role in securing user assets within the blockchain ecosystem. They provide a robust layer of security by enabling authentication, identity verification, and data integrity \cite{lu2019blockchain}. Moreover, cryptographic keys facilitate safe user interaction within the network. This limits the access and modification of sensitive data to authorized users only. Blockchain networks, such as Ethereum \cite{tikhomirov2018ethereum} and Bitcoin \cite{manimuthu2019literature}, extensively use public-private cryptographic keys to protect user assets. Furthermore, while anyone can access public keys, the secrecy of private keys is paramount since they are used to prove ownership, notably giving the owner access to all digital assets. Accordingly, a compromised private key can result in significant losses, as it grants an attacker full control over all assets in the associated account \cite{guri2018beatcoin}.

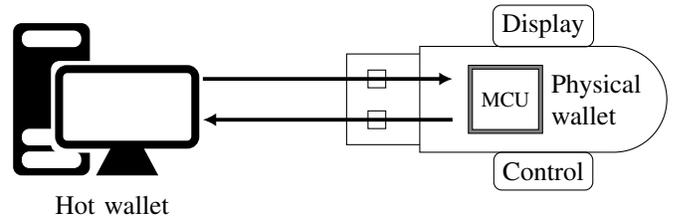
\begin{figure}[t]
    \centering
    \resizebox{0.5\textwidth}{!}{
            \begin{tikzpicture}
        %\definecolor{lightblue}{HTML}{FFF2CC} \fill[lightblue] (-2,-4) rectangle (10,5);
        \node[fill=black, rectangle, rounded corners=3pt, minimum width=1cm, minimum height=2cm, draw](cpu){};%cpu
        \node[fill=white, rectangle, rounded corners=3pt, minimum width=0.8cm, minimum height=0.05cm, draw](cpu0) at ($(cpu)+(0,0.8)$){};
        \node[fill=white, rectangle, rounded corners=3pt, minimum width=0.8cm, minimum height=0.05cm, draw](cpu1) at ($(cpu)+(0,-0.5)$){};
        \node[fill=white, rectangle, rounded corners=3pt, minimum width=0.8cm, minimum height=0.05cm, draw](cpu2) at ($(cpu)+(0,-0.8)$){};
        \node[white, fill=white, rectangle, rounded corners=3pt, minimum width=2cm, minimum height=1.2cm, draw](mon0) at ($(cpu)+(1,-0.1)$){};%mon0
        \node[line width=3pt, fill=white, rectangle, rounded corners=3pt, minimum width=1.8cm, minimum height=1cm, draw](mon1) at ($(cpu)+(1,-0.1)$){};%mon1
        \node[trapezium, fill=black, trapezium stretches = true, minimum width=0.8cm, minimum height=0.4cm, draw](stand) at ($(mon1)+(0,-0.7)$){};%stand
        \node[rectangle, minimum width=1.2cm, minimum height=1.2cm, draw](sqr0) at ($(stand)+(3.5, 0.8)$){};%sqr0
        \node[fill=white, rounded rectangle, rounded rectangle west arc=0pt, minimum width=3.5cm, minimum height=1.4cm, draw](pys_wal) at ($(stand)+(5.5, 0.8)$){};%pys_wal
        \node[rectangle, rounded corners=3pt, minimum width=1cm, minimum height=0.5cm, draw](disp) at ($(pys_wal)+(0,0.97)$){Display}; %disp
        \node[rectangle, rounded corners=3pt, minimum width=1cm, minimum height=0.5cm, draw](cont) at ($(pys_wal)+(0,-0.95)$){Control}; %cont
        \node[rectangle, minimum width=0.1cm, minimum height=0.1cm, draw](sqr1) at ($(sqr0)+(-0.2, 0.27)$){};%sqr1
        \node[rectangle, minimum width=0.1cm, minimum height=0.1cm, draw](sqr2) at ($(sqr0)+(-0.2, -0.27)$){};%sqr2
        \node[fill=gray, rectangle, minimum width=0.98cm, minimum height=0.9cm, draw](proc0) at ($(pys_wal)+(-0.5, 0)$){};
        \node[fill=white, rectangle, minimum width=0.8cm, minimum height=0.8cm, draw](proc) at ($(pys_wal)+(-0.5, 0)$){\footnotesize MCU};
        \node[](hw) at ($(pys_wal)+(0.7, 0)$){\shortstack[l]{Physical \\wallet}};
        \node[](hw) at ($(cpu)+(0.8, -1.4)$){Hot wallet};

        %arrows
        \draw[line:thick] ($(pys_wal)+(-4.5, 0.27)$) -- ($(pys_wal)+(-1.2, 0.27)$);
        \draw[line:thick] ($(pys_wal)+(-1.2, -0.27)$) -- ($(pys_wal)+(-4.5, -0.27)$);
        
    \end{tikzpicture}
    }
    \caption{A cold cryptocurrency physical wallet. It manages keys offline and is indirectly connected to the blockchain network via a hot wallet to enhance the security of the keys.}
    \label{fig:lit_wallet}
\end{figure}

Blockchain users utilize \gls{cryp} wallets to store and track the keys securely. \Gls{cryp} wallets are devices or programs that generate, store, and manage public and private cryptographic keys \cite{suratkar2020cryptocurrency}. These wallets are considered \say{hot} when based online. Users use them to sign transactions, buy and sell \gls{cryp} assets, as well as generate and store cryptographic keys \cite{9705033}. Some notable examples of hot wallet applications include MetaMask \cite{meta}, Coinbase \cite{CISJHHF}, and Edge \cite{edgjsjsi}. Conversely, \say{cold} wallets generate, store, and manage keys offline. Also, the wallets do not directly interact with the user's requests to sign transactions; instead, they exchange keys with hot wallets, which in turn interact with the user's requests \cite{ khan2019security, dfgrewq}. Consequently, cold wallets are considered the safest type of \gls{cryp} wallets. \autoref{fig:lit_wallet} shows a common cold physical \gls{cryp} wallet. Physical wallets are termed \say{hardware wallets} in literature, not because they possess a hardware architecture but because they are physical devices. The figure shows that the wallet communicates with a hot wallet during utilization. Some common physical \gls{cryp} wallets include Trezor T by SatoshiLabs \cite{hkddooee}, Ledger Nano S by Ledger \cite{jppoerj}, and KeepKey by ShapeShift \cite{gsgbsr}.

 The \gls{cryp} wallets discussed above can either be \gls{nd} or \gls{hd} \cite{kim2020secure}. The former type generates one pair of corresponding private and public keys, while the latter uses one master key to generate almost infinite public and private keys. \gls{hd} wallets are currently the most popular type in the market due to their better key management property \cite{kim2020secure}.

 As previously discussed, \gls{cryp} keys, stored in \gls{cryp} wallets, are vital in securing user assets on the blockchain network. Consequently, the ability of \gls{cryp} wallets to securely generate, store, and manage \gls{cryp} keys is equally crucial. Several techniques in the literature propose different methods of enhancing the security of \gls{cryp} wallets.  For instance, CryptoVault is a platform that generates and maintains keys inside an Intel Software Guard Extension (SGX) enclave. This allows users to utilize private keys in a process highly isolated from other processes executing in the same environment \cite{lehto2021cryptovault}. Additionally, CryptoVault presents a secure approach for storing and retrieving a backup key from an external repository. Likewise, the secure blockchain lightweight wallet based on TrustZone (SBLWT) is a \gls{cryp} wallet that utilizes isolation to safeguard private keys \cite{dai2018sblwt}. The wallet is designed to secure simplified payment verifications (SPV) used in mobile devices that store partial blockchains due to resource constraints. Moreover, various \gls{cryp} wallets in the industry, such as  Ellipal Titan \cite{CISJHHFclelipd} and COLDCARD \cite{CISJHHFcld}, claim to utilize isolation (air-gap) technology to secure the keys. In contrast, the hot-cold hybrid decentralized exchange (HCHDEX) method stores \gls{cryp} wallet data locally on personal devices. It enables direct transactions between two devices with no dependence on a central server \cite{azman2020hch}. Furthermore, the HCHDEX method employs a secure two-way authentication technique, utilizing robust handshaking between e-wallets and lightweight distributed ledger technology (DLT) nodes. 
 
 It is worth noting that the literature works discussed above propose \gls{cryp} wallets that are software implementations running off \glspl{mcu}. This observation is also true for various market-leading \gls{cryp} wallets. For instance, COLDCARD \cite{CISJHHFcld}, Ledger Nano X \cite{jppoerj}, and Trezor Model T \cite{hkddooee} run on an STM32 \gls{mcu}. Therefore, the risk of malware attacks is often persistent \cite{ivanov2021ethclipper}. Also, attackers who gain physical access to the device can exploit physical attacks, such as the \gls{sca}, to retrieve private keys by leveraging unintended information leakage from power consumption, \gls{em} emissions, or timing variations \cite{park2024cloning}. In many attacks, adversaries target critical algorithms in the wallet, such as the \gls{ecc} and HMACSHA-512 algorithms, to extract the private keys.

 \Gls{ecc} is a fundamental algorithm used by \gls{cryp} wallets to generate public keys given private keys and to sign every transaction authorized by the owner. Moreover, it is the most complex and computationally intensive algorithm in a wallet. It is also sensitive to branching and conditional operations, which can lead to leakage of private data,  making it a prime target for attackers seeking to exploit vulnerabilities in the wallet's security \cite{renes2016complete}. Blockchains like Ethereum and Bitcoin utilize the SECP256K1 algorithm, an \gls{ecc} variant whose detailed explanation is provided in \autoref{subsec:pbkdf}. 
 
 Various security breaches are reported in the literature where SECKP256K1 within the \gls{cryp} wallets has been targeted. For example, private keys were extracted from a Trezor one wallet using a \gls{spa} attack \cite{park2023stealing, trezone, san2019side}. Furthermore, \gls{sca} was used to successfully attack the STM32 \gls{mcu} used by various commercially available wallets \cite{zhijian2019side, ngo2022side}. Also, an adversarial attack was modeled to extract private keys from an isolated wallet in seconds by infecting it with malicious code \cite{guri2018beatcoin}. As a result, wallet hacks play a significant role in the billions of dollars lost to \gls{cryp} theft \cite{cointelegraph, decrypt}.

Hence, a hardware architecture-based \gls{cryp} wallet can offer distinct advantages. It can integrate cryptographic operations directly into the hardware, rather than relying on general-purpose software environments. \Gls{fpga} devices are designed to function on a physical level, where configurations are essentially embedded into the hardware, resembling a \say{hard-wired} setup. This makes it extremely challenging for an attacker to alter the configuration of an \gls{fpga} implementation in a structured and predictable manner \cite{zhao2019sok}. This tailored approach not only isolates sensitive processes but also ensures that the wallet is executed in a dedicated, tamper-resistant environment, minimizing exposure to malware.

 Moreover, an \gls{ecc} algorithm secure against \gls{sca} attacks could further enhance security by preventing adversaries from extracting private keys or sensitive information through power and time analysis. Integrating an \gls{sca}-resistant \gls{ecc} within the wallet architecture ensures the protection of private keys even under physical proximity attacks, making it a robust solution for securing \gls{cryp} wallets. 

 \subsection{Contributions}
This work proposes EthVault, the first complete hardware architecture of an Ethereum \gls{hd} cold wallet, along with its \gls{fpga} implementation. Moreover, it introduces a SECP256K1 architecture resistant to \gls{sca} and the first hardware architecture of the \gls{ckd} function. Also, the design emphasizes minimal resource requirements for algorithms used in the wallet to meet the market demand for a small, portable \gls{cryp} wallet. The algorithms include:

\begin{itemize}
    \item The \gls{pa} protocol using complete addition equations.
    \item The Montgomery ladder algorithm using the \gls{pa} and the \gls{pd} to perform \gls{pm}.
    \item The \gls{bia} used to convert projective to affine coordinates.
    \item The \gls{hmac}-\gls{sha}-512 algorithm.
    \item The \gls{ckd} function utilized by the \gls{hd} wallet to generate child keys.
    \item The Ethereum checksum algorithm used to compute the Ethereum checksummed address.
    \item The \gls{pbkdf} used to generate mnemonics in \gls{hd} wallets.
    \item The \gls{ecdsa} for digitally signing transaction data.
\end{itemize}

To guide the design process, we set quantifiable design goals: a logic utilization of under 70\,k \glspl{lut}, and a minimum throughput of 10\,kbps sufficient for real-time signing. These targets were derived from our analysis of existing \gls{fpga} implementations of the cryptographic building blocks within the wallet, as well as the performance requirements of the current Ethereum blockchain, discussed in detail in \autoref{sec:impResults}.

%https://ieeexplore.ieee.org/stamp/stamp.jsp?arnumber=9614649

\subsection{Outline}
The subsequent sections of this work are structured as follows. \autoref{sec_relim} discusses key algorithms in an Ethereum \glsfirst{hd} wallet, notably, their functionality and role in the wallet. \autoref{sec: hw_arch} describes the hardware architecture of EthVault, including that of key algorithms, while \autoref{sec:impResults} discusses the implementation results on an \gls{fpga} Xilinx board. Moreover, \autoref{sec:Threats} outlines potential threats to the wallet and discusses possible mitigation techniques, whereas \autoref{sec:Limitations} addresses the limitations of EthVault and directions for future work. Finally, \autoref{sec:Conclusion} concludes this work with a summary.

\section{Preliminaries}
\label{sec_relim}

`This section delves into the workings of an Ethereum \gls{hd} wallet, focusing on the key algorithms that underpin its functionality. Each algorithm is discussed in detail to provide a thorough understanding of its operations and significance, to help readers gain a comprehensive understanding of how these components work together to enable the wallet’s features. \autoref{tab_nota} shows the notations used in this work.

The next section highlights the structural arrangement and functionality of an Ethereum \gls{hd} \gls{cryp} wallet.

% Table of notations
\begin{table}[t]
\centering
\caption{Summary of mathematical and logical notations used in this work.}
\begin{tabular}{ c c } 
 \hline
 \hline
 \textit{a}$\gg k$ & Shift $a$ to the right by $k$ bit\\
 $\lll k$ & Rotate left by $k$ bit\\
 $a \parallel b$ & $a$ is concatenated with $b$ \\
 $\oplus$ & Modulo-2 addition\\
 $\Rightarrow$ & Transforms to\\
 $\neq$ & Not equal to\\
 $e_x$ & An array of zeros, with a 1 at index $x$\\
 \hline
 \hline
\end{tabular}
\label{tab_nota}
\end{table}

\subsection{The Ethereum Hierarchically Deterministic Wallet}
\label{subsec:hd_wal}

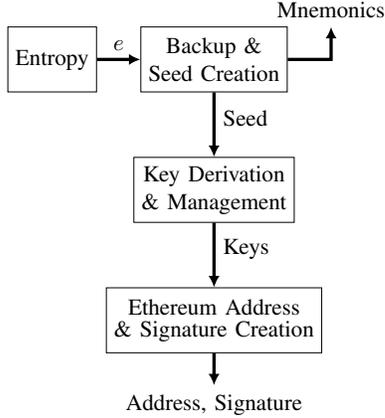
\begin{figure}[t]
\centering
    \resizebox{0.3\textwidth}{!}{
        \begin{tikzpicture}
    \node[rectangle, minimum width=1cm, minimum height=1cm, draw](RNG){Entropy};%rng
    \node[rectangle, minimum width=1.5cm, minimum height=1cm, draw](bip39) at ($(RNG.east)+(1.8,0)$){\shortstack{Backup \& \\ Seed Creation}}; %bip39
    \node[rectangle, minimum width=2cm, minimum height=1cm, draw](bip3239) at ($(bip39.south)+(0,-1.5)$){\shortstack{Key Derivation \\ \& Management}}; %bip3239
    \node[rectangle, minimum width=1.5cm, minimum height=1cm, draw](addr) at ($(bip3239.south)+(0, -1.5)$){\shortstack{Ethereum Address\\  \& Signature Creation}}; %bip39

    \draw[line:thick] (RNG)--node[pos=0.5, above]{$e$}(bip39);
    \draw[line:thick] (bip39)--node[right]{Seed}++(0,-1.3)--(bip3239);
    \draw[line:thick] (bip3239)--node[right]{Keys}++(0,-1.3)--(addr);
    \draw[line:thick] (addr)--($(addr)+(0, -1)$)node[below]{Address, Signature};
    \draw[line:thick] (bip39)-|($(bip39)+(1.8, 0.5)$)node[above]{Mnemonics}; 
\end{tikzpicture}
    }
    \caption{A high-level architecture of a \gls{hd} \gls{cryp} wallet. The wallet can generate Ethereum cryptographic keys, addresses, and signatures.}
    \label{eth_wallet} 
\end{figure}

The Ethereum \gls{hd} \gls{cryp} wallet operates through four main processes as illustrated in \autoref{eth_wallet}. The processes, which include entropy creation, human-readable backup and seed creation, key derivation and management, and blockchain address creation, work together to ensure the security, usability, and compatibility of the wallet with blockchain standards.

\begin{figure}[t]
    \resizebox{0.5\textwidth}{!}{
    \begin{tikzpicture}
        %nodes
         \node[ckd_node, pattern=dots, label={$e$}](s) at (0,0) {}; %s
         \node[anchor=north] at (s.south) {\shortstack{Entropy\\source}};
         \node[] at ($(s)+(-0.3,4.5)$) {\shortstack{Master \\ Seed}}; 
         \node[ckd_node, label={$m$}](m) at ($(s)+(1.5,0)$) {}; %m
         \node[] at ($(m)+(0,4.8)$) {\shortstack{Depth=0}}; 
         \node[] at ($(m)+(1.8,4.8)$) {\shortstack{Depth=1}};
         \node[] at ($(m)+(3.5,4.8)$) {\shortstack{Depth=2}};
         \node[] at ($(m)+(5.8,4.8)$) {\shortstack{Depth=3}};
         \node[ckd_node, label={$m/0$}](m0) at ($(m)+(2,2.5)$) {}; %m0
         \node[ckd_node, label={$m/1$}](m1) at ($(m)+(2,0.5)$) {}; %m1
         \node[ckd_node, label={$m/i$}](mi) at ($(m)+(2,-2.5)$) {}; %mi
         \node[branch](b1)at ($(m1)+(0,-0.8)$){};
         \node[branch](b2)at ($(b1)+(0,-0.5)$){};
         \node[branch](b3)at ($(b2)+(0,-0.5)$){};
         \node[ckd_node, label={$m/0/0$}](m00) at ($(m0)+(1.5,0.5)$) {}; %m00
         \node[ckd_node, label={$m/0/1$}](m01) at ($(m0)+(1.5,-0.5)$) {}; %m01
         \node[ckd_node, label={$m/1/0$}](m10) at ($(m1)+(1.5,0.5)$) {}; %m10
         \node[ckd_node, label={$m/1/1$}](m11) at ($(m1)+(1.5,-0.5)$) {}; %m11
         \node[ckd_node, label={$m/i/0$}](mi0) at ($(mi)+(1.5,0.5)$) {}; %mi0
         \node[ckd_node, label={$m/i/1$}](mi1) at ($(mi)+(1.5,-0.5)$) {}; %mi1
         \node[ckd_node, label={$m/0/0/0$}](m000) at ($(m00)+(1.5,0.8)$) {}; %m000
         \node[ckd_node, label={$m/0/0/k$}](m00k) at ($(m000)+(1.5,0)$) {}; %m00k
         \node[ckd_node, label={$m/0/1/0$}](m010) at ($(m01)+(1.5,0.3)$) {}; %m010
         \node[ckd_node, label={$m/0/1/k$}](m01k) at ($(m010)+(1.5,0)$) {}; %m01k
         \node[ckd_node, label={$m/i/1/0$}](mi10) at ($(mi1)+(1.5,0.5)$) {}; %mi10
         \node[ckd_node, label={$m/i/1/k$}](mi1k) at ($(mi10)+(1.5,0)$) {}; %mi1k
         \node[branch](m110)at ($(m11)+(2.3,0.5)$){};%m011...
         \node[branch](m111)at ($(m110)+(0,-0.5)$){};
         \node[branch](m112)at ($(m111)+(0,-0.5)$){};
         \node[branch](m0010)at ($(m000)+(0.5,0)$){};%m000...
         \node[branch](m0011)at ($(m0010)+(0.2,0)$){};
         \node[branch](m0012)at ($(m0011)+(0.2,0)$){};
         \node[branch](m0110)at ($(m010)+(0.5,0)$){}; %m010...
         \node[branch](m0111)at ($(m0110)+(0.2,0)$){};
         \node[branch](m0112)at ($(m0111)+(0.2,0)$){};
         \node[branch](mi100)at ($(mi10)+(0.5,0)$){}; %mi10...
         \node[branch](mi101)at ($(mi100)+(0.2,0)$){};
         \node[branch](mi102)at ($(mi101)+(0.2,0)$){};

         %lines
         \draw[line:thin] (s)--node[pos=0.3, above, xshift=0.2cm]{HMC}(m); %gls{hmac}\gls{sha}-512
         \draw[line:thin] (m)--node[left]{\gls{ckd}}(m0);
         \draw[line:thin] (m)--node[above]{\gls{ckd}}(m1);
         \draw[line:thin] (m)--node[left]{\gls{ckd}}(mi);
         \draw[line:thin] (m0)--node[above]{\gls{ckd}}(m00);
         \draw[line:thin] (m0)--node[below, xshift=-1mm]{\gls{ckd}}(m01);
         \draw[line:thin] (m1)--node[above]{\gls{ckd}}(m10);
         \draw[line:thin] (m1)--node[below, xshift=-1mm]{\gls{ckd}}(m11);
         \draw[line:thin] (mi)--node[above]{\gls{ckd}}(mi0);
         \draw[line:thin] (mi)--node[below, xshift=-1mm]{\gls{ckd}}(mi1);
         \draw[line:thin, bend right] (mi1) to (mi10.south);
         \draw[line:thin, bend right=30] (mi1) to (mi1k.south);
         \draw[line:thin, bend right] (m01) to (m010.south);
         \draw[line:thin, bend right=30] (m01) to (m01k.south);
         \draw[line:thin, bend right=15] (m00) to (m000.south);
         \draw[line:thin, bend right=15] (m00) to (m00k.south);
         \draw[dashed] ($(s)+(0.7,5)$) to ($(s)+(0.7,-4)$);
        \draw[dashed] ($(m)+(1,5)$) to ($(m)+(1,-4)$);
        \draw[dashed] ($(m)+(2.7,5)$) to ($(m)+(2.7,-4)$);
        \draw[dashed] ($(m)+(4.3,5)$) to ($(m)+(4.3,-4)$);   
\end{tikzpicture}
    }
    \caption{The \gls{hd} wallet structure for a 3-level key derivation path (e.g $m/0'/0'/k'$) as outlined by the \gls{bip}-32 standard, adapted from \cite{bip32}.}
    \label{fig:bip32}
\end{figure}
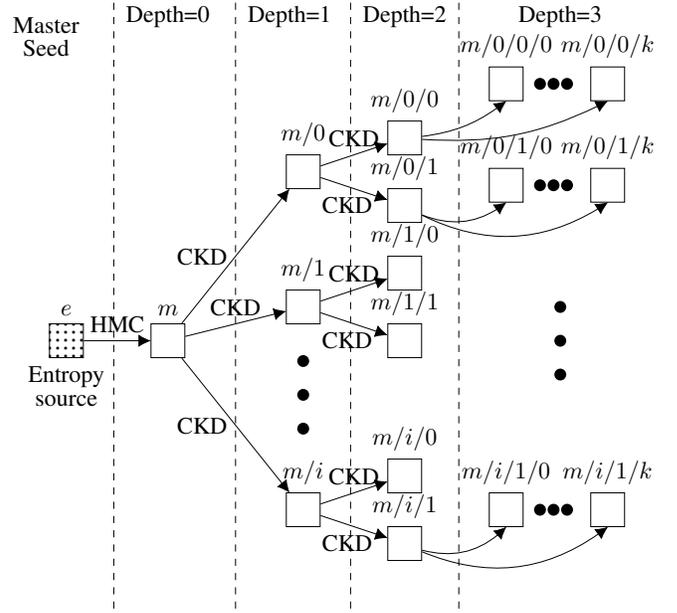

\subsubsection{Entropy Creation} In an Ethereum \gls{hd} \gls{cryp} wallet, the process begins with an \gls{rng} that generates a random value $e$, which serves as the source of entropy for creating the master key $m$. A high-entropy \gls{rng} is crucial, as it ensures a more secure master key. 

\subsubsection{Backup and Seed Creation} Next, the \gls{bip}-39 protocol utilizes $e$, an optional \gls{pin} input from the user, and the phrase “mnemonics” in \gls{ascii} format to produce random mnemonic phrases and a 512-bit seed number \cite{bip39}. This protocol employs the \gls{pbkdf}, a function based on the \gls{hmac}\gls{sha}-512 and \gls{sha}-256 hash algorithms. The mnemonic phrases are 12 or 24 randomly selected human-readable words that serve as a backup and recovery mechanism for \gls{hd} wallet keys and addresses \cite{bip39}.

\subsubsection{Key Derivation and Management} In this part, the seed value and the phrase “Bitcoin seed" in \gls{ascii} format are used as inputs to the \gls{hmac}\gls{sha}-512 function to generate the master private key $m$ and chain code $c$. Following the \gls{bip}-32 and \gls{bip}-44 standards, the \gls{ckd} function uses $m$ and $c$ alongside \gls{hmac}\gls{sha}-512 and SECP256K1 algorithms to derive child private and public keys. 

The \gls{bip}-32 standard serves as the foundation for all \gls{hd} wallets, defining a deterministic structure for generating child private and public keys from a single master key \cite{bip32}. \autoref{fig:bip32} illustrates the hierarchical structure of \gls{bip}-32-based \gls{hd} wallets. At the root (depth zero), \gls{hmac}\gls{sha}-512 (denoted by HMC ) utilizes a random number $e$ to generate the master key $m$. Moreover, a \gls{ckd} function is used to generate the other nodes of the hierarchical tree from depth one to three. Notably, the nodes at depth three correspond to blockchain addresses.

Building on \gls{bip}-32, the \gls{bip}-44 standard provides a practical application tailored to various \glspl{cryp} \cite{bip44}. It defines a specific path consisting of five hierarchical levels, each comprising constants and variables executed sequentially within the \gls{bip}-32 framework. These unique paths allow \gls{bip}-44 protocol to support multiple \glspl{cryp}, ensuring compatibility across different blockchain ecosystems.

\begin{figure*}[t]
\centering
    \resizebox{1\textwidth}{!}{
        \begin{tikzpicture}
     \node (hmacmk) [rectangle, minimum width=1cm, minimum height=1cm, draw]{HMAC};
     \node (sha512mk) [rectangle, minimum width=1.5cm, minimum height=1cm, draw] at ($(hmacmk.south)+(0, -0.6)$){SHA512};
     \draw[dashed] ($(hmacmk.north)+(-1.2,0.7)$) -- ($(hmacmk.south)+(-1.2,-1.5)$);
     \draw[dashed] ($(hmacmk.north)+(1.2,0.7)$) -- ($(hmacmk.south)+(1.2,-1.5)$);
     \node at ([yshift=0.5cm]hmacmk.north) {\shortstack{\textbf{Master key}\\ \textbf{and Chaincode}}};
     
     \node (hmacma) [rectangle, minimum width=1cm, minimum height=1cm, draw] at ($(hmacmk.west)+(-1.5, 0)$) {HMAC};
     \node (sha512ma) [rectangle, minimum width=1.5cm, minimum height=1cm, draw] at ($(hmacma.south)+(-0.8, -0.6)$){SHA512};
     \node (sha256ma) [rectangle, minimum width=1.5cm, minimum height=1cm, draw] at ($(hmacma.west)+(-0.85, 0)$){SHA256};
     \draw[dashed] ($(sha256ma.north)+(-1.2,0.7)$) -- ($(sha256ma.south)+(-1.2,-1.5)$);
     \node at ([xshift=0.5cm, yshift=0.6cm]sha256ma.north) {\textbf{Mnemonic and Seed}};

     \node (rng) [rectangle, minimum width=1cm, minimum height=1cm, draw] at ($(sha256ma.west)+(-1.3, -0.5)$) {RNG};
     \node at ([yshift=1cm]rng.north) {\textbf{Entropy}};

     \node (hmaccp) [rectangle, minimum width=1cm, minimum height=1cm, draw] at ($(hmacmk.east)+(1.5, 0)$) {HMAC};
     \node (sha512cp) [rectangle, minimum width=1.5cm, minimum height=1cm, draw] at ($(hmaccp.south)+(0.8, -0.6)$){SHA512};
     \node (secpcp) [rectangle, minimum width=2cm, minimum height=1cm, draw] at ($(hmaccp.east)+(1.1, 0)$){SECP256K1};
     \draw[dashed] ($(secpcp.north)+(1.3,0.7)$) -- ($(secpcp.south)+(1.3,-1.5)$);
     \node at ([xshift=-0.5cm, yshift=0.5cm]secpcp.north) {\shortstack{\textbf{Child Private}\\ \textbf{and Public Keys}}};
     \node at ([xshift=0.1cm, yshift=-0.2cm]sha512cp.south) {Executed $n$ times};

     \node (kecc256) [rectangle, minimum width=2cm, minimum height=1cm, draw] at ($(secpcp.east)+(1.7, -0.5)$){KECCAK256};
     \draw[dashed] ($(secpcp.north)+(4.2,0.7)$) -- ($(secpcp.south)+(4.2,-1.5)$);
     \node at ([yshift=1cm]kecc256.north) {\textbf{Address}};
     \node at ([yshift=-0.7cm]kecc256.south) {Executed $n$ times};

     \node (secps) [rectangle, minimum width=2cm, minimum height=1cm, draw] at ($(kecc256.east)+(1.7, 0)$){SECP256K1};
     \node at ([yshift=1cm]secps.north) {\textbf{Signature}};
     
\end{tikzpicture}
    }
    \caption{Execution flow of the Ethereum \gls{hd} wallet, illustrating the sequential stages from entropy generation to transaction signing, along with the cryptographic algorithms applied at each stage. $n$ is the number of child keys needed by the user.}
    \label{fig:allAlgs} 
\end{figure*}
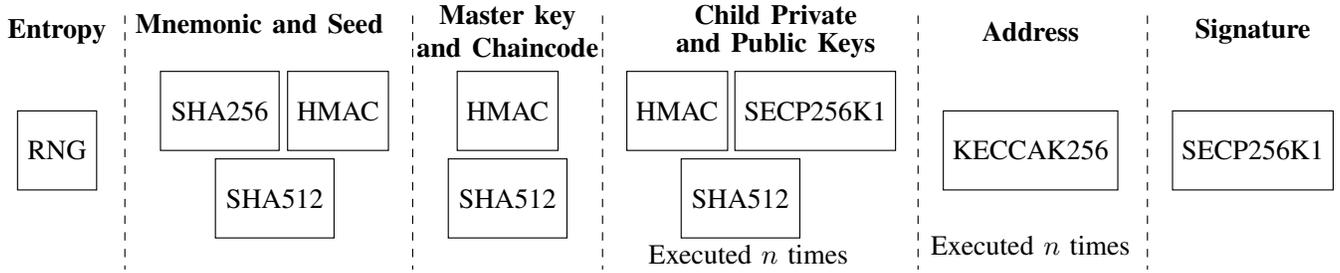

The \gls{bip}-44 path $m$/ $purpose'$/ $ coin\_type'$/ $account'$/ $change$/ $address\_index$, contains the root $m$, representing the master private key and the five levels. Moreover, $address\_index$ is a 32-bit variable defining the index of the key generated by the wallet. The remaining levels are 32-bit constants unique to different \glspl{cryp}. For instance, Ethereum follows the path $m/44'/60'/0'/0/address\_index$. The apostrophe ($'$) in the path indicates the use of the \gls{bip}-32 hardened derivation method within the \gls{ckd} function.

Hardened and non-hardened key derivation methods, as defined in \gls{bip}-32, provide users with the flexibility to balance trade-offs between security, backup and recovery, and transaction convenience \cite{bip32}. Non-hardened keys are calculated as taking the \gls{hmac}\gls{sha}-512 hash of (Parent private key $\|$ Index), where index $\geq 2^{31}$, wheres hardened keys are calculated as taking the \gls{hmac}\gls{sha}-512 hash of (Parent public key $\|$ Index), where index $< 2^{31}$ \cite{bip39}. As shown in \autoref{fig:bip32}, the \gls{bip}-44 path also facilitates the precise location of blockchain addresses within a \gls{hd} wallet structure.

\subsubsection{Ethereum Address and Signature Creation} At this stage, the address-generation process applies the Keccak hash function to create uniquely checksummed Ethereum addresses. This sequence ensures both security and adherence to cryptographic standards when generating \gls{cryp} keys. Furthermore, the wallet employs the \gls{ecdsa} algorithm to digitally sign and authorize transactions using the derived child private keys.

This section outlined the stages of the Ethereum \gls{hd} wallet and the cryptographic algorithms employed in each step. \autoref{fig:allAlgs} expands this process into six distinct sequential stages based on their outputs and specifies the algorithms applied at each stage. The first three stages are executed once for every new entropy generated by the \gls{rng}, while the child private–public key and address stages are repeated $n$ times, where $n$ is the number of child keys required by the user. The signature stage is executed each time a transaction is authorized. EthVault implements all stages and is designed to efficiently reuse the algorithms to minimize resource utilization.

The following section introduces the SECP256K1 algorithm, a key cryptographic element used in Ethereum wallets.

\subsection{The SECP256K1 Algorithm}
\label{sec:secp}
This subsection discusses the SECP256K1 elliptic curve, its arithmetic properties, and its vulnerabilities to \gls{sca} in the context of scalar multiplication.

\subsubsection{Elliptic Curve Parameters}

SECP256K1 is a specific \gls{ec} among the diverse variants used in \gls{ecc}. These variants include the Weierstrass, Edwards, Hessian, and Koblitz curves \cite{ashraf2012alternate}. Moreover, \gls{ecc} is a form of public key cryptography based on an \gls{ec} over a finite \gls{gf} \cite{pirotte2019balancing}. There are two main types of finite fields commonly used in \gls{ecc}: prime fields, denoted by $GF(\mathbb{F}_p)$, where $p$ is a large prime number and binary extension fields, denoted by $GF( \mathbb{F}_{2^m})$, where $2^m$ is the number of elements in the field and $m$ is a positive integer \cite{pirotte2019balancing}. The \gls{ec} is defined by the cubic equation:

\begin{equation}
y^2 + xy = x^3 + ax + b,
\label{eq_metr_dscf}
\end{equation}
where $x$ and $y$ are coordinates on the \gls{ec}, and $a$ and $b$ are constants that define the curve. After a linear change of variables, \eqref{eq_metr_dscf} is transformed into \eqref{eq_2}, expressed in standard short Weierstrass form. It returns a public key solution comprising $(x, y)$ for variables $a$, $b$ in the \gls{gf}\cite{kapoor2008elliptic}.
\begin{equation}
y^2 = x^3 + ax + b.
\label{eq_2}
\end{equation}

 \subsubsection{Elliptic Curve Operations} 
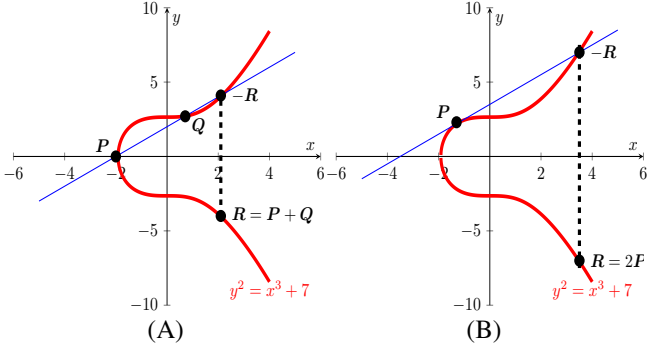
\begin{figure}[t]
\centering
    \resizebox{0.5\textwidth}{!}{
        \begin{tikzpicture}[xscale=0.6, yscale=0.7]
    \node[](a) at (3.4,-0.5){(A)};
    \node[](a) at (10.5,-0.5){(B)};
    \begin{axis}[ xmin=-6, xmax=6, ymin=-10, ymax=10,
    axis lines=middle, xlabel=$x$, ylabel=$y$]
        \addplot[line width=2pt, color=red, samples=1001, domain=-4:4]{sqrt(x^3 + 7)};
        \addplot[line width=2pt, color=red, samples=1001, domain=-4:4]{-sqrt(x^3 + 7)};
        \addplot[color=blue, no markers]{x+2};
        \addplot +[color=black, dashed, mark=none, line width=2pt] coordinates {(2.1, -4) (2.1, 4)};
        \node[circle, fill, scale=0.7, color=black,label={[xshift=0.1cm,yshift=0.2cm]left:$\bm{P}$}] at (axis cs:-2,0){};
        \node[circle, fill, scale=0.7, color=black,label={[xshift=-0.1cm,yshift=-0.2cm]right:$\bm{Q}$}] at (axis cs:0.71,2.7){};
        \node[circle, fill, scale=0.7, color=black,label=right:$-\bm{R}$]at (axis cs:2.1,4.1){};
        \node[circle, fill, scale=0.7, color=black,label={right:$\bm{R}=\bm{P}+\bm{Q}$}] at (axis cs:2.1,-4){};
        \node[text=red](eq) at (4,-9){$y^2=x^3+7$};
        %\addplot[color=blue, samples=50, domain=-4:4, dashed, mark=*]{1-x^2};
    \end{axis}
    
 \begin{scope}[xshift=7.2cm]
    \begin{axis}[ xmin=-6, xmax=6, ymin=-10, ymax=10,
    axis lines=middle, xlabel=$x$, ylabel=$y$]
        \addplot[line width=2pt, color=red, samples=1001, domain=-4:4]{sqrt(x^3 + 7)};
        \addplot[line width=2pt, color=red, samples=1001, domain=-4:4]{-sqrt(x^3 + 7)};
        \addplot[color=blue, no markers]{x+3.5};
        \addplot +[color=black, dashed, mark=none, line width=2pt] coordinates {(3.5, -7.5) (3.5, 7.5)};
        \node[circle, fill, scale=0.7, color=black,label={[xshift=0.1cm,yshift=0.2cm]left:$\bm{P}$}] at (axis cs:-1.3,2.3){};
        \node[circle, fill, scale=0.7, color=black,label=right:$-\bm{R}$]at (axis cs:3.5,7){};
        \node[circle, fill, scale=0.7, color=black,label={right:$\bm{R}=2\bm{P}$}] at (axis cs:3.5,-7){};
        \node[text=red](eq) at (4,-9){$y^2=x^3+7$};
        %\addplot[color=blue, samples=50, domain=-4:4, dashed, mark=*]{1-x^2};
    \end{axis} 
 \end{scope}
\end{tikzpicture}
    }
\caption{ (A) \gls{pa} is the addition of two points ($\bm{P}$ and $\bm{Q}$) on the elliptic curve. (B) \gls{pd} is the addition of a point $\bm{P}$ on the elliptic curve with itself. Adapted from \cite{kapoor2008elliptic}.}
\label{fig:papd}
\end{figure}

\Glsfirst{pa} and \glsfirst{pd} are arithmetic operations used to compute public keys on the \gls{ec} \cite{panchbhai2015implementation}. \gls{pa} defines adding two points on the curve, as shown in \autoref{fig:papd} (A). Given points $\bm{P}=(x_0, y_0)$ and $\bm{Q}=(x_1, y_1)$ on the \gls{ec}, \gls{pa} comprises two processes. First, draw a straight line through points $\bm{P}$ and $\bm{Q}$. The line intersects with the curve at point $-\bm{R}=(x_2, y_2)$. Second, reflect the point $\bm{-R}$ by the x-axis to obtain the results of the \gls{pa} as shown in \eqref{eq:pa}. 

\begin{equation}
\bm{R} = \bm{P} + \bm{Q}.
\label{eq:pa}
\end{equation}

Conversely, \gls{pd} defines adding a point on the \gls{ec} with itself, as shown in \autoref{fig:papd} (B). Given point  $\bm{P}=(x_0, y_0)$ on the curve, \gls{pd} also comprises two steps. First, draw a tangential line to the curve at point $\bm{P}$. The line intersects with the curve at the point $-\bm{R}=(x_1, y_1)$. Second, reflect the point $\bm{-R}$ by the x-axis to obtain the results of the \gls{pd} as shown in \eqref{eq:pd}

\begin{equation}
\bm{R} = 2\bm{P}.
\label{eq:pd}
\end{equation}

\Gls{pa} and \gls{pd} are used to compute the scalar \glsfirst{pm} on the \gls{ec}. Scalar \gls{pm} is an integral \gls{ecc} operation as it is the primary process used to calculate the public key. Scalar \gls{pm} has the form $\bm{R} = k\cdot\bm{P}$. It is the sum of $k$ copies of $\bm{P}$, such that:
\begin{equation}
\bm{R} = k\cdot\bm{P} = \sum_{i=1}^{k}\bm{P},
\label{eq:pmult}
\end{equation}
where $k$ is a positive integer, and $\bm{R}$ and $\bm{P}$ is a points on the curve. This work will use the Montgomery Ladder algorithm to compute \gls{pm} \cite{pirotte2019balancing}. \autoref{sec: hw_arch} further explains the Montgomery Ladder algorithm.

The Koblitz Curve \gls{ecc} variant over GF($\mathbb{F}_p$), known as the \gls{secp}, is an \gls{ec} whose $a$ and $b$ parameters of \eqref{eq_2} are 0 and 7, respectively \cite{renes2016complete}. Moreover, other parameters such as the generator $\bm{G}$, synonymous to $\bm{P}$ in \eqref{eq:pmult}, and the base point of $\bm{G}$ denoted as $n$ are specified. SECP256K1 is the core algorithm used by the Ethereum \gls{cryp} wallet to generate a public key from a private key and sign transactions.

\subsubsection{Side Channel Attack Attack on SECP256K1} In protocols that utilize \gls{ecc}, $k$ in \eqref{eq:pmult} is usually considered a private key. Hence, a successful attack correctly derives $k$ via unauthorized means. For example, an \gls{sca} can exploit the current drawn or \gls{em} waves emitted by an \gls{ecc} device while processing $k$. The attacks rely on the variations in power consumption when bit value 1 or 0 of $k$ is processed (i.e., $k_i$ where $i$ is the index) \cite{park2023stealing, trezone, san2019side}. 

 The Montgomery Ladder algorithm depicted in \autoref{alg:montg} is popularly used to calculate \gls{pm} \cite{pirotte2019balancing}. It details the bitwise processing of the secret key $k$ from \gls{msb} to \gls{lsb}. The algorithm is balanced because the sequence of mathematical operations is independent of the private key. Hence, the literature considers the algorithm safe against simple \gls{sca} attacks \cite{kabin2020breaking}. Nevertheless, the algorithm still contains inconsistencies that potentially make it susceptible to \gls{dpa} and timing attacks.

 The \gls{pd} in each branch of the \textit{if} statement is performed on different registers. When $k_i$ is 1, \gls{pd} is performed on $\bm{R_1}$. Conversely, when $k_i$, is 0 \gls{pd} is performed on $\bm{R_0}$. These differences create distinct power consumption patterns and execution time discrepancies due to the use of different registers, memory locations, and data paths. Variations in power consumption, delays, and propagation times among these hardware resources can be exploited by attackers to extract the scalar $k$ \cite{cointelegraph,decrypt,kabin2020breaking}. Moreover, the conventional Weierstrass \gls{ec} addition operation involves branching when performing \gls{pa}, \gls{pd}, or handling a point at infinity. The branching introduces timing variability, which can be exploited to compromise the secret key \cite{renes2016complete}.
 
 Various works in literature have proposed ways to protect the Montgomery Ladder algorithm against \gls{sca}. The work in \cite{pirotte2018design} proposed a method to randomize the sequence of writing $\bm{Q_0}$ and $\bm{Q_1}$ inside the loop. However, the addressing did not change, making the risk of \gls{sca} persistent. Moreover, using complete addition formulas removes the branching in the Weierstrass \gls{ec} addition operation \cite{renes2016complete}. However, since the Montgomery Ladder algorithm is vulnerable to \gls{sca}, employing the equations still makes the threat prevalent. 
 
 This work employs temporary registers, parallel processing, and complete \gls{pa} formulas to prevent variations during \gls{pm}. A detailed explanation of the proposed algorithm is provided in \autoref{subsec:arch_secp}. 
 
The following section presents the \gls{bia} as implemented within the SECP256K1 \gls{ec} cryptography scheme.
%Montgomery ladder algorithm
 \begin{algorithm}[t]
 \caption{Montgomery ladder algorithm, adapted from \cite{montgomery1987speeding}.}
 \label{alg:montg}
 \begin{algorithmic}[1]\small
 \renewcommand{\algorithmicrequire}{\textbf{Input:}}
 \renewcommand{\algorithmicensure}{\textbf{Output:}}
 \REQUIRE $\bm{P} \in (x, y, z), k=(k_{t-1}, \cdots, k_0)$ with $k_{t-1}=1$
 \ENSURE  $\bm{R}=k\bm{P}$
 \\ \textit{Initialisation} :
  \STATE $\bm{R_0} \leftarrow \bm{P}$ \label{montg:init-r0}
  \STATE $\bm{R_1} \leftarrow 2\bm{P}$ \label{montg:init-r1}
 \\ \textit{LOOP Process} :
  \FOR {$i = t-2$ : $0$}\label{montg:for-st}
  \IF {$k_i = 1$}\label{montg:if-st}
    \STATE $\bm{R_0} \leftarrow \bm{R_0} + \bm{R_1}$\label{montg:if-st-r0}
    \STATE $\bm{R_1} \leftarrow 2\bm{R_1}$\label{montg:if-st-r1}
  \ELSE
    \STATE $\bm{R_1} \leftarrow \bm{R_0} + \bm{R_1}$\label{montg:if-else-r1}
    \STATE $\bm{R_0} \leftarrow 2\bm{R_0}$\label{montg:if-else-r0}
  \ENDIF
  \ENDFOR \label{montg:end-for}
  \STATE $\bm{R} \leftarrow \bm{R_0}$ \label{montg:return}
  \RETURN $\bm{R}$
 \end{algorithmic} 
 \end{algorithm}

\subsection{The Binary Inversion Algorithm}

The \glsfirst{bia} computes the multiplicative inverse of elements in an \gls{ecc}’s finite field. In SECP256K1, for instance, it can convert the projective coordinates back to the affine coordinate system. We analyze the arithmetic operations performed in the utilized \gls{ecc} to understand the significance of \gls{bia}.

SECP256K1 executes modular arithmetic operations, including addition, subtraction, multiplication, and division in an affine coordinate system, i.e., GF($\mathbb{F}_p$) where $ \mathbb{F}_p \in (x,y)$ \cite{panchbhai2015implementation}. However, modular inversion/division is the most expensive in complexity, area, and execution time \cite{guo2023efficient, hossain2015high}. Nevertheless, transforming the coordinates from affine to projective reduces the number of modular division operations performed by SECP256K1 \cite{panchbhai2015implementation} (i.e., GF($\mathbb{F}_p$) where $ \mathbb{F}_p \in (x,y,z)$). 

Therefore, \autoref{alg:compEqs} depicts a set of equations used to compute the complete \gls{pa} in the projective coordinate system over prime-order elliptic curves \cite{renes2016complete}. However, SECP256K1 must perform one final modular division to return the final results to affine coordinates, i.e $(x, y, z) \Rightarrow (xz^{-1}, yz^{-1})$. SECP256K1 utilizes the \gls{bia} to compute the modular inversion $z^{-1}$. The algorithm shown in \autoref{alg:binInv} is based on the \gls{eea} which calculates the multiplicative inverse of an integer $z \in \mathbb{F}_p$ by calculating two variables $r$ and $q$ that satisfy:
\begin{equation}
zr + pq = \text{gcd}(z, p) = 1,
\label{eq_utq}
\end{equation}
where gcd is a function used to calculate the greatest common divisor of two numbers \cite{hossain2015high}. 

The following section introduces the \gls{pbkdf} algorithm used in the human-readable backup and seed creation part of the wallet.

% Complete projective equations
\begin{algorithm}[t]
 \caption{Equations for complete, projective \gls{pa} for SECP256K1. Taken from \cite{renes2016complete}.}
 \label{alg:compEqs}
 \begin{algorithmic}[1]\small
 \renewcommand{\algorithmicrequire}{\textbf{Input:}}
 \renewcommand{\algorithmicensure}{\textbf{Output:}}
 \REQUIRE $\bm{P} = (X_1, Y_1, Z_1), \bm{Q} = (X_2, Y_2, Z_2)$ on $E : Y^2Z = X^3 + bZ^3$ and $b_3 = 3 \cdot b.$
 \ENSURE  $(X_3, Y_3, Z_3) = \bm{P} + \bm{Q}$;
  \begin{multicols}{3}
    \STATE $t_0 \leftarrow X_1 \cdot X_2$
    \STATE $t_1 \leftarrow Y_1 \cdot Y_2$
    \STATE $t_2 \leftarrow  Z_1 \cdot Z_2$
    \STATE $t_3 \leftarrow  X_1 + Y_1$
    \STATE $t_4 \leftarrow X_2 + Y_2$
    \STATE $t_3 \leftarrow t_3 \cdot t_4$
    \STATE $t_4 \leftarrow t_0 + t_1$
    \STATE $t_3 \leftarrow t_3 - t_4$
    \STATE $t_4 \leftarrow Y_1 + Z_1$
    \STATE $X_3 \leftarrow Y_2 + Z_2$ 
    \STATE $t_4 \leftarrow t_4 \cdot X_3$ 
    \STATE $X_3 \leftarrow t_1 + t_2$
    \STATE $t_4 \leftarrow t_4 - X_3$  
    \STATE $X_3 \leftarrow X_1 + Z_1$ 
    \STATE $Y_3 \leftarrow X_2 + Z_2$
    \STATE $X_3 \leftarrow X_3 \cdot Y_3$  
    \STATE $Y_3 \leftarrow t_0 + t_2$ 
    \STATE $Y_3 \leftarrow X_3 - Y_3$
    \STATE $X_3 \leftarrow t_0 + t_0$  
    \STATE $t_0 \leftarrow X_3 + t_0$ 
    \STATE $t_2 \leftarrow b_3 \cdot t_2$
    \STATE $Z_3 \leftarrow t_1 + t_2$  
    \STATE $t_1 \leftarrow t_1 - t_2$ 
    \STATE $Y_3 \leftarrow b_3 \cdot Y_3$
    \STATE $X_3 \leftarrow t_4 \cdot Y_3$  
    \STATE $t_2 \leftarrow t_3 \cdot t_1$ 
    \STATE $X_3 \leftarrow t_2 - X_3$
    \STATE $Y_3 \leftarrow Y_3 \cdot t_0$  
    \STATE $t_1 \leftarrow t_1 \cdot Z_3$ 
    \STATE $Y_3 \leftarrow t_1 + Y_3$
    \STATE $t_0 \leftarrow t_0 \cdot t_3$  
    \STATE $Z_3 \leftarrow Z_3 \cdot t_4$ 
    \STATE $Z_3 \leftarrow Z_3 + t_0$
\end{multicols}
 \end{algorithmic}
 \end{algorithm}

% BIA
\begin{algorithm}[t]
 \caption{Binary Inversion Algorithm. Adapted from \cite{hossain2015high}.}
 \label{alg:binInv}
 %\begin{multicols}{2}
 \begin{algorithmic}[1]\small
 \renewcommand{\algorithmicrequire}{\textbf{Input:}}
 \renewcommand{\algorithmicensure}{\textbf{Output:}}
 \REQUIRE ${z} \in [1, {p}], {p}$
 \ENSURE  ${r}$ = ${z}^{-1}$ mod  ${p}$
 \\ \textit{Initialisation} :
  \STATE ${u} \leftarrow {z}; {v} \leftarrow {p}; {x} \leftarrow e_0; {y} \leftarrow {0}$
 \\ \textit{LOOP Process} :
 \WHILE{${u} \neq 0$}  \label{otl:0}
    \WHILE{${u}(0) = 0$} \label{inl:0}
        \STATE$ {u} \leftarrow {u} \gg 1 $
        \STATE $\textbf{if}\ x(0)=0\ \textbf{then}\ {x} \leftarrow {x} \gg 1 \ \textbf{else}\ {x} \leftarrow ({x}+{p}) \gg 1\ \textbf{end if}$
    \ENDWHILE \label{inl:1}
    \WHILE{${v}(0) = 0$} \label{inl:2}
        \STATE$ {v} \leftarrow {v} \gg 1 $
        \STATE$\textbf{if}\ {y}(0)=0\ \textbf{then}\ {y} \leftarrow {y} \gg 1\ \textbf{else}\ {y} \leftarrow ({y} + {p}) \gg 1\ \textbf{end if}$
    \ENDWHILE  \label{inl:3}
    \IF{${u} \geq {v}$}
     \STATE ${u} \leftarrow {u} - {v}$
        \STATE $\textbf{if}\ {x} > {y}\ \textbf{then}\ {x} \leftarrow {x} - {y}\ \textbf{else}\ {x} \leftarrow {x} + {p} - {y}\ \textbf{end if}$
    \ELSE
    \STATE ${v} \leftarrow {v} - {u}$
    \STATE $\textbf{if}\ {y} > {x}\ \textbf{then}\ {y} \leftarrow {y} - {x}\ \textbf{else}\ {y} \leftarrow {y} + {p} - {x}\ \textbf{end if}$ 
    \ENDIF
 \ENDWHILE \label{otl:1}
 \STATE$\textbf{if}\ {u}=1\ \textbf{then}\ {r} \leftarrow \bmod({x}, {p}) \textbf{else}\ {r} \leftarrow \bmod({y}, {p})\ \textbf{end if}$
 \RETURN ${r}$
 \end{algorithmic}
 %\end{multicols}
 \end{algorithm}

\subsection{The Password-based Key Derivation Function-2}
\label{subsec:pbkdf}

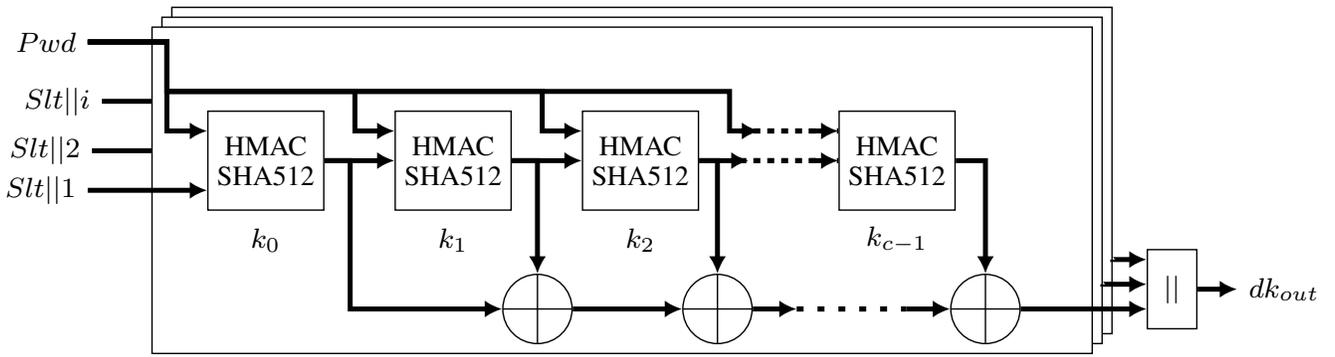
\begin{figure*}[t]
\centering
    \resizebox{1\textwidth}{!}{
    \footnotesize
\begin{tikzpicture}
%nodes
    \node[fill=white, rectangle, minimum width=9.5cm, minimum height=3.3cm, draw] at (3.8, -0.1) (rec2){};%rec2
    \node[fill=white, rectangle, minimum width=9.5cm, minimum height=3.3cm, draw] at (3.7, -0.2) (rec1){};%rec1
        \draw[line:thick] ($(rec2.west)+(-0.7, 0.7)$)node[left]{$Slt || i$}-|($(rec2.west)+(0, 0.7)$);
        \draw[line:thick] ($(rec1.west)+(-0.7, 0.3 )$)node[left]{$Slt || 2$}-|($(rec1.west)+(0, 0.3)$);
    \node[fill=white, rectangle, minimum width=9.5cm, minimum height=3.3cm, draw] at (3.6, -0.3) (rec0){};%rec0
    \node[rectangle, minimum width=1cm, minimum height=1cm, font=\footnotesize, draw](hmac0){\shortstack{HMAC \\ SHA512}};%hmac0
    \node at ($(hmac0)+(0,-0.8)$) {$k_0$};
    \node[rectangle, minimum width=1cm, minimum height=1cm, font=\footnotesize, draw](hmac1) at ($(hmac0.east)+(1.3,0)$){\shortstack{HMAC \\ SHA512}};%hmac1
    \node at ($(hmac1)+(0,-0.8)$) {$k_1$};
    \node[rectangle, minimum width=1cm, minimum height=1cm, font=\footnotesize, draw](hmac2) at ($(hmac1.east)+(1.3,0)$){\shortstack{HMAC \\ SHA512}};%hmac2
    \node at ($(hmac2)+(0,-0.8)$) {$k_2$};
    \node[rectangle, minimum width=1cm, minimum height=1cm, font=\footnotesize, draw](hmacn) at ($(hmac2.east)+(2,0)$){\shortstack{HMAC \\ SHA512}};%hmacn
    \node at ($(hmacn)+(0,-0.8)$) {$k_{c-1}$};
    \node [circle, minimum size=0.7cm, font=\footnotesize, draw]  at ($(hmac1)!0.45!(hmac2)+(0,-1.5)$) (xor0) {}; %xor0
        \draw ($(xor0)+(0, 0.35)$) -- ($(xor0)+(0, -0.35)$); % verticle xor0
        \draw ($(xor0)+(-0.35, 0)$) -- ($(xor0)+(0.35, 0)$); % horizantle xor0
    \node [circle, minimum size=0.7cm, font=\footnotesize, draw]  at ($(hmac2)!0.3!(hmacn)+(0,-1.5)$) (xor1) {}; %xor1
    \node [circle, minimum size=0.7cm, font=\footnotesize, draw]  at ($(hmacn.east)+(0.3,-1.5)$) (xorn) {}; %xorn
    \node[rectangle, minimum width=0.5cm, minimum height=0.8cm, font=\footnotesize, draw](conc) at ($(rec0.east)+(0.8, -1)$){$||$}; %conc

%lines
    \draw[line:thick] ($(hmac0)+(-1.8, 1.2)$)node[left]{$Pwd$}-|($(hmac0)+(-1, 0.7)$)|-($(hmac0.west)+(0, 0.3)$);%Psw
    \draw[line:thick] ($(hmac0)+(-1.8, -0.3)$)node[left]{$Slt || 1$}--($(hmac0.west)+(0, -0.3)$); % Slt
    \draw[line:thick] ($(hmac0)+(-1.8, 1.2)$)-|($(hmac0)+(-1, 0.7)$)-| ($(hmac0)+(0.9, 0.3)$)--($(hmac1.west)+(0, 0.3)$);
    \draw[line:thick] ($(hmac0)+(-1.8, 1.2)$)-|($(hmac0)+(-1, 0.7)$)-| ($(hmac1)+(0.9, 0.3)$)--($(hmac2.west)+(0, 0.3)$);
    \draw[line:thick] ($(hmac0)+(-1.8, 1.2)$)-|($(hmac0)+(-1, 0.7)$)-| ($(hmac2)+(0.9, 0.3)$)--($(hmacn.west)+(-0.8, 0.3)$);
    \draw[line:thick] (hmac0)--(hmac1);
    \draw[line:thick] (hmac1)--(hmac2);
    \draw[line:thick] (hmac2)--($(hmac2.east)+(0.5, 0)$);
    \draw[line:thick] (hmac0)--($(hmac0)!0.45!(hmac1)$)|-(xor0);
    \draw[line:thick] (hmac1)-|(xor0);
    \draw[line:thick] (hmac2)-|(xor1);
    \draw[line:thick] (xor0)--(xor1);
    \draw[line:thick] (hmacn)-|(xorn);
    \draw[line:thick] ($(hmacn.west)+(-0.3, 0.3)$)--($(hmacn.west)+(0, 0.3)$);
    \draw[-, line width=1.5pt, dash pattern=on 2pt off 2pt] ($(hmacn.west)+(-1, 0.3)$) -- ($(hmacn.west)+(0, 0.3)$);
    \draw[line:thick] ($(hmacn.west)+(-0.3, 0)$)--(hmacn.west);
    \draw[-, line width=1.5pt, dash pattern=on 2pt off 2pt] ($(hmac2)+(1, 0)$) -- ($(hmacn.west)+(0, 0)$);
    \draw[line:thick] (xor1)--($(xor1)+(0.8, 0)$);
    \draw[line:thick] ($(xorn)+(-0.8, 0)$)--(xorn);
    \draw[-, line width=1.5pt, dash pattern=on 2pt off 4pt] ($(xor1)+(0.8, 0)$) -- ($(xorn)+(-0.8, 0)$);
    \draw[line:thick] (xorn.east)|-($(conc.west)+(0, -0.2)$);
    \draw[line:thick] ($(rec1.east)+(0, -1.05)$)|-($(conc.west)+(0, 0.05)$);
    \draw[line:thick] ($(rec2.east)+(0, -0.9)$)|-($(conc.west)+(0, 0.3)$);
    \draw[line:thick] (conc.east)--($(conc.east)+(0.4, 0)$)node[right]{$dk_{out}$};
    \draw ($(xor1)+(0, 0.35)$) -- ($(xor1)+(0, -0.35)$); % verticle xor1
    \draw ($(xor1)+(-0.35, 0)$) -- ($(xor1)+(0.35, 0)$); % horizantle xor1
    \draw ($(xorn)+(0, 0.35)$) -- ($(xorn)+(0, -0.35)$); % verticle xorn
    \draw ($(xorn)+(-0.35, 0)$) -- ($(xorn)+(0.35, 0)$); % horizantle xorn
\end{tikzpicture}
    }
    \caption{\Glsfirst{pbkdf} using \gls{sha}-512. \gls{pbkdf} is used by \gls{bip}-39 to generate the seed ($dk_{out}$) used by the \gls{hd} wallet given the mnemonics and salt as $Pwd$ and $Slt$ respectively.}
    \label{fig:pbkdf2} 
\end{figure*}

The \glsfirst{pbkdf} is a widely utilized cryptographic hash algorithm for generating secure keys given a password \cite{choi2021optimization}. The algorithm takes a user-defined password and other variables to generate a unique key $dk_{out}$ as follows:

\begin{equation}
    dk_{out}= PBKDF2_{PRF}(Pwd, Slt, c, dk_{Len}),
    \label{eq:pkdf2}
\end{equation}
where $Pwd$ is the user-defined password, $Slt$ is a salt variable used to further strengthen the security of the key, $PRF$ is the preferred cryptographic hash function such as \gls{sha}-256 or \gls{sha}-512, $c$ is the number of iterations, and $dk_{Len}$ is the desired size of the output. 

\autoref{fig:pbkdf2} illustrates how to compute the digest of \gls{pbkdf}. In the figure, $i$ is the resulting quotient after dividing the desired size of output $dk_{Len}$ by the output size of $PRF$. Hence, in the case where $PRF$ is \gls{sha}-512 and the desired output size is 512, $i$=1. $Slt$ is concatenated with  $i$ and used as the input to the first instance of \gls{hmac}\gls{sha}-512. From the second instance to instance $c-1$, the previous \gls{hmac}\gls{sha}-512 digest is used as the input to the current instance. The algorithm uses $Pwd$ as the second input to all the \gls{hmac}\gls{sha}-512 computations. Moreover, it calculates the exclusive OR ($\oplus$) of the output of each \gls{hmac}\gls{sha}-512 digest and concatenates the output of each $i$ operation to get $dk_{out}$. \gls{bip}-39 uses \gls{pbkdf} with $c=2048$ to compute the seed used in the \gls{hd} wallet.

The following section talks about the Keccak hash function used to generate Ethereum addresses.

\subsection{The Keccak Hash Function}

 The Keccak hash function, like many others, is designed to offer robust security by preventing collision attacks and other vulnerabilities \cite{tikhomirov2018ethereum}. The core of the Keccak hash function is the sponge construction technique, which operates in two phases: absorbing and squeezing. During the absorbing phase, the input message is divided into blocks, and a permutation function iteratively processes these blocks, integrating the input message into the function's state. In the squeezing phase, the function extracts the output from its state by repeatedly applying the same permutation function until the desired output size is achieved \cite{homsirikamol2012security}. This flexible sponge construction enables Keccak to produce digests of varying sizes, making it suitable for a wide range of applications.

The Keccak family includes four primary hash functions, categorized by the size of their digests: Keccak-224, Keccak-256, Keccak-384, and Keccak-512 \cite{sideris2023novel}. In blockchain technology, different variants of Keccak are utilized in various system components. For example, Stellar employs Keccak-512 in its consensus protocol \cite{stellar}, while Ethereum uses Keccak-256 in its address generation process. Specifically, Ethereum generates an address by hashing the public key with Keccak-256. Additionally, Keccak is used to create a checksummed Ethereum address, ensuring greater security and integrity. In this work, we utilize an open-source Keccak-256 hardware implementation provided by the Keccak group \cite{frree3}.

In the next section, we introduce the \gls{hmac}\gls{sha}-512 algorithm used in various stages of the key generation process. 

\subsection{The HMACSHA-512 Algorithm}

The \glsfirst{hmac} based on \gls{sha}-512 (\gls{hmac}\gls{sha}-512) is an algorithm proposed by the \gls{nist} to ensure data integrity and authenticity \cite{juliato2011fpga, 9842174}. The \gls{hmac}\gls{sha}-512 takes two inputs called key $k$ and message $m$ and outputs a 512-bit digest as follows:
\begin{equation}
   \text{HMAC}(k,m) = \text{H}(k\oplus opad \parallel \text{H}(k\oplus ipad \parallel m)),
    \label{hmac_eqn}
\end{equation}
where H($\cdot$) denotes the \gls{sha}-512 hash function. Moreover, $opad$ is the outer padding, which is 0x36 repeated 64 times, and $ipad$ is the inner padding, which is 0x5C also repeated 64 times\cite{juliato2011fpga}. \gls{hmac} uses $ipad$ and $opad$ to modify the key and message before applying the hash function to enhance security. \autoref{subsec:hd_wal} explains that \gls{hmac}\gls{sha}-512 is used to create the master seed and mnemonics in an Ethereum \gls{hd} wallet. \autoref{sebsec:hmac} provides further details on the proposed hardware architecture of the \gls{hmac}\gls{sha}-512 algorithm.

The next section introduces the \gls{ckd} function, which is integral to the hierarchical structure of the wallet.

\subsection{The Child Key Derivation Function}
\label{subsec:ckdf}

 The \glsfirst{ckd} function in the \gls{bip}-32 standard for \gls{hd} wallets (see \autoref{fig:bip32}) allows deriving child keys from a parent key in a secure and reproducible manner. The \gls{ckd} function utilizes SECP256K1 and \gls{hmac}\gls{sha}-512 hash algorithms. \autoref{ckdf_algo} depicts the execution process \gls{ckd} function. 
 
 The inputs $k$, $c$, $n$, and $p$ are the private key, chain code, child number, and the prime field modulus for SECP256K1, respectively. The private key and the chain code are digests from the \gls{hmac}\gls{sha}-512 hash function, where 256 \glspl{lsb} of the digest are the chain code, and 256 \glspl{msb} of the digest are the private key. Lines \autoref{ckdA:1}-- \autoref{ckdA:2} describe how the $if$ condition specifies the creation of hardened keys. It appends $x00$ at the beginning of $k$ if $n$ $\geq$ $2^{31}$. The $else$ statement in lines \autoref{ckdA:3} -- \autoref{ckdA:6} describe how the \gls{ckd} function creates normal keys. First, SECP256K1 generates a child public key $b$ using $k$ as input. Second, the \gls{ckd} function executes a serialization function, taking $b$ as input. The serialization process takes 256 \glspl{msb} of the 512-bit long $b$ value as the public key.  It then prepends $x02$ if the 256-bit \gls{lsb} integer $b$ is even or $x03$ if the value is odd.

% Child key derivation function
 \begin{algorithm}[t]
 \caption{The child key derivation (CKD) function as described in BIP-32. Adapted from \cite{bip32}.}
 \label{ckdf_algo}
 \begin{algorithmic}[1]\small
 \renewcommand{\algorithmicrequire}{\textbf{Input:}}
 \renewcommand{\algorithmicensure}{\textbf{Output:}}
 \REQUIRE $k$, $c$, $n$, $p$
  \IF {$n \geq 2^{31}$} \label{ckdA:1}
    \STATE $h  \leftarrow  00 || k$\label{ckdA:2}
  \ELSE\label{ckdA:3}
    \STATE $b \leftarrow \textsc{secp256k1}(k)$
	\STATE $h \leftarrow  \textsc{serialize}(b)$
  \ENDIF\label{ckdA:6}
  \STATE $\hat{h} \leftarrow h || n$ \label{ckdA:7}
  \STATE $l$, $r$ $\leftarrow  \textsc{hmacsha512}(c, \hat{h})$ \label{ckdA:8}
  \STATE $\hat{k} \leftarrow \bmod(l + k$, $p)$ \label{ckdA:9}
  \STATE $\hat{c}\leftarrow r$
  \RETURN $\hat{k}, \hat{c}$
 \end{algorithmic} 
 \end{algorithm}

After the $if-else$ statement, the \gls{ckd} function creates a hardened or normal key, $h$. Line \autoref{ckdA:7} creates $\hat{h}$ by concatenating $h$ and the child number $n$. Line \autoref{ckdA:8} then uses \gls{hmac}\gls{sha}-512 with $c$ and $\hat{h}$ as key and message, respectively, to generate $l$ and $r$. Here, $l$ is the 256 \glspl{msb} of the \gls{hmac}\gls{sha}-512 digest while $r$ is the remaining 256 \glspl{lsb} of the digest. Line \autoref{ckdA:9} performs modulo $p$ addition of $l$ and $k$. The \gls{ckd} function returns $\hat{c}$ and $\hat{k}$ as child chain code and child private key, respectively. 

The following section discusses how the Ethereum address is generated.

\subsection{The Ethereum Address}
\label{subsec:EthAddr}
The Ethereum address is derived by computing the Keccak-256 hash of the public key and extracting the 160 \glspl{lsb} of the resulting digest. To improve readability and reduce the risk of input errors, a checksummed address is then generated according to the \gls{eip}-55 protocol, as illustrated in \autoref{alg_1} \cite{EIP55}. This process takes the lowercase hexadecimal Ethereum address, denoted $a$, and computes its Keccak-256 hash, producing a digest $d$. For each alphabetical character in $a$, if the corresponding nibble in $d$ is greater than 7, the character is converted to uppercase using the function \textsc{capital}(). This results in a mixed-case Ethereum address that incorporates a checksum, enabling basic error detection during manual entry.

The following section discusses the \glsfirst{ecdsa}.
 \begin{algorithm}[t]
 \caption{Pseudo algorithm for Ethereum checksum. Adapted from \cite{EIP55}.}
 \label{alg_1}
 \begin{algorithmic}[1]\small
 \renewcommand{\algorithmicrequire}{\textbf{Input:}}
 \renewcommand{\algorithmicensure}{\textbf{Output:}}
 \REQUIRE $a$, $d\leftarrow$keccak256($a$)
 \ENSURE  $\hat{d}$
 \\ \textit{Initialisation} :
  \STATE $\hat{d}\leftarrow a$
  \STATE $j\leftarrow \textsc{length}(a)/4$
  \STATE $k\leftarrow 3$
 \\ \textit{LOOP Process} :
  \FOR {$i = j-1$ to $0$}
  \IF {$ a(k : k-3) > 9$}
  \IF {$ d(k : k-3) > 7$}
  \STATE $\hat{d}(k : k-3) \leftarrow \textsc{capital}(\hat{d}(k : k-3))$
  \ENDIF
  \ENDIF
  \ENDFOR
  \RETURN $\hat{d}$
 \end{algorithmic} 
 \end{algorithm}

\subsection{The Elliptic Curve Digital Signature Algorithm}
\label{subsec:ecdsa}

\Gls{ecdsa} is a cryptographic algorithm based on \gls{ec} arithmetic that generates digital signatures to ensure both data integrity and authenticity. It is widely deployed and standardized by organizations such as the \gls{iso}, the \gls{ansi}, and the \gls{nist} \cite{johnson2001elliptic}.

The \gls{ecdsa} process consists of three stages: key generation, signature generation, and verification. In the key generation stage, a private key is multiplied by the base point of the chosen \gls{ec} to produce the corresponding public key, as discussed in \autoref{sec:secp}. In the signature generation stage, the private key, the message hash, and auxiliary variables are used to compute the digital signature. Finally, in the verification stage, the receiver uses the message and public key to reconstruct the signature and compare it with the received signature. If they match, the message is considered valid \cite{kieu2022low}.

Ethereum employs \gls{ecdsa} to generate digital signatures for transaction authorization. Notably, it uses the SECP256K1 \gls{ec} to perform the signing. In practice, physical \gls{cryp} wallets typically implement the key generation and signature generation stages to manage private keys and signing operations securely. The signature generation procedure is summarized in \autoref{alg:ecdsa}.

On \autoref{alg:ecdsa}, the \gls{ecdsa} algorithm takes as input the order of the generator point $G$ (denoted as $n$), the private key $d$, and the \gls{sha}-256 hash digest of the transaction data $z$. In line\,\autoref{ln0}, a nonce $k$ is generated, either uniformly at random from the interval $[1, n-1]$ or deterministically using the RFC 6979 protocol \cite{RFC}. On line\,\autoref{ln1}, the corresponding elliptic curve point $kG$ is computed. Line\,\autoref{ln2} sets the $x$-coordinate of this point, reduced modulo $n$, as the signature component $r$. On line\,\autoref{ln3}, the second signature component $s$ is derived as $k^{-1}(z + dr) \bmod n$. Finally, line\,\autoref{ln4} outputs the signature pair $(r, s)$.

The following section provides an in-depth exploration of the hardware design and implementation details of EthVault.

 \begin{algorithm}[t]
  \caption{Pseudo algorithm for signature-generation in \gls{ecdsa}. Adapted from \cite{kieu2022low}}
  \label{alg:ecdsa}
  \begin{algorithmic}[1]\small
    \renewcommand{\algorithmicrequire}{\textbf{Input:}}
    \renewcommand{\algorithmicensure}{\textbf{Output:}}
    \REQUIRE Curve prime $p$, order of $G$ $n$, private key $d\in[1,n-1]$, $z\leftarrow$ SHA256($msg$)
    \ENSURE Signature $(r,s)$
      \STATE Select nonce $k \leftarrow [1, n-1]$\label{ln0}
      \STATE Compute $(x_1,y_1) \gets k\cdot G$ over $\mathbb{F}_p$\label{ln1}
      \STATE $r \gets x_1 \bmod n$\label{ln2}
      \STATE $s \gets k^{-1}\,(z + d\cdot r) \bmod n$\label{ln3}
    \RETURN $(r,s)$\label{ln4}
  \end{algorithmic}
\end{algorithm}

\section{Proposed Hardware Architecture of EthVault}
\label{sec: hw_arch}
This section presents the proposed EthVault architecture. It begins with an overview of the wallet’s design, optimizations, and core functionalities. It then details the proposed architectures of the integrated cryptographic and key-management algorithms, including SECP256K1, \gls{bip}-39, \gls{bia}, \gls{hmac}\gls{sha}-512, \gls{ckd}, Ethereum checksum, and the \gls{ecdsa}.

For each architecture, we highlight the specific optimization strategies implemented, as well as the testing and validation techniques employed, including the sources of the test vectors. Additionally, the coverage report for each algorithm was generated using the Vivado code coverage tool \cite{vivadoUG937ConcurrentAssertion}, with coverage evaluated for statement, branch, and condition metrics. This tool produces coverage reports for each sub-module of a design rather than just the top level. Therefore, we compute the average coverage across all sub-modules and report it as the statement, branch, and condition coverage for the entire module.
%Functional coverage metrics were not generated due to VHDL limitations.

\begin{figure*}[t]
\centering
    \resizebox{1\textwidth}{!}{
        \input{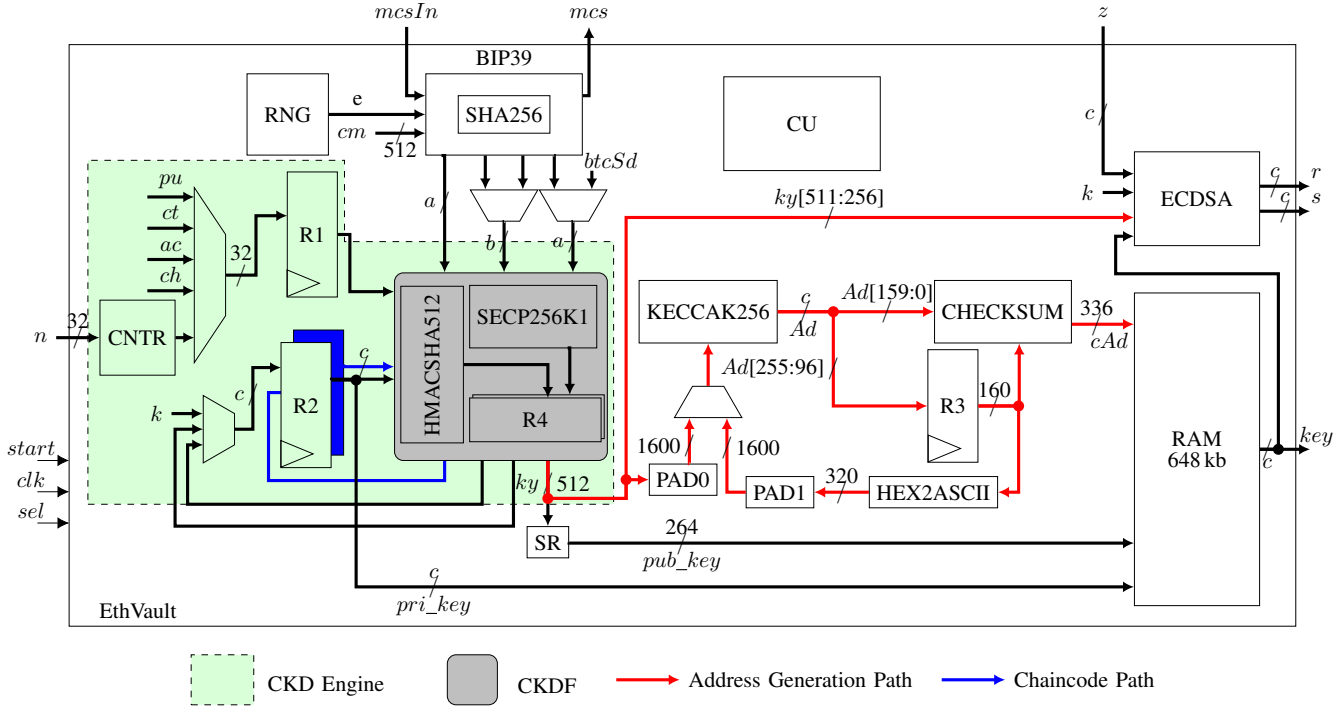}
    }
    \caption{Proposed hardware architecture of EthVault. The constants $pu$, $ct$, $ac$, and $ch$ represent the parameters of the \gls{bip}-44 path, corresponding to $purpose$, $coin\_type$, $account$, and $change$, respectively. The variable $n$ denotes the number of child keys, which determines the $address\_index$ of the path through the counter CNTR. The path widths are defined as $a = 1024$, $b = 512$, and $c = 256$.}
    \label{fig:ethWalArch} 
\end{figure*}

\subsection{Architecture of EthVault}
\label{subsec:ETHwallet}

\autoref{fig:ethWalArch} illustrates the proposed architecture of EthVault. The design includes several \gls{io} interfaces that facilitate communication with both the end user and a hot wallet, enabling data exchange and transaction approval. The specifics of this interaction are further detailed in \autoref{subsec:sysInteg}.

The input signal $n$ specifies the number of child key pairs (private/public) and their corresponding Ethereum addresses to be generated. The $start$ signal initiates the key generation process, while the $sel$ input selects which key pair is to be used for subsequent operations. Moreover, $z$ is the \gls{sha}-256 hash digest of the transaction data to be signed. Also, EthVault operates using a single synchronous clock, denoted as $clk$.

Additionally, the output $mcs$ represents the mnemonic phrase generated by the wallet, while the input $mcsIn$ allows users to provide a mnemonic phrase to recover previously generated keys. The $key$ output contains the public key and the Ethereum address selected by $sel$. Finally, $r$ and $s$ represent the components of the transaction data signature.

\subsubsection{Module Functions of EthVault}

\autoref{fig:ethWalArch} also highlights the main components integrated within the EthVault architecture. Registers R1 to R4 are used to store intermediate and final values during the key generation process. The CKDF component implements the \gls{ckd} function, which includes two submodules: HMACSHA512, responsible for executing the \gls{hmac}\gls{sha}-512 algorithm, and SECP256K1, which performs \gls{ec} operations on the SECP256K1 curve.

The RNG module supplies entropy for mnemonic generation, while the BIP39 module uses this entropy to generate mnemonics. The BIP39 module also internally invokes the \gls{ckd} function to derive the master and intermediate keys. The CNTR module functions as a counter, iterating from 0 to $n$–1 to produce the $address\_index$ variable of the derivation path.

The SR component handles the serialization of the public key into a compressed format. The KECCAK256 and CHECKSUM components implement the Keccak-256 hash function and the \gls{erp}-55 checksum algorithm, respectively. The HEX2ASCII module converts the hexadecimal Ethereum address into its \gls{ascii} format.

Modules PAD0 and PAD1 append zero bits to their respective inputs to produce a 1\,600-bit output, as required by the Keccak sponge function. The RAM module serves as storage for the generated key pairs, allowing efficient retrieval and selection, while the ECDSA module implements the \gls{ecdsa} algorithm to sign transaction data. Finally, the \gls{cu} module implements a \gls{fsm} that coordinates the operation of all modules by generating control signals to ensure correct sequencing and timing throughout the key generation process.

\subsubsection{BIP44-Compliant HD Wallet Architecture}

EthVault fully implements the \gls{hd} structure specified in the Ethereum \gls{hd} wallet standard, as detailed in \autoref{subsec:hd_wal}. The wallet adheres to the \gls{bip}-44 specification by executing all derivation steps along the standardized path. The constants used in the \gls{bip}-44 derivation path are denoted in \autoref{fig:ethWalArch} as ${pu}$ (purpose), ${ct}$ (coin type), ${ac}$ (account), ${ch}$ (change), and ${n}$ (address index), corresponding to the hardened path ${m}/44'/60'/0'/0/address\_index$ for Ethereum.

The green region highlights the \gls{ckd} engine, which performs \gls{ecc} arithmetic to compute child keys at each level. These operations follow the hardened and non-hardened derivation rules defined by \gls{bip}-32, which \gls{bip}-44 builds upon.

\subsubsection{Datapaths and Key Generation}

\autoref{fig:ethWalArch} illustrates various internal signals and paths with different widths. To enhance readability, some widths that can be easily inferred are omitted. For instance, while the input width of a register or multiplexer may be indicated, the corresponding output width may not appear. However, it can be directly deduced from the input. Additionally, certain widths are denoted by letters: $a$ corresponds to 1024, $b$ to 512, and $c$ to 256. The signal $cm$ represents the concatenation of the chaincode and the master private key, while $btcSd$ denotes the string “Bitcoin seed” in \gls{ascii} format, padded with zeros for use during the master private key and chaincode generation stage. Moreover, $k$ is the random value employed by \gls{ecdsa} during signature generation.

The red data path represents the Ethereum address generation process. Here, the public key is hashed using KECCAK256 and truncated to produce a 160-bit Ethereum address. In the figure, the 160 \glspl{lsb} of \textit{Ad} represents this raw Ethereum address. The corresponding checksummed address, \textit{cAd}, is computed as described in \autoref{alg_1}, following the \gls{eip}-55 specification \cite{EIP55}. Moreover, the compressed public key (\textit{pub\_key}), the private key (\textit{priv\_key}) and the checksummed (\textit{cAd}) addresses are stored within the RAM module as shown in the figure.

To generate keys within EthVault, the user inputs the number of keys to be generated using $n$ and commences key generation using the $start$ input. The \gls{rng} then generates a random number, $e$, which the BIP39 module uses to produce a seed and the corresponding mnemonic phrase through $mcs$. The user securely stores this phrase, which can be re-entered via the $mcsIn$ input to recover the associated keys. HMACSHA512 then processes the seed to compute the master key, $m$, as shown in \autoref{fig:bip32}. Using this master key and the \gls{bip}-44 path, the wallet derives $n$ private/public key pairs. Each public key maps to a unique checksummed Ethereum address (\textit{cAd}), and the generated keys are stored in the \gls{ram}. The user can select any key pair from memory using $sel$ to send or receive \glspl{cryp}.

\subsubsection{Optimizations}
To enhance throughput in EthVault, the derivation path $m / purpose' / coin\_type' / account' / change $ $/ ddress\_index$ is optimized by avoiding redundant computations. Since part of this path is repeatedly executed during the generation of multiple keys, the output of the \gls{ckd} function is stored in registers after the first execution. Specifically, the result of the \gls{ckd} function for the partial path $m / purpose' / coin\_type' / account' / change$ is cached in registers and reused for subsequent key generations. As a result, only the $address\_index$ needs to be computed for each new child key.

Additionally, the key derivation process involves repeatedly executing algorithms such as \gls{hmac}\gls{sha}-512, SECP256K1, and \gls{sha}-512 at various stages, as outlined in \autoref{subsec:hd_wal} and demonstrated by \autoref{fig:allAlgs}. To reduce the wallet's size, the CKDF module provides dedicated paths to use each of these algorithms, enabling their reuse across different stages.

\subsubsection{Validation and Testing}
To generate test vectors for validation, an Ethereum \gls{hd} wallet implemented in Python was employed to produce entropy $e$ and the corresponding child private and public keys and their derived addresses. Additionally, an online implementation was utilized to generate random mnemonic phrases, from which the corresponding private keys, public keys, and addresses were obtained \cite{mnem}.

Moreover, we tested edge cases such as when $e$ = all 1s, $e$ = all 0s, which represent the maximum and minimum possible entropy values, respectively. Similarly, we tested $mcsIn$ = all 1s, and $mcsIn$ = all 0s covering boundary scenarios for mnemonic-to-seed conversion.

The following section highlights the proposed architecture of the SECKP256K1 algorithm. 

\subsection{Architecture of SECP256K1}
\label{subsec:arch_secp}

As discussed in \autoref{sec:secp}, SECP256K1 consists of three key processes: \gls{pa}, \gls{pd}, and \gls{pm}. The \gls{pa} operation calculates the sum of two distinct points on the \gls{ec}, defined as $ \bm{R} = \bm{P} + \bm{Q}$. In contrast, \gls{pd} represents the doubling of a point on the curve ($ \bm{R} = 2\bm{P}$). For efficient implementation, we compute \gls{pd} by applying the \gls{pa} operation to the same point twice, setting $\bm{P} = \bm{Q}$ to yield $ \bm{R} = \bm{P} + \bm{P} = 2\bm{P}$. Additionally, scalar \gls{pm} computes the product of a point with a scalar integer, denoted by $\bm{R} = k\cdot\bm{P}$ ($\bm{P}$ is set as the generator point $\bm{G}$ of SECP256K1). This \gls{pm} operation is implemented using the Montgomery ladder algorithm, which incorporates both \gls{pa} and \gls{pd} steps, as shown in \autoref{alg:montg}. To avoid costly inversion operations, we perform \gls{pm} in projective coordinates, requiring only one final inversion to return to affine coordinates, i.e., ($xz^{-1}, yz^{-1}$) $\Rightarrow$ ($x$, $y$).

Although physical \gls{cryp} wallets are vulnerable to a wide range of attacks, including power and timing analysis, \gls{ema}, \gls{fia}, memory attacks, and brute-force attacks \cite{arapinis2019formal, guri2018beatcoin}, this work specifically focuses on mitigating \gls{dpa} and timing-based \gls{sca}. In particular, we propose a modified Montgomery ladder algorithm with a temporary register, $\bm{R_t}$, as shown in \autoref{alg:montgNew}. This register ensures that $\bm{R_0}$ is accessed when $k_i = 1$ and when $k_i = 0$ $\bm{R_1}$ is accessed. Hence, maintaining consistent power and timing patterns. $\bm{R_t}$ ensures uniform access patterns by always involving both $\bm{R_0}$ and $\bm{R_1} $ in computations, regardless of the value of $k_i$. Specifically, when $k_i = 1$ or $k_i = 0$, \gls{pd} is performed on both $\bm{R_0}$ and $\bm{R_1}$. This design enforces consistent power and timing patterns by maintaining uniform memory access and data path utilization, thereby minimizing the side-channel information leaked to an attacker.

Furthermore, we employ the complete \gls{pa} equations from \autoref{alg:compEqs} in the modified algorithm. These equations eliminate the conditional branching typically associated with traditional \gls{pa} processes on Weierstrass curves \cite{renes2016complete}. To further obscure any computational patterns, all operations within each branch (addition and doubling) are executed in parallel, creating a uniform control structure that conceals the order of operations.

% Montgomery algorithm with temporary registers
 \begin{algorithm}[t]
 \caption{Montgomery Ladder algorithm with temporary registers.}
 \label{alg:montgNew}
 \begin{algorithmic}[1]\small
 \renewcommand{\algorithmicrequire}{\textbf{Input:}}
 \renewcommand{\algorithmicensure}{\textbf{Output:}}
 \REQUIRE $\bm{P} \in (x, y, z), k=(k_{t-1}, \cdots, k_0)$ with $k_{t-1}=1$
 \ENSURE  $\bm{R}=k\bm{P}$
 \\ \textit{Initialisation} :
  \STATE $\bm{R_0} \leftarrow \bm{P}$ \label{montg:init-r0}
  \STATE $\bm{R_1} \leftarrow 2\bm{P}$ \label{montg:init-r1}
 \\ \textit{LOOP Process} :
  \FOR {$i = t-2$ : $0$}\label{montg:for-st}
  \IF {$k_i = 1$}\label{montg:if-st}
    \STATE $\bm{R_0} \leftarrow \bm{R_0} + \bm{R_1}$\label{montg:if-st-r0}
    \STATE $\bm{R_1} \leftarrow 2\bm{R_1}$\label{montg:if-st-r1}
    \STATE $\textcolor{blue}{\bm{R_t} \leftarrow 2\bm{R_0}}$\label{montg:rt0}
  \ELSE
    \STATE $\bm{R_1} \leftarrow \bm{R_0} + \bm{R_1}$\label{montg:if-else-r1}
    \STATE $\bm{R_0} \leftarrow 2\bm{R_0}$\label{montg:if-else-r0}
    \STATE $\textcolor{blue}{\bm{R_t} \leftarrow 2\bm{R_1}}$\label{montg:rt1}
  \ENDIF
  \ENDFOR \label{montg:end-for}
  \STATE $\bm{R} \leftarrow \textsc{BIA}(\bm{R_0})$ \label{montg:return}
  \RETURN $\bm{R}$
 \end{algorithmic} 
 \end{algorithm}

\begin{figure}[t]
\centering
    \resizebox{0.5\textwidth}{!}{
    \input{pm}
    }
    \caption{The hardware architecture of the proposed Montgomery Ladder algorithm used to perform \gls{pm}. PA0 and PA1 can perform either \gls{pa} or \gls{pd} in parallel, depending on the status of $k_i$. }
    \label{fig:eth_pm}
\end{figure}

\autoref{fig:eth_pm} illustrates the proposed hardware architecture for the modified Montgomery ladder algorithm presented in \autoref{alg:montgNew}, which implements elliptic curve operations over SECP256K1. The light blue dashed modules represent the \gls{pa} hardware architecture, which executes the equations in \autoref{alg:compEqs}. This Montgomery ladder architecture utilizes two \gls{pa} modules executing in parallel, with a \gls{cu} module processing the bit values $k_i$ of the private key to control all select and enable signals for multiplexers and registers. 

The light orange module in the figure corresponds to the \gls{bia} architecture. Since \gls{bia} is executed at the end of the \gls{pm} process, it reuses $\bm{R_1}$  and  $\bm{R_t}$ registers from the preceding \gls{pa} operations to optimize resource utilization. The rugged light red region highlights this reuse.

Each \gls{pa} module executes the complete \gls{pa} formulas detailed in \autoref{alg:compEqs}. All operations in \autoref{alg:compEqs} are computed modulo $p$, where $p$ is the prime number specific to SECP256K1. Also, the \gls{bia} module performs modulo $p$ operations as shown in \autoref{alg:binInv}. Accordingly, we design a \gls{malu} that performs modular addition, subtraction, and multiplication. A shift-and-add algorithm is used to execute these modular operations \cite{opencores0}.

The inputs and outputs of the \autoref{fig:eth_pm} correspond to those defined in \autoref{alg:montgNew}. Specifically, $x$, $y$, and $z$ denote the 256-bit projective coordinates of the generator point $\bm{G}$, $k$ is the 256-bit private key input, $b_3$ represents the value defined in \autoref{alg:compEqs}, and $\bm{R}$ is the output in affine coordinates. The output of the \gls{bia} block is a bus containing the contents of all five registers used by the \gls{bia} architecture. This bus has a width of 1\,280 bits, denoted as $b$. Furthermore, the red and blue paths represent data buses carrying the outputs of registers $\bm{R_0}$ and $\bm{R_1}$, respectively.

\subsubsection{Optimizations}: In addition to mitigating \gls{dpa} and timing \gls{sca} attacks using temporary registers, parallel operations, and optimized \gls{pa} equations, we aim to minimize resource utilization. Notably, the architecture in \autoref{fig:eth_pm} employs only two \gls{pa} modules, instead of four, to achieve a more efficient and compact design. Also, the \gls{bia} implementation shares registers with the implementation of the \gls{pm} algorithm, as indicated by the rugged region in the figure, further optimizing hardware resources.

\subsubsection{Validation and Testing}

Software implementations of the \gls{pa} and \gls{bia} modules were developed in Python to generate test vectors for verifying the corresponding hardware architectures. Additionally, the official SECP256K1 implementation used by Bitcoin \cite{secp}, along with other widely adopted implementations available in Python libraries, was utilized to generate reference test vectors and validate the correctness of the proposed SECP256K1 architecture.

The test vectors for \eqref{eq:pmult} of SECP256K1 include edge cases shown it \autoref{table:testSecp} such as $k = 0$, $k = n-1$, $k = n$, $k = n+1$, $k = 2^{256}-1$, and $k = 2^{255}$, where $n$ is the order of the SECP256K1 generator $\bm{G}$. Also, other random values of $k$ were used. Functional verification achieved a coverage of 98\% for statements, 96\% for branches, and 97\% for conditions. This indicates that the test bench thoroughly exercised the design, leaving only a small fraction of rarely triggered code untested.

The following section introduces the proposed architecture of the \gls{bip}-39 protocol.

\begin{table}[t]
\centering
\caption{SECP256K1 test vectors and edge-case purposes.}
%\resizebox{5cm}{!}{
\begin{tabular}{ll}
    \hline \hline
    $k$ & Purpose of the test \\
    \hline \hline
    $0$ & Multiplication produces the point at infinity \\
    $n-1$ & Point negation and handling of scalars just below the group order \\
    $n$ & Scalars equal to group order reduce to zero and are rejected \\
    $n+1$ & Verifies modular reduction wraps around; behaves like $k=1$\\
    $2^{256}-1$ & Tests very large scalars and reduction modulo $n$ \\
    $2^{255}$ & Tests handling of scalars with the highest bit set. \\
    \hline \hline
\end{tabular}
%}
\label{table:testSecp}
\end{table}

\begin{figure}
    \centering
    \resizebox{0.5\textwidth}{!}{
        \begin{tikzpicture}
        %doted 
        \coordinate(a) at (-3.8,3.6);
        \coordinate(b) at (1.2,3.6);
        \coordinate(c) at (1.2,-2.0);
        \coordinate(d) at (-7.7,-2.0);
        \coordinate(e) at (-7.7,2);
        \coordinate(f) at (-3.8,2);
        
        \fill[red!10] (a) -- (b) -- (c) -- (d) -- (e) -- (f) -- cycle;
        \draw[red, line:thick_dot] (a) -- (b) -- (c) -- (d) -- (e) -- (f) -- cycle;
    
        \node [label={CKDF}, fill=lightgray, rounded corners, rectangle, minimum width=2.4cm, minimum height=2cm, draw](ckd){}; %ckd 
        \node [rectangle, minimum width=1.5cm, minimum height=0.5cm, draw](sha512){SHA512}; %sha512
        \node [label={\gls{hmac}\gls{sha}512}, rectangle, minimum width=2cm, minimum height=1cm, draw](HMAC){}; %HMAC
        \node [rectangle, minimum width=0.8cm, minimum height=1.5cm, draw](reg0)[above left=-1 and 1.5 of ckd]{R0}; %reg0 close to ckd
        \node [isosceles triangle, minimum width=2mm, minimum height=0.8mm, draw] (clk0)[below left=-0.4cm and -0.25cm of reg0]{}; %clk0
        \node [rectangle, minimum width=0.8cm, minimum height=0.5cm, draw](pad0)[below right=0.15 and -0.3 of reg0]{pad0}; %pad0
        \node [rectangle, minimum width=0.8cm, minimum height=1.5cm, draw](reg1)[left=0.3 of reg0]{R1}; %reg1 behind reg0
        \node [isosceles triangle,  minimum width=2.8mm, minimum height=0.8mm, draw] (clk1)[below left=-0.4cm and -0.25cm of reg1]{}; %clk1
        \node [rectangle, minimum width=0.8cm, minimum height=1.5cm, draw](reg3)[above=0.5 of reg0]{R3}; %reg3
        \node [isosceles triangle,  minimum width=2.8mm, minimum height=0.8mm, draw] (clk3)[below left=-0.4cm and -0.25cm of reg3]{}; %clk3
        \node [rectangle, minimum width=0.8cm, minimum height=0.8cm, draw](mng)[above left=-0.2 and 1.8 of reg3]{MNG}; %mng
        \node [rectangle, minimum width=1cm, minimum height=1cm, draw](fsm) at ($(mng.south)+(-0.5, -0.7)$){CU};
        \node [rectangle, minimum width=0.5cm, minimum height=0.5cm, draw](conc)[above=0.3  of mng]{$||$}; %conc
        \node [rectangle, minimum width=0.8cm, minimum height=0.8cm, draw](sha)[above right=0 and 0.8 of conc]{SHA256}; %sha 
        \node [rectangle, minimum width=0.8cm, minimum height=0.8cm, draw](counter)[right=2 of reg3]{CNTR}; %counter 
        \node [rectangle, minimum width=0.8cm, minimum height=0.8cm, draw] at ($(conc)+(-1.8,0.7)$) (qrng){RNG}; %qrng
        \node [trapezium, draw, rotate=-90, minimum width=1cm, minimum height=0.5cm](mux0)[below=1 of reg1]{}; %mux0
        \node [trapezium, draw, rotate=180, minimum width=1cm, minimum height=0.5cm] at ($(reg0)!0.45!(ckd)+(0, 1.5)$) (mux1){}; %mux1 above ckd
        \node [trapezium, draw, rotate=-180, minimum width=1cm, minimum height=0.5cm](mux2) at ($(mng.east)+(1, -0.5)$) {}; %mux0
        \node [rectangle, minimum width=0.8cm, minimum height=1.5cm, draw](reg2)[left=1.9 of mux0]{R2}; %reg2 behind xor
        \node [isosceles triangle,  minimum width=2.8mm, minimum height=0.8mm, draw] (clk2)[below left=-0.4cm and -0.25cm of reg2]{}; %clk2
        \node [circle, minimum size=0.7cm, font=\footnotesize, draw]  at ($(reg1)!0.45!(reg2)+(0,-0.3)$) (xor0) {}; %xor0
            \draw ($(xor0)+(0, 0.35)$) -- ($(xor0)+(0, -0.35)$); % verticle xor0
            \draw ($(xor0)+(-0.35, 0)$) -- ($(xor0)+(0.35, 0)$); % horizontal xor0
        % Legend
        \node (leg_red)[red, fill=red!10, label={[right=0.6cm, yshift=-0.25cm]PBKDF2}, line width=0.3mm, rectangle, decorate, decoration={snake, amplitude=.2mm, segment length=1mm}, minimum width=1cm, minimum height=0.5cm, draw] at ($(reg2.south)+(-0.2,-1.3)$){};

        \draw[line:thick] ($(reg2.east)+(0,0.49)$)|-(xor0);
        \draw[line:thick] ($(reg2.east)+(0,0.43)$) -| node[branch]{}($(reg2)+(0.6,0.1)$) |- ($(mux0)+(-0.25,0.3)$);
         \draw[line:thick] ($(mux0)+(-0.8,-0.3)$)node[left]{$Slt$}--($(mux0)+(-0.25,-0.3)$);
        \draw[line:thick] (mux0)-| node[pos=0.3]{$/$}node[pos=0.3,below]{256}($(reg1)!0.5!(reg0)+(2.2,-1.5)$)|-($(ckd)+(-1.2,-0.6)$);
        \draw[line:thick] (ckd) |- ($(reg0)!0.25!(ckd)+(-1.43,-2.5)$)node[branch]{} |- ($(reg0)+(-0.4, 0.2)$);
        \draw[line:thick] (ckd)|-node[pos=0.2]{$/$}node[pos=0.2, right]{512}($(reg0)!0.25!(ckd)+(0,-2.5)$)-|($(reg2)+(-0.7, 0.45)$)--($(reg2)+(-0.4, 0.45)$);
        \draw[line:thick] ($(reg1)+(0.4, 0.45)$)-|($(reg1)!0.5!(reg0)+(0,1)$) -| node[pos=0.8]{$/$}node[pos=0.8, left]{512}node[branch]{}(xor0);
        \draw[line:thick] ($(reg1)+(0.4, 0.45)$)-|($(reg1)!0.5!(reg0)+(0,0.5)$)|-($(xor0)!-0.5!(reg1)+(0,2.355)$)node[left]{$dk_{out}$};
        \draw[line:thick] (xor0)-|($(reg1)+(-0.7, 0.45)$)--($(reg1)+(-0.4, 0.45)$);
        \draw[line:thick] (pad0.east)|-node[pos=0.7]{$/$}node[pos=0.7, above]{1024}($(ckd)+(-1.2,-0.4)$);
        \draw[line:thick] ($(reg0)+(0.4, 0.35)$)-|node[pos=0.6, xshift=0.37cm]{$Pwd$}(pad0);
        \draw[line:thick] (mux1)|-node[pos=0.2]{$/$}node[pos=0.2, right]{1024}($(ckd)+(-1.2, 0.8)$);
        \draw[line:thin] ($(ckd)+(-1.5, 0.6)$)node[left]{$j$}--($(ckd)+(-1.2, 0.6)$);
        \draw[line:thin] ($(ckd)+(-1.5, 0.3)$)node[left]{$k$}--($(ckd)+(-1.2, 0.3)$);
        \draw[line:thick] (qrng.east) |- node[pos=0.8]{$/$}node[pos=0.8, below]{256}node[pos=0.8, above]{$e$}($(qrng.east)+(0.8,0)$)node[branch]{} |-(conc);
        \draw[line:thick] (qrng.east)|-(sha);
        \draw[line:thick] (sha)|-node[pos=0.8]{$/$}node[pos=0.8, below]{8}(conc);
        \draw[line:thick] (conc)--(mng);
        \draw[line:thick] ($(mng.east)+(0,0.05)$)-|($(mng)+(0.8,-0.6)$)node[pos=1.1, below]{$mcs$};
        \draw[line:thick] ($(mux2.south)+(0.3, 0.7)$)node[above]{$mcsIn$} -- ($(mux2.south)+(0.3, 0)$);
        \draw[line:thick] ($(mng.east)+(0,0.05)$) -| node[pos=0.1]{$/$}node[pos=0.3, above]{2048}($(mux2.south)+(-0.3, 0)$);% ($(reg3)+(-0.4, 0.45)$);
        \draw[line:thick] ($(reg3)+(0.4, 0.45)$) -| node[branch]{}node[pos=0.2, above]{$d$}node[pos=0.8, left]{$dL$}($(mux1)+(-0.3, 0.25)$);
        \draw[line:thick] ($(reg3)+(0.4, 0.45)$)-|node[pos=0.8, right]{$dR$}($(mux1)+(0.3, 0.25)$); 
        \draw[line:thick] ($(mux2.north)+(0,0)$) |- (reg3.west);  
        \draw[line:thick] ($(fsm.west)+(-0.8,0)$)node[left]{$clk$} -- ($(fsm.west)+(-0.4,0)$);
        
    \end{tikzpicture}
    }
    \caption{Proposed hardware architecture of \glsfirst{bip}-39 used to generate the seed and mnemonics. The architecture uses \gls{hmac}\gls{sha}-512 module inside the \gls{ckd} function and \gls{sha}-512 inside the \gls{hmac}\gls{sha}-512 module, reducing the size of the device.}
    \label{fig:bip39RTL}
\end{figure}

\subsection{Architecture of BIP39}
\label{subsec:bip39arch}

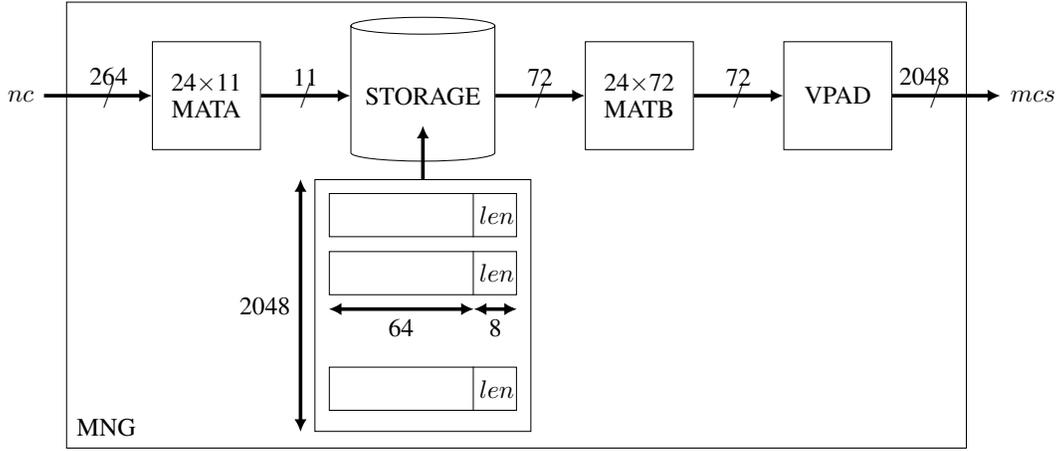
\begin{figure*}[t]
    \centering
    \resizebox{0.8\textwidth}{!}{
    \begin{tikzpicture}
    \node (mng) [rectangle,  minimum width=12.5cm, minimum height=6.2cm, draw] at (1.3,-1.8){};
    \node at ($(mng.south)+(-5.7,0.3)$){MNG};
    \node (memory) [cylinder,  rotate=90, draw,minimum height=2cm,minimum width=2cm] {};
    \node at ([yshift=0cm]memory) {STORAGE};
    \node (matrixA) [rectangle,  minimum width=1.5cm, minimum height=1.5cm, draw] at ($(memory.north)+(-2,0)$){\shortstack{24$\times$11\\MATA}};
    \node (matrixB) [rectangle,  minimum width=1.5cm, minimum height=1.5cm, draw] at ($(memory.south)+(2,0)$){\shortstack{24$\times$72\\MATB}};
    \node (vpad) [rectangle,  minimum width=1.5cm, minimum height=1.5cm, draw] at ($(matrixB.east)+(2,0)$){VPAD};
    \node (mem) [rectangle,  minimum width=3cm, minimum height=3.5cm, draw] at ($(memory.west)+(0,-2)$){};
    \node (len1) [rectangle,  minimum width=2.6cm, minimum height=0.6cm, draw] at ($(mem.north)+(0,-0.5)$){};
        \draw ($(len1.north)+(0.7,0)$) -- ($(len1.south)+(0.7,0)$);
        \node at ($(len1.east)+(-0.28,0)$){$len$};
    \node (len2) [rectangle,  minimum width=2.6cm, minimum height=0.6cm, draw] at ($(len1.south)+(0,-0.5)$){};
        \draw ($(len2.north)+(0.7,0)$) -- ($(len2.south)+(0.7,0)$);
        \node at ($(len2.east)+(-0.28,0)$){$len$};
    \node (len3) [rectangle,  minimum width=2.6cm, minimum height=0.6cm, draw] at ($(len2.south)+(0,-1.3)$){};
        \draw ($(len3.north)+(0.7,0)$) -- ($(len3.south)+(0.7,0)$);
        \node at ($(len3.east)+(-0.28,0)$){$len$};

    \draw[line:thickAr]($(mem.north)+(-1.7,0)$) -- node[pos=0.5,left]{2048}($(mem.south)+(-1.7,0)$);
    \draw[line:thickAr]($(len2.west)+(0,-0.5)$) -- node[pos=0.5,below]{64}($(len2.east)+(-0.6,-0.5)$);
    \draw[line:thickAr]($(len2.east)+(-0.6,-0.5)$) -- node[pos=0.5,below]{8}($(len2.east)+(0,-0.5)$);
    \draw[line:thick](mem.north) -- ($(memory.west)+(0,0.5)$);
    \draw[line:thick](matrixA.east) -- node[pos=0.5,above]{11}node[pos=0.5]{$/$}(memory.north);
    \draw[line:thick](memory.south) -- node[pos=0.5,above]{72}node[pos=0.5]{$/$}(matrixB.west);
    \draw[line:thick](matrixB.east) -- node[pos=0.5,above]{72}node[pos=0.5]{$/$}(vpad.west);
    \draw[line:thick](vpad.east) -- node[pos=0.3,above]{2048}node[pos=0.4]{$/$}($(vpad.east)+(1.5,0)$)node[right]{$mcs$};
    \draw[line:thick]($(matrixA.west)+(-1.5,0)$)node[left]{$nc$} -- node[pos=0.6,above]{264}node[pos=0.6]{$/$}(matrixA.west);
\end{tikzpicture}
    }
    \caption{Architecture of the MNG module. The input $nc$ denotes the checksummed entropy $e$. The memory stores 2,048 English mnemonic words, each with a bit-length specified by $len$. The MATA (24×11) and MATB (24×72) blocks perform word indexing and selection, while the VPAD unit creates the mnemonic using valid words. The resulting mnemonic word is provided at the output $mcs$.}
    \label{fig:mng}
\end{figure*}

\autoref{fig:bip39RTL} presents the proposed hardware architecture for executing the \gls{bip}-39 protocol. The protocol relies on the random number $e$ generated by the \gls{rng}, the \gls{pbkdf} shown in \autoref{fig:pbkdf2}, and the \gls{sha}-256 algorithm to generate both the mnemonic codes ($mcs$) and the seed value ($dk_{out}$).

The MNG module computes the mnemonic using the \gls{bip}-39 English wordlist, which contains 2048 words \cite{bip39wordlist}. Its architecture is shown in \autoref{fig:mng}. In this design, $nc$ represents the checksummed entropy used to calculate the mnemonic indices. It is derived using the following equations:

\begin{equation}
CS = \frac{ENT}{32}, \quad
MS = \frac{ENT + CS}{11},
\label{eq:mnc}
\end{equation}

where $CS$ is the checksum length in bits, $ENT$ is the entropy length in bits, and $MS$ is the number of mnemonic words. Moreover, $ENT$ can take values of 128, 160, 192, 224, or 256 \cite{bip39mnc}. The checksum is generated by taking the \gls{sha}-256 hash of $e$.

In EthVault, $ENT$ is set to 256. Hence, $nc$ in \autoref{fig:mng} is 256 bits, and $MS$ equals 24. Accordingly, MATA is created with 24 indices, each 11 bits long. The wordlist is stored in STORAGE with 64-bit word sizes. The size corresponds to the longest word in the list. Since there are shorter words, the number of valid bits in the word is appended to each word using 8 bits, denoted as $len$ in \autoref{fig:mng}.

Using MATA as the index of selected mnemonic words in STORAGE, MATB is formed, which contains the selected mnemonic words. Then, by applying $len$ to collect the valid bits, we construct $mcs$, the mnemonic phrase arranged according to the indices. Accordingly, $mcs$ must contain consecutive valid bits, since it is later hashed using \gls{hmac}\gls{sha}-512. Any extra bit would lead to an incorrect hash.

The output, $mcs$ in \autoref{fig:bip39RTL}, is padded with zeros to ensure its length is a multiple of 128 bytes, which corresponds to the block size of the \gls{sha}-512 hash function. In the proposed architecture, $mcs$ is padded to 2048 bits and stored in register R3, whose output is denoted as $d$. The data in R3 is then divided into two equal blocks: $dR = d[1023:0]$ and $dL = d[2047:1024]$, each of which is hashed using \gls{sha}-512. The resulting hash digest is stored in R0, which serves as the $Pwd$ input to the \gls{pbkdf} function in \autoref{fig:pbkdf2}. Likewise, the $Slt$ input in \autoref{fig:bip39RTL} corresponds to the $Slt$ input defined in \autoref{fig:pbkdf2}.

In an Ethereum \gls{hd}, $Slt$ is represented by the \gls{ascii} string “mnemonics” which is padded with an optional 416-bit \gls{pin} chosen by the user to enhance security. Since EthVault currently does not require a \gls{pin}, $Slt$ is instead padded with zeros. After storing the \gls{sha}-512 digest in R0, the \gls{pbkdf} computation begins. A counter, CNTR, tracks the execution of 2048 \gls{hmac}\gls{sha}-512 operations, with each digest being XORed and stored in R1. The seed generated in R1 ($dk_{out}$) is then used in the subsequent child key derivation process.

\subsubsection{Optimization} The proposed architecture uses the \gls{sha}-512 instance inside the \gls{hmac}\gls{sha}-512 module. Also, the architecture uses the \gls{hmac}\gls{sha}-512 module within the \gls{ckd} function using control signals $j$ and $k$. By reusing these modules, the architecture minimizes resource usage, as only one instance of each is required.

\subsubsection{Validation and Testing}
The SHA256 module was validated using official \gls{nist} test vectors, which include a variety of edge cases \cite{sha512}. In addition, the PBKDF2 module employing the \gls{sha}-512 hash function, highlighted in the light red rugged region of \autoref{fig:bip39RTL}, was independently tested and verified using a Python implementation of its architecture. This Python implementation generated multiple input-output test vectors with varying passwords, salts, and iteration counts to ensure correct functionality across different scenarios. The validated edge cases are shown in \autoref{table:edgepbsk2}.

\begin{table}[t]
\centering
\caption{PBKDF2 edge cases for functional verification.}
\begin{tabular}{ll}
\hline \hline
Parameter & Edge Cases \\
\hline \hline
Password ($d$) & All $0$s, All $1$s, repeating patterns \\
Salt ($Slt$) & All $0$s, All $1$s, reused salts across different passwords \\
Iteration Count & All $0$s, All $1$s, very large (stress test, e.g., $2^{32}$) \\
\hline \hline
\end{tabular}
\label{table:edgepbsk2}
\end{table}

Finally, the complete BIP39 implementation was tested and verified against the standard test vectors in \cite{bip39}, as well as additional entropy and mnemonic combinations generated by \cite{mnem}. These tests included edge cases for entropy ($e$), such as all zeros, all ones, repeated patterns, small values, and large values, and for mnemonic input ($mcsIn$), such as all zeros and all ones. Moreover, the functional verification achieved a coverage of 99\% for statements, 98\% for branches, and 93\% for conditions, indicating the thoroughness of the test vectors.

The following section discusses the proposed architecture of the \gls{hmac}\gls{sha}-512 hash algorithm.
%https://ieeexplore.ieee.org/stamp/stamp.jsp?tp=&arnumber=7396499

 \subsection{Architecture of HMACSHA512}
 \label{sebsec:hmac}

\autoref{fig:eth_hmac} depicts the architecture of the proposed universal \gls{hmac}\gls{sha}-512. This single architecture handles \gls{hmac}\gls{sha}-512 hashing in various key derivation stages in the wallet by selecting different inputs. Specifically, either \textit{k\_0} and \textit{m\_0} or \textit{k\_1} and \textit{m\_1} are chosen for $k$ and $m$ in \eqref{hmac_eqn}, respectively, depending on the task (e.g., creating the seed value or generating the master key).

The PDC module prepares the input for \gls{sha}-512 by padding it to a multiple of 1024 bits and dividing it into 1024-bit blocks. For instance, when calculating the \gls{sha}-512 digest of a 1536-bit input, denoted as $\bm{x}$, the input is first padded to 2048 bits, then split into two 1024-bit blocks. The \gls{sha}-512 algorithm processes the first block, retaining intermediate variables for subsequent computation on the second block. The final output is the \gls{sha}-512 hash digest of $\bm{x}$.

To demonstrate the operation of the proposed architecture, let’s consider an example where \textit{k\_0} and \textit{m\_0} are selected as inputs, corresponding to \textit{k} and \textit{m} in \eqref{hmac_eqn}. First, these inputs are XORed with the padding values \textit{opad} and \textit{ipad} and concatenated to form the initial \gls{hmac} input blocks. They are then padded and divided into 1024-bit chunks, making them ready for processing by the \gls{sha}-512 algorithm. The \gls{sha}-512 algorithm subsequently performs four rounds of hashing, with an intermediate register (R0) storing the partial results after each step. Each pair of rounds corresponds to a separate instance of \gls{sha}-512, as depicted in \eqref{hmac_eqn}. This operation demonstrates how the architecture efficiently processes data in stages, producing the final \gls{hmac}\gls{sha}-512 output through layered hash calculations and intermediate result handling. 

\subsubsection{Optimizations} 
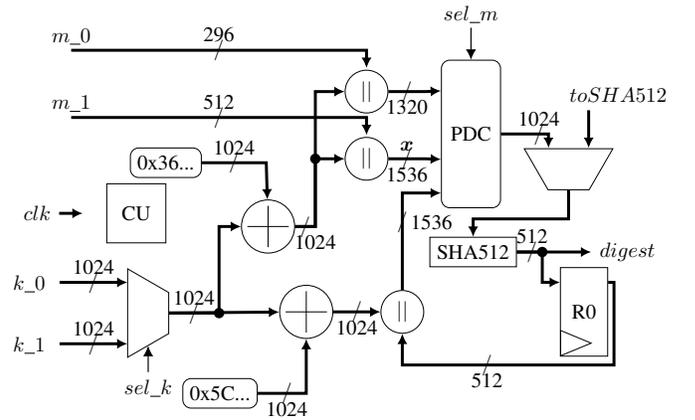
\begin{figure}[t]
\centering
    \resizebox{0.5\textwidth}{!}{
            \begin{tikzpicture}
    %nodes
    \node [circle, draw](and2){$||$}; %and2
    \node [circle, draw](and3)[below=0.4cm of and2]{$||$}; %and3
    \node [rectangle, rounded corners, minimum width=1cm, minimum height=2.5cm, draw](pad)[below right=-0.8cm and 1cm of and2]{PDC}; %pad
    \node [trapezium, draw, rotate=180, minimum width=1.5cm, minimum height=0.5cm](mux1)[above right=-1.5cm and 1.5cm of pad]{}; %mux1 close to sha512
    \node [rectangle, minimum width=1cm, minimum height=0.5cm, draw](sha)[below right=1.5cm and -2cm of mux1]{\gls{sha}512}; %sha
    \node [rectangle, minimum width=0.8cm, minimum height=1.5cm, draw](dff)[below right=1cm and 1cm of pad]{R0}; % dff
    \node[isosceles triangle,  minimum width=2.8mm, minimum height=0.8mm, draw] (clk1)[below left=-0.35cm and -0.25cm of dff]{}; %clk1
    \node [circle, draw](and0)[left=2.3cm of dff]{$||$}; %and0 xor
    \node [circle, minimum size=0.9cm, font=\footnotesize, draw](and1)[left=0.8cm of and0]{}; %and1 xor
        \draw ($(and1)+(0, 0.35)$) -- ($(and1)+(0, -0.35)$); % verticle xor0
        \draw ($(and1)+(-0.35, 0)$) -- ($(and1)+(0.35, 0)$); % horizontal xor0
    \node [circle, minimum size=0.9cm, font=\footnotesize, draw](and4)[above left=0.8cm and 0cm of and1]{}; %and4
        \draw ($(and4)+(0, 0.35)$) -- ($(and4)+(0, -0.35)$); % verticle xor0
        \draw ($(and4)+(-0.35, 0)$) -- ($(and4)+(0.35, 0)$); % horizontal xor0
    \node [rectangle, rounded corners, minimum width=1cm, minimum height=0.5cm, draw](0x5c)[below left=0.8cm and 0.5cm of and1]{0x5C...}; %Ox5c
    \node [trapezium, draw, rotate=-90, minimum width=1.5cm, minimum height=0.5cm](mux0)[below left=0.05cm and 2cm of and1]{}; %mux0
    \node [rectangle, rounded corners, minimum width=1cm, minimum height=0.5cm, draw](0x36)[above left=0.5cm and 0.8cm of and4]{0x36...}; %Ox36
    \node [rectangle, minimum width=1cm, minimum height=1cm, draw](cu) at ($(0x36.south)+(-0.5, -0.6)$) {CU}; %k0
    %\node [rectangle, minimum width=0.8cm, minimum height=0.5cm, draw](k1)[below left=0.6cm and 1.3cm of mux0]{k$||$pad}; %k1
    
    %lines
    \draw[line:thick](mux1)|-($(mux1)!0.6!(sha)$)-|(sha.north);
    \draw[line:thick](mux0)--(and1);
    \draw[line:thick](and1)--(and0)node[midway]{$/$}node[midway, below]{1024};
    \draw[line:thick](and2)--($(and2)+(1, 0)$)node[midway]{$/$}node[midway, below]{1320}--($(pad.west)+(0, 0.73)$);
    \draw[line:thick](and3)--($(and3)+(1, 0)$)node[midway]{$/$}node[midway, below]{1536}node[midway, above]{$\bm{x}$}--($(pad.west)+(0, -0.43)$);
    \draw[line:thick]($(pad.east)+(0, 0)$)-|node[pos=0.4]{$/$}node[pos=0.4, above]{1024}($(mux1)+(-0.34,0.36)$);
    \draw[line:thick]($(mux1)+(0.34,1)$)node[xshift=5mm, above]{$toSHA512$}--($(mux1)+(0.34,0.36)$);
    \draw[line:thick]($(dff.east)+(0,0.5)$)-|($(dff.east)+(0.1,-0.9)$) -| (and0.south)node[pos=0.3]{$/$}node[pos=0.3, below]{512};
    \draw[line:thick]($(mux0)+(-1.5, 0.52)$)node[left=0.1cm]{$k\_0$}--($(mux0)+(-0.35, 0.52)$)node[midway]{$/$}node[midway, above]{1024};
    \draw[line:thick]($(mux0)+(-1.5, -0.52)$)node[left=0.1cm]{$k\_1$}--($(mux0)+(-0.35, -0.52)$)node[midway]{$/$}node[midway, above]{1024};
    \draw[line:thick](mux0)--($(mux0)+(1.2, 0)$)node[midway]{$/$}node[midway, above]{1024}node[branch]{}|-(and4);
    \draw[line:thick](0x5c)-|node[pos=0.3]{$/$}node[pos=0.3, below]{1024}(and1);
    \draw[line:thick](0x36)--($(0x36)+(1.7, 0)$)node[midway]{$/$}node[midway, above]{1024}--(and4);
    \draw[line:thick](and4)--($(and4)+(0.8, 0)$)node[midway]{$/$}node[right, below]{1024}|-(and2);
    \draw[line:thick](and4)--($(and4)+(0.8, 0)$) |- node[branch]{}(and3);
    \draw[line:thick]($(and3)+(-5, 0.7)$)node[left, above]{$m\_1$}--($(and3)+(0, 0.7)$)node[midway]{$/$}node[midway, above]{512}--($(and3)+(0, 0.4)$);
    \draw[line:thick]($(and2)+(-5, 0.7)$)node[left, above]{$m\_0$}--($(and2)+(0, 0.7)$)node[midway]{$/$}node[midway, above]{296}--($(and2)+(0, 0.4)$);
    \draw[line:thick](and0)|-node[pos=0.35]{$/$}node[pos=0.35, right]{1536}($(pad.west)+(0, -1)$);
    \draw[line:thick](sha.east) -| node[branch]{}node[pos=0.3]{$/$}node[pos=0.3, above]{512}($(dff.west)+(-0.3,0.9)$)|-($(dff.west)+(0,0.5)$);
    \draw[line:thick](sha)--($(sha)+(2, 0)$)node[right]{$digest$};
    %\draw[line:thick]($(k0)+(-2, 0)$)--(k0)node[midway]{$/$}node[midway, above]{256}node[left=2cm]{k\_0};
    %\draw[line:thick]($(k1)+(-2, 0)$)--(k1)node[midway]{$/$}node[midway, above]{96}node[left=2cm]{k\_1};
    %\draw[line width=0.2mm, arrows = {-Latex[width=5pt, length=5pt]}]($(dff)+(1, -0.5)$)node[right]{en\_dff}--($(dff)+(0.4, -0.5)$);
    \draw[line width=0.2mm, arrows = {-Latex[width=5pt, length=5pt]}]($(pad)+(0, 1.8)$)node[above]{$sel\_m$}--(pad);
    \draw[line width=0.2mm, arrows = {-Latex[width=5pt, length=5pt]}]($(mux0)+(0, -1)$)node[below]{$sel\_k$}--(mux0); 
    \draw[line:thick]($(cu.west)+(-0.8,0)$)node[left]{$clk$} -- ($(cu.west)+(-0.4,0)$);
\end{tikzpicture}%}
    }
    \caption{The hardware architecture of the proposed \gls{hmac}\gls{sha}-512 designed for an Ethereum \gls{hd} \gls{cryp} wallet. The architecture has two pairs of key ($k\_0$, $k\_1$) and message ($m\_0$, $m\_1$) inputs that enable it to be used in all the stages of the \gls{hd} wallet.}
    \label{fig:eth_hmac}
\end{figure}

As shown in \eqref{hmac_eqn}, the \gls{hmac}\gls{sha}-512 process involves executing the \gls{sha}-512 function (denoted by \( H(\cdot) \)) twice. To optimize resource usage, however, the proposed hardware architecture, illustrated in \autoref{fig:eth_hmac}, implements only a single instance of the \gls{sha}-512 function for the entire \gls{hmac}\gls{sha}-512 operation. This design reduces resource consumption while still achieving the necessary functionality.

Furthermore, as outlined in \autoref{subsec:hd_wal}, the \gls{hmac}\gls{sha}-512 function is applied at various stages of the Ethereum \gls{cryp} wallet’s key generation process. To support this multi-stage requirement, inputs \textit{k\_0}, \textit{m\_0}, \textit{k\_1}, and \textit{m\_1} are introduced, allowing the creation of a universal \gls{hmac}\gls{sha}-512 instance that can be shared across all key generation stages.

Additionally, \autoref{subsec:bip39arch} explains the role of the \gls{sha}-512 hashing algorithm in the seed generation process. Therefore, the architecture includes a dedicated $toSHA512$ input which enables the wallet to use \gls{sha}-512 independently within the broader \gls{hmac}\gls{sha}-512 framework.

\subsubsection{Validation and Testing}

The \gls{hmac}\gls{sha}-512 and \gls{sha}-512 architectures were independently tested and validated using test vectors containing edge cases from \cite{hmacTest} and \cite{sha512}, respectively. Additionally, further edge cases were evaluated, in which the inputs $k\_0$, $k\_1$, $m\_0$, and $m\_1$ consisted entirely of zeros or ones, and the control signals $sel\_k$, $sel\_m$ toggled to select different internal and external padding.

The functional verification of \gls{hmac}\gls{sha}-512 achieved 100\% coverage for statements, 100\% for branches, and 97\% for conditions, while the functional verification of \gls{sha}-512 achieved 100\% coverage for statements, 100\% for branches, and 98\% for conditions.

The next section discusses the proposed hardware architecture of the \gls{ckd} function.

\subsection{Architecture of the CKDF Module}
\label{subsec:ckdf:ARCH}

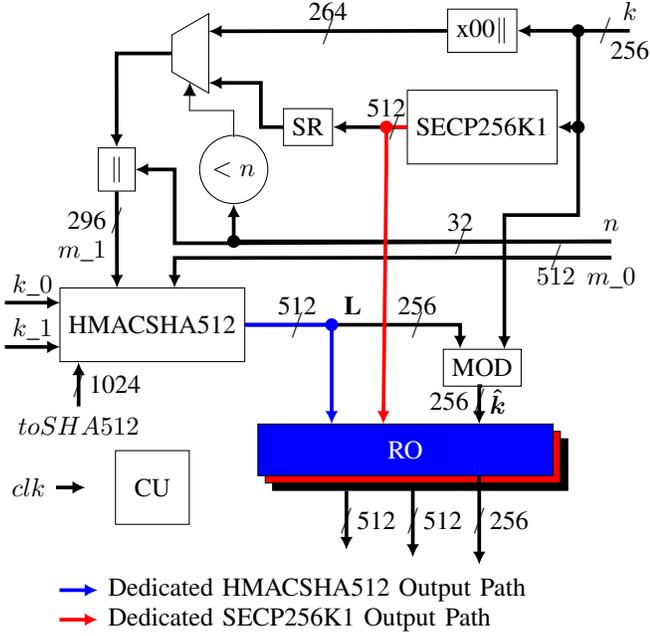
\begin{figure}[t]
\centering
    \resizebox{0.5\textwidth}{!}{
    \begin{tikzpicture}
    %\node [rectangle, minimum width=4cm, minimum height=2cm, draw](bod){}; %mmalu
    \node [rectangle, minimum width=2cm, minimum height=1cm, draw](secp){SECP256K1}; %secp
    \node [rectangle, minimum width=0.5cm, minimum height=0.5cm, draw](compres)[left=1cm of secp]{SR}; %cmprs
    \node [rectangle, minimum width=1cm, minimum height=0.5cm, draw](md)[below=2.5cm of secp]{MOD}; %md
    \node [trapezium, draw, rotate=90, minimum width=1cm, minimum height=0.5cm](mux0)[above left=1cm and 1cm of compres]{}; %mux0
    \node [fill=black, rectangle, minimum width=4cm, minimum height=0.7cm, draw](out1)[below left=0.7cm and -1.7 of md]{}; %outreg1
    \node [fill=red, rectangle, minimum width=4cm, minimum height=0.7cm, draw](out0)[below left=0.6cm and -1.6 of md]{}; %outreg0
    \node [fill=blue, text=white, rectangle, minimum width=4cm, minimum height=0.7cm, draw](out)[below left=0.5cm and -1.5 of md]{RO}; %outreg
    %\node[isosceles triangle,  minimum width=2.8mm, minimum height=0.8mm, draw] (clk1)[left=-0.55cm of out]{}; %clk1
    \node [circle, draw](gt)[below right=1.5cm and 0cm of mux0]{$<n$}; %gt
    \node [rectangle, minimum width=0.5cm, minimum height=0.5cm, draw](mg)[left= 0.86cm of gt]{$\parallel$}; %mg
    \node [rectangle, minimum width=2.5cm, minimum height=1cm, draw](hmac)[below left=1.3cm and -2cm of mg]{HMACSHA512}; %hmac
    \node [rectangle, minimum width=0.5cm, minimum height=0.5cm, draw](mg0)[above=0.5cm of secp]{x00$\parallel$}; %mg0
    \node [rectangle, minimum width=1cm, minimum height=1cm, draw](cu) at ($(hmac.south)+(0, -1.7)$){CU}; %cu

    %Legend
    \draw[blue,line:thick]($(hmac.west)+(0, -3.6)$) -- ($(hmac.east)+(-2, -3.6)$)node[right, black]{Dedicated HMACSHA512 Output Path};
    \draw[red,line:thick]($(hmac.west)+(0, -4)$) -- ($(hmac.east)+(-2, -4)$)node[right, black]{Dedicated SECP256K1 Output Path};

    %line
    \draw[line:thick]($(hmac.south)+(-1, -0.6)$)node[below]{$toSHA512$}--node[pos=0.5]{$/$}node[pos=0.5,right]{1024}($(hmac.south)+(-1, 0)$);
    \draw[line:thick](mg0)--node[pos=0.5]{$/$}node[pos=0.5, above]{264}($(mux0)+(0.25, 0.3)$);
    \draw[line:thick](secp)--(compres);
    \draw[line:thick]($(compres.west)+(0, 0)$)--($(compres.west)+(-0.3, 0)$) |-($(mux0)+(0.25, -0.4)$);
    \draw[line:thick](mux0)--($(mux0)+(-1, 0)$)--(mg);
    \draw[line:thick](mg)--node[pos=0.6,left]{$m\_1$}node[pos=0.3]{$/$}node[pos=0.3,left]{296}($(hmac)+(-0.48, 0.5)$);
    \draw[line:thin]($(gt)+(0, 0.44)$)--($(gt)+(0, 0.8)$)--($(gt)+(-0.6, 0.8)$)--(mux0);
    \draw[line:thick]($(secp)+(1.75, -1.55)$)--($(gt)+(0, -1)$)--($(gt)+(-0.8, -1)$)--($(gt)+(-0.8, 0)$)--($(mg)+(0.25, 0)$);
    \draw[line:thick]($(secp)+(1.75, -1.55)$)node[right, above]{$n$} -| node[branch]{}node[pos=0.2]{$/$}node[pos=0.2, above]{32}(gt);
    \draw[line:thick]($(secp)+(1.75, -1.769)$)--node[below] {$m\_0$} ++(0,0)--node[midway] {$/$} ++(-1.5,0)--node[below] {$512$} ++(1.5,0)--($(gt)+(-0.8, -1.2)$)--($(hmac)+(0.3, 0.5)$);%%%--node[above] {en} ++(0.35,0)
    \draw[line:thick]($(md)+(0,-0.25)$)-|node[pos=0.7, right]{$\bm{\hat{k}}$}node[pos=0.7]{$/$}node[pos=0.7, left]{256}($(out)+(1, 0.35)$);
    \draw[line:thick](hmac)-|node[pos=0.25, above]{$\textbf{L}$}node[pos=0.4]{$/$}node[pos=0.4, above]{256}($(md.north)+(-0.3,0)$);
    \draw[blue, line:thick](hmac) -| node[branch]{}node[black,pos=0.3]{$/$}node[black, pos=0.3, above]{512}($(out)+(-1,0.35)$);
    \draw[red, line:thick](secp) -| node[branch]{}($(secp)+(-1.3,-0.5)$)node[black, pos=0.3]{$/$}node[black, pos=0.5, above]{512}--($(out)+(-0.3,0.35)$);
    \draw[line:thick]($(mg0)+(2, 0)$)node[right, above]{$k$}node[right, below=0cm and -0.8cm]{256}--node[right=0.2]{$/$}(mg0);
    \draw[line:thick]($(mg0)+(1.3, 0)$)--($(mg0)+(1.3, -2.5)$)--($(mg0)+(0.3, -2.5)$)--($(md)+(0.3,0.25)$);
    \draw[line:thick]($(mg0)+(1.3, 0)$)node[branch]{} |- node[branch]{}(secp);
    \draw[line:thick]($(hmac)+(-2, -0.3)$)--node[midway, above]{$k\_1$}($(hmac)+(-1.25, -0.3)$);
    \draw[line:thick]($(hmac)+(-2, 0.3)$)--node[midway, above]{$k\_0$}($(hmac)+(-1.25, 0.3)$); 
    \draw[line:thick]($(out.south)+(1, 0)$)--node[midway]{$/$}node[midway, right]{256}($(out.south)+(1, -1.2)$);
    \draw[line:thick]($(out0.south)+(0,0)$)--node[midway]{$/$}node[midway, right]{512}($(out0.south)+(0, -1)$);
    \draw[line:thick]($(out1.south)+(-1,0)$)--node[midway]{$/$}node[midway, right]{512}($(out1.south)+(-1, -0.8)$);
    \draw[line:thick]($(cu.west)+(-0.8,0)$)node[left]{$clk$} -- ($(cu.west)+(-0.4,0)$);
\end{tikzpicture}
    }
    \caption{The proposed hardware architecture of the \glsfirst{ckd} function. SR is the serialization of the public key to a compressed format. $k\_0$, $k\_1$, and $toSHA512$ are 1024-bit inputs. }
    \label{fig:ckdf_fig}
\end{figure}

\autoref{fig:ckdf_fig} illustrates the proposed architecture of the CKDF module in \autoref{fig:ethWalArch}, as described in \autoref{ckdf_algo}. The inputs, $k\_0$ and $k\_1$, represent the key inputs for the \gls{hmac}\gls{sha}-512 function, while the $m\_0$ input is message data. The signal $toSHA512$ is the input to the SHA-512 function within the HMACSHA512 module. The parameters $k$ and $n$ are the private key and child number, respectively, provided as inputs to the \gls{ckd} function. The modulus module (MOD) performs modular addition with respect to $p$, the 256-bit prime number defined by SECP256K1.

As described in \autoref{ckdf_algo}, the inputs $k$, $c$, and $n$ are loaded at the start of the \gls{ckd} function. Notably, $c$ is provided through the $m\_0$ input, as illustrated in \autoref{fig:ckdf_fig}. The CKDF module then executes according to the steps in \autoref{ckdf_algo}, processing these inputs to generate the child private key and chain code. These outputs are subsequently stored in the output registers (R0) for further use.

\subsubsection{Optimizations} The algorithms employed by the \gls{ckd} function also perform other functions outside the \gls{ckd} function. For instance, \gls{hmac}\gls{sha}-512 is utilized within the \gls{pbkdf} of the \gls{bip}-39 protocol to generate seed values and is also involved in creating the master private key and master chain code outside the \gls{bip}-39 and \gls{ckd} functions. Additionally, the \gls{sha}-512 algorithm, which operates within the \gls{pbkdf}, computes various digests as outlined in \autoref{subsec:bip39arch}. Similarly, after performing child key generation within the \gls{ckd} function, SECP256K1 is used again outside the \gls{ckd} function to compute public keys.

To reduce the resources needed by the device, we optimized the \gls{ckd} function by modifying its algorithm and reusing components for multiple processes, as depicted in the proposed hardware architecture illustrated in \autoref{fig:ckdf_fig}. This design allows each algorithm to be accessed directly through dedicated inputs and output paths, enhancing resource efficiency. The inputs $k\_0$, $k\_1$, $toSHA512$, and $m\_0$ allow \gls{hmac}\gls{sha}-512 and \gls{sha}-512 to operate independent of the \gls{ckd} function and store their outputs in dedicated registers via the blue data path. Furthermore, the SECP256K1 output is routed through the red data path to an output register, enabling its reuse in external processes via the input $k$.

\subsubsection{Validation and Testing}
To verify the correctness of the CKDF module, a software reference model was developed in Python. Using a known seed, the script generated a master key and chain code, followed by the derivation of several child private keys. The generated keys were verified against the outputs of an online \gls{hd} wallet generator \cite{mnem}.

The CKDF hardware module was then tested using the master key and chain code as inputs and validated by comparing its output (child public and private keys) with the Python-generated reference. Additionally, the internal cryptographic modules (\gls{hmac}\gls{sha}-512, \gls{sha}-512, and SECP256K1) were tested independently using control signals, as they had already been validated in earlier stages.

Edge cases of the \gls{ckd} function were also tested, including both hardened and non-hardened key derivation (i.e., $n \geq 2^{31}$ and $n < 2^{31}$), as well as invalid derivation scenarios resulting from edge conditions in the modular arithmetic, such as scalar overflow, resulting in $k \bmod n = 0$, or attempts to derive a child key that would produce a point at infinity. Moreover, the functional verification of the CKDF module achieved 96\% coverage for statements, 96\% for branches, and 90\% for conditions.

The following section discusses the implementation of the Ethereum checksum encoding.

\subsection{ Architecture of Ethereum Checksum}

A checksummed Ethereum address consists of a mix of numbers and both uppercase and lowercase letters. To create a hardware implementation of the uppercase conversion described in \autoref{subsec:EthAddr}, the function $\textsc{capital()}$ converts the hexadecimal address to \gls{ascii} format. This conversion is optimized by using fixed offsets for each 4-bit group of $a$, avoiding the use of \glspl{lut} and reducing resource utilization.

\subsubsection{Validation and Testing}

This implementation was validated against the reference software implementation proposed by Ethereum \cite{EIP55}. In addition to the standard test vectors provided in \cite{EIP55}, edge cases such as inputs composed entirely of numbers, all zeros, and only letters were also tested and successfully passed. Also, the functional verification of the CKDF module achieved 100\% coverage for statements, 100\% for branches, and 94\% for conditions.

The following section discusses the proposed architecture of the \gls{ecdsa} algorithm.

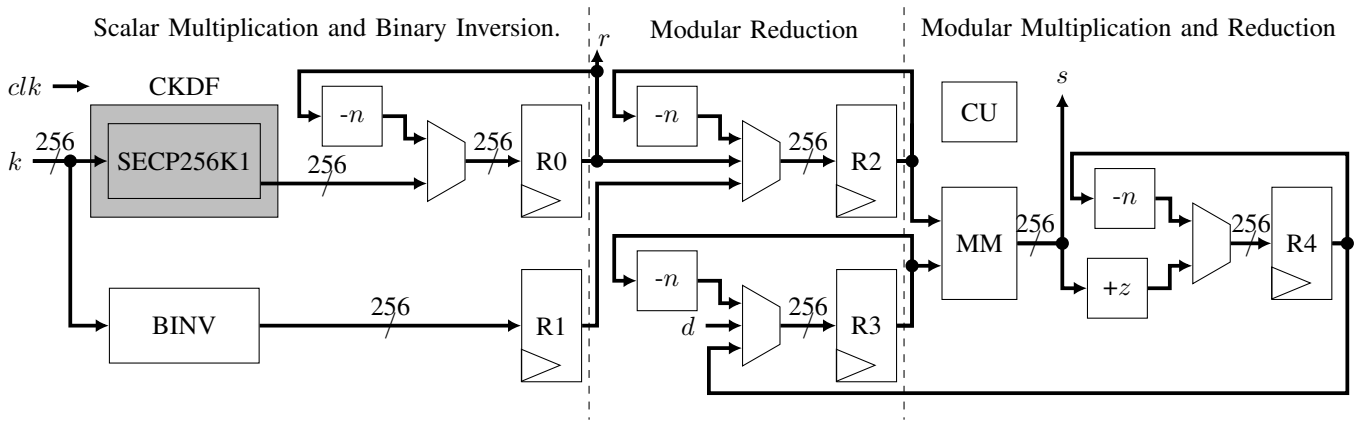
\begin{figure*}
    \centering
    \resizebox{\textwidth}{!}{\begin{tikzpicture}
    \node (mm) [rectangle, minimum width=1cm, minimum height=1.5cm, draw] at (0,0){MM};
    \node (r2) [rectangle, minimum width=0.8cm, minimum height=1.5cm, draw] at ($(mm.west)+(-1,1.1)$){R2};
    \node (clkr2) [clk, draw] at ($(r2.west)+(0.12,-0.53)$){};
    \node (r3) [rectangle, minimum width=0.8cm, minimum height=1.5cm, draw] at ($(mm.west)+(-1,-1.1)$){R3};
    \node (clkr3) [clk, draw] at ($(r3.west)+(0.12,-0.53)$){};
    \node (mux2) [trapezium, draw, rotate=-90, minimum width=1cm, minimum height=0.5cm] at ($(r2.west)+(-1, 0)$){}; 
    \node (mux3) [trapezium, draw, rotate=-90, minimum width=1cm, minimum height=0.5cm] at ($(r3.west)+(-1, 0)$){};
    \node (n2) [rectangle, minimum width=0.8cm, minimum height=0.8cm, draw] at ($(mux2.south)+(-1, 0.6)$){-$n$};
    \node (n3) [rectangle, minimum width=0.8cm, minimum height=0.8cm, draw] at ($(mux3.south)+(-1, 0.6)$){-$n$};
    \node (r1) [rectangle, minimum width=0.8cm, minimum height=1.5cm, draw] at ($(r3.west)+(-3.8,0)$){R1};
    \node (clkr1) [clk, draw] at ($(r1.west)+(0.12,-0.53)$){};
    \node (r0) [rectangle, minimum width=0.8cm, minimum height=1.5cm, draw] at ($(r2.west)+(-3.8,0)$){R0};
    \node (clkr0) [clk, draw] at ($(r0.west)+(0.12,-0.53)$){};
    \node (mux0) [trapezium, draw, rotate=-90, minimum width=1cm, minimum height=0.5cm] at ($(r0.west)+(-1, 0)$){};
    \node (n0) [rectangle, minimum width=0.8cm, minimum height=0.8cm, draw] at ($(mux0.south)+(-1, 0.6)$){-$n$};
    \node (ckdf) [rectangle, fill=lightgray, label={CKDF}, minimum width=2.5cm, minimum height=1.5cm, draw] at ($(r0.west)+(-4.5,0)$){};
    \node (secp) [rectangle, minimum width=2cm, minimum height=1cm, draw] at (ckdf){SECP256K1};
    \node (binv) [rectangle, minimum width=2cm, minimum height=1cm, draw] at ($(r1.west)+(-4.5,0)$){BINV};    
    \node (r4) [rectangle, minimum width=0.8cm, minimum height=1.5cm, draw] at ($(mm.east)+(3.8,0)$){R4};
    \node (clkr4) [clk, draw] at ($(r4.west)+(0.12,-0.53)$){};
    \node (mux4) [trapezium, draw, rotate=-90, minimum width=1cm, minimum height=0.5cm] at ($(r4.west)+(-0.8, 0)$){};
    \node (n4) [rectangle, minimum width=0.8cm, minimum height=0.8cm, draw] at ($(mux4.south)+(-0.9, 0.6)$){-$n$};
    \node (p4) [rectangle, minimum width=0.8cm, minimum height=0.8cm, draw] at ($(mux4.south)+(-1, -0.6)$){+$z$};
    \node (cu) [rectangle, minimum width=1cm, minimum height=0.8cm, draw] at ($(mm.north)+(0, 1)$){CU};
    \node at ($(r0.north)+(-3,1)$){Scalar Multiplication and Binary Inversion.};
    \node at ($(r2.north)+(-1.5,1)$){Modular Reduction};
    \node at ($(r2.north)+(3.5,1)$){Modular Multiplication and Reduction};

    \draw[line:thick] (binv) -- node[pos=0.5]{$/$}node[pos=0.5, above]{256}(r1);
    \draw[line:thick] (mux0) -- node[pos=0.5]{$/$}node[pos=0.5, above]{256}(r0);
    \draw[line:thick] (mux2) -- node[pos=0.5]{$/$}node[pos=0.5, above]{256}(r2);
    \draw[line:thick] (mux3) -- node[pos=0.5]{$/$}node[pos=0.5, above]{256}(r3);
    \draw[line:thick] (mux4) -- node[pos=0.5]{$/$}node[pos=0.5, above]{256}(r4);
    \draw[line:thick] (r0) -- (mux2);
    \draw[line:thick] ($(secp.east)+(0,-0.3)$) |- node[pos=0.7]{$/$}node[pos=0.7, above]{256}($(mux0.south)+(0,-0.3)$);
    \draw[line:thick] (n0.east) -| ($(n0.east)+(0.3, -0.1)$) |- ($(mux0.south)+(0,0.3)$);
    \draw[line:thick] (r0.east) -| node[branch]{}($(r0.east)+(0.2, 0.4)$) |- node[branch]{}($(n0.west)+(-0.3, 0.6)$) |- (n0.west);
    \draw[line:thick] (n2.east) -| ($(n2.east)+(0.3, -0.1)$) |- ($(mux2.south)+(0,0.3)$);
    \draw[line:thick] (r2.east) -| ($(r2.east)+(0.2, 0.4)$) |- ($(n2.west)+(-0.3, 0.6)$) |- (n2.west);
    \draw[line:thick] (n3.east) -| ($(n3.east)+(0.3, -0.1)$) |- ($(mux3.south)+(0,0.3)$);
    \draw[line:thick] (r3.east) -| ($(r3.east)+(0.2, 0.4)$) |- ($(n3.west)+(-0.3, 0.6)$) |- (n3.west);
    \draw[line:thick] (r1.east) -| ($(r1.east)+(0.2, 1.3)$) |- ($(mux2.south)+(0,-0.3)$);
    \draw[line:thick] (n4.east) -| ($(n4.east)+(0.2, -0.1)$) |- ($(mux4.south)+(0,0.3)$);
    \draw[line:thick] (r4.east) -| ($(r4.east)+(0.2, 0.4)$) |- ($(n4.west)+(-0.3, 0.6)$) |- (n4.west);
    \draw[line:thick] (r3.east) -| ($(r3.east)+(0.2,0.5)$) |- node[branch]{}($(mm.west)+(0,-0.3)$);
    \draw[line:thick] (r2.east) -| node[branch]{}($(r2.east)+(0.2,0.5)$) |- ($(mm.west)+(0,0.3)$);
    \draw[line:thick] ($(mux3.north)+(-1,0)$)node[left]{$d$} -- ($(mux3.south)+(0,0)$);
    \draw[line:thick] (mm.east) -| node[pos=0.2]{$/$}node[pos=0.2, above]{256}($(mm.east)+(0.6,-0.1)$) |- (p4.west);
    \draw[line:thick] (p4.east) -| ($(mux4.south)+(-0.3, -0.3)$) -- ($(mux4.south)+(0, -0.3)$);
    \draw[line:thick] (r4.east) -| node[branch]{}($(r4.east)+(0.2, -1)$) |- ($(mux3.east)+(-0.7, -0.5)$) |- ($(mux3.south)+(0,-0.3)$);
    \draw[line:thick] ($(secp.west)+(-1,0)$)node[left]{$k$} -- node[pos=0.3]{$/$}node[pos=0.3, above]{256}(secp.west);
    \draw[line:thick] ($(secp.west)+(-1,0)$) -| node[branch]{}($(secp.west)+(-0.5,-1)$) |- (binv);
    \draw[line:thick] (mm.east) -| node[branch]{}($(mm.east)+(0.6,2)$)node[above]{$s$};
    \draw[line:thick] (r0.east) -| ($(r0.east)+(0.2, 1.5)$)node[above,yshift=-0.1cm, xshift=0.1cm]{$r$};
    \draw[line:thick] ($(ckdf.west)+(-0.5,1)$)node[left]{$clk$} -- ($(ckdf.west)+(0,1)$);
    \draw[dashed] ($(r0.north)+(0.5,1.3)$) -- ($(r1.south)+(0.5,-0.5)$);
    \draw[dashed] ($(r2.north)+(0.5,1.3)$) -- ($(r3.south)+(0.5,-0.5)$);
\end{tikzpicture}}
    \caption{Proposed hardware architecture of the \gls{ecdsa} algorithm. The SECP256K1 elliptic curve operations are encapsulated within the CKDF module. The design integrates modular arithmetic blocks, including modular inversion, multiplication, and addition, to efficiently compute the signature values $r$ and $s$.}
    \label{fig:ecdsa}
\end{figure*}

\begin{table*}[t]
\centering
\caption{ECDSA edge case test scenarios.}
\label{tab:ecdsa_edge}
\begin{tabular}{ll}
\hline \hline
Edge Case & Purpose of Test \\ 
\hline \hline
$k = 0$, $d = 0$, $z = 0$ & Tests invalid zero values. \\ 
$k$, $d$, $z$ = all $1$s (maximum bit patterns) & Ensures correct handling of maximum scalar values. \\ 
Small values of $k$, $d$, $z$ (e.g., $1,2,3$) & Verifies correctness in minimal input scenarios. \\ 
Large values of $k$, $d$, $z$ (close to $n$) & Tests boundaries near the curve order $n$. \\ 
$k > n$ & Verifies modular reduction of ephemeral key scalar. \\
$d > n$ & Ensures private key modular reduction is correctly applied. \\ 
$z > n$ & Confirms message hash is reduced modulo $n$ when required. \\
\hline \hline
\end{tabular}
\end{table*}

\subsection{Architecture of \gls{ecdsa}}
\label{subsec:ecdsaArch}

\autoref{fig:ecdsa} illustrates the proposed hardware architecture of the \gls{ecdsa} algorithm described in \autoref{alg:ecdsa}. In the figure, R0 to R4 denote registers, BINV represents the \gls{bia} algorithm defined in \autoref{alg:binInv}, and MM corresponds to the shift-and-add modular multiplication algorithm proposed in \cite{opencores0}. The architecture employs three subtractors and one adder. The inputs $k$, $d$, and $z$ denote the random number, the private child key, and the \gls{sha}-256 digest of the message to be signed, respectively. The parameter $n$ is the order of the SECP256K1 \gls{ec}, while $r$ and $s$ are the resulting signature components. The subtractors perform modular reduction with respect to $n$ on the values stored in the registers before further processing is carried out.

Once the wallet generates the public and private child keys and stores them in RAM, the \gls{ecdsa} architecture is used to authorize Ethereum transactions. Specifically, the child private key is used as $d$, a constant random number is used for $k$ (uniformly, where $k \in [1, n-1]$), and the \gls{sha}-256 hash of the transaction data is used as $z$. The corresponding public key is then transmitted to the receiver for use in the verification stage of \gls{ecdsa}. For multiple signature generations, pipelining is employed: the computation of the next signature begins immediately after the SECP256K1 and BINV modules complete their operations for the current signature, thereby improving throughput.

\subsubsection{Optimization}
To optimize the \gls{ecdsa} implementation, the SECP256K1 algorithm is embedded within the CKDF module. This integration minimizes resource utilization by avoiding the need for additional instances of the algorithm. Furthermore, the SECP256K1 and BINV modules are executed in parallel, along with the two subtractors in the middle section and the subtractor–adder pair at the rightmost part of the architecture. This parallel execution significantly reduces the latency of the signing process. In addition, pipelining is employed to further enhance the overall performance of the wallet.

\subsubsection{Validation and Testing}
The \gls{ecdsa} module was first developed, tested, and verified independently before being integrated into the EthVault architecture.

To ensure correctness, a reference software implementation was developed in Python, which generated various test vectors (i.e., $k$, $z$, $d$, $r$, and $s$) that were applied as inputs and compared against the outputs of the proposed hardware architecture. In addition, test vectors were obtained from an online \gls{ecdsa} implementation \cite{learnmeabitcoin_ecdsa} as well as from Project Wycheproof \cite{wycheproof}. The functional verification of the
ECDSA module achieved 94\% coverage for statements, 94\% for branches, and 87\% for conditions.

Several edge cases were evaluated, including settings where $k$, $d$, and $z$ were assigned all 0’s, all 1’s, small values, and large values. In addition, scenarios with $k > n$, $d > n$, and $z > n$ were tested, as detailed in \autoref{tab:ecdsa_edge}, to ensure the robustness of the design under atypical input conditions.

The next section details the implementation results and discussions of the proposed EthVault on \gls{fpga}.

\section{Implementation Results and Analysis of EthVault on FPGA}
\label{sec:impResults}

This section presents the implementation results of EthVault on an \gls{fpga}. To the best of our knowledge, no comparable hardware-based \gls{cryp} wallet architectures have been reported in the literature. As a result, the evaluation focuses on comparing the critical building blocks of our design with similar components from existing implementations.

\subsection{Target Platforms and Development Tools}

EthVault is implemented on a Xilinx ZCU104 Evaluation Kit (Part number xczu7ev-ffvc1156-2-e), which features a Zynq UltraScale+ (US) ZU7EV MPSoC. The ZU7EV includes a \gls{pl} section with 230,400 \glspl{lut}, 28,800 \glspl{clb}, 460,800 registers, and 44.2 Mb of \gls{ram}. Also, it features a \gls{ps} with a quad-core ARM Cortex-A53 processor. However, EthVault implementation resides entirely within the \gls{pl}.  

SECP256K1 and \gls{hmac}\gls{sha}-512 architectures are also implemented on an Artix-7 \gls{fpga} (Part number xc7a200tfbg676-2) for fair comparison against other 7-series \gls{fpga} implementations. All modules are described in VHDL (version 2008 or later). Synthesis and implementation are carried out in Xilinx Vivado 2022.2. A constraint file defines the clock period, start and reset signals, as well as the \gls{fpga} internal clock pin mapping. Verification is carried out through simulation in Vivado, with outputs validated against a reference software implementation \cite{shabir2023qwallet}.

\subsection{Methodology and Metrics for Evaluation}
Comparative results for SECP256K1 and \gls{hmac}\gls{sha}-512 are shown in Tables \ref{table:secp} and \ref{table:hmac}, respectively. The \say{Platform} column in these tables specifies the types of \gls{fpga} devices used in the corresponding references. It is important to note that the listed platforms differ in capabilities and implementation methods, which may limit the fairness of direct comparisons. However, metrics such as the number of \glspl{lut} utilized are deliberately used as they provide a more consistent estimate of the utilized area across platforms (since the listed platforms feature \glspl{lut} with similar input sizes). The area metric comprises \glspl{lut}, \gls{dsp} blocks, \gls{ram} blocks, and registers. 

Furthermore, the estimated power consumption of EthVault was analyzed in \autoref{table:power}, and its efficiency metrics were benchmarked against the Trezor One physical wallet \cite{donjonTREZ}.

Latency is reported in both milliseconds (ms) and \gls{cc}. Throughput is computed as (Frequency $\div$ CC) $\times$ $k$, where $k$ is the size of the output in bits \cite{romel2023fpga}. The system clock constraints are defined in the \gls{xdc} file to enforce the target operating frequency.

It is important to note that during timing analysis, Xilinx Vivado’s timing engine automatically evaluates timing using device-specific libraries that model the worst-case process, voltage, and temperature (PVT) conditions corresponding to the selected \gls{fpga} speed grade. This ensures that the reported timing margins and slack values reflect guaranteed operating limits under all supported environmental variations.

%SECP256K1 comparative results
\begin{table*}[t]
    \centering
    \caption{Comparing implementation results of the SECP256K1 algorithm.}
    \begin{threeparttable}
    \begin{tabular}{l l l l l l l l l l l}
    \hline \hline
        Work & Platform & \multicolumn{4}{c}{Area} & Frequency & \multicolumn{2}{c}{Latency}  & Throughput\tnote{a} \\
        & & kLUT & DSP & RAM (kbits) & Registers & (MHz) & (k\gls{cc}) & (ms) & (bits/k\gls{cc})\\
        \hline\hline
         \textbf{This work} & \textbf{Zynq-US} & \textbf{21.48} & \textbf{0} & \textbf{0} & \textbf{13\,881} & \textbf{250.00} & \textbf{1\,887.52} & \textbf{7.55} & \textbf{0.27}\\
         \textbf{This work} & \textbf{Artix-7} & \textbf{24.00} & \textbf{0} & \textbf{0} & \textbf{13\,385} & \textbf{90.00} & \textbf{1\,887.52} & \textbf{20.97} & \textbf{0.27}\\
         Mehrabi et al.\cite{mehrabi2020elliptic} & Virtex-7 &  46.90 & 560 & 0 & 29\,742 & 125.00 & N/A & 0.25 & N/A\\
         Asif et al.\cite{asif2018fully} & Virtex-7 & 18.80 & 1\,036 & 828 & N/A & 86.60 & 63.20 & 0.73 & 8.10\\
         Islam et al.\cite{islam2019fpga} & Virtex-7 & 35.60 & N/A & N/A & N/A & 177.70 & 262.70 &1.48 & 1.95\\
         Romel et al.\cite{romel2023fpga} & Virtex-7 & 51.64 & 0 & N/A & 15\,263 & 122.33 & 65.78 & 0.54 & 7.78\\
         Arunachalam et al.\cite{arunachalam2022fpga} & Virtex-5 & 32.92 & N/A & N/A & N/A & 192.00 & 232.20 & 1.21  & 2.20\\
         Roy et al.\cite{roy2012implementation} & Virtex-5 & 39.68 & 0 & N/A & N/A & 43.00 & 25.70 &0.60 & 19.92\\
         Wang et al.\cite{wang2023fpga} & Virtex-7 & 23.10 & N/A & N/A & N/A & 105.30 & N/A &  0.08& N/A\\
         Yang et al.\cite{yang2020optimized} & Virtex-7 & 22.94 & 64 & N/A & N/A & 123.27 & 13.65 & 0.15 & 37.51\\ 
         Asif et al.\cite{asif2017high} & Virtex-7 & 96.90 & 2799 & 7\,452 & N/A & 72.90 & 215.90 & 2.96 & 2.37\\
         Javeed et al.\cite{javeed2017high} & Virtex-6 & 22.15 & N/A & N/A & N/A & 95.00 & 220.10 & 2.30 & 2.33\\
        \hline \hline
    \end{tabular}
     \begin{tablenotes}
            %\item[a]  AT =Area $\times$ Time (\gls{lut} $\times$ ms). %Xilinx Virtex-7 (XC7VX690T) BRAM are 36 Kb in size.
            \item[a] Throughput is estimated by authors as: $\frac{\text{512}}{\text{kCC}}$.
        \end{tablenotes}
    \end{threeparttable}
    \label{table:secp}
\end{table*}

\subsection{Submodule Implementation Overview}
 \label{subsec:submodImpResDis}

 \autoref{sub:secp} presents the implementation results of the proposed SECP256K1 algorithm, including the \gls{sca} performed on the design deployed on an \gls{fpga} and a discussion of the findings. Similarly, \autoref{sub:hmac} details the implementation results of the \gls{hmac}\gls{sha}-512 algorithm. \autoref{subsec:bipCkdf} provides results for the implementation of the \gls{bip}-39 and \gls{ckd} algorithms. Furthermore, \autoref{sub:eth_imp} discusses the overall implementation results of EthVault. \autoref{subsec:EthvaultVsTrez} compares EthVault with the Trezor One \gls{cryp} wallet. \autoref{subsec:powEv} evaluates the estimated power consumption of EthVault, and \autoref{subsec:sysInteg} explores potential integration strategies with hot wallets, along with considerations for real-world deployment.

\subsection{SECP256K1}
\label{sub:secp}

The comparison in \autoref{table:secp} evaluates the proposed SECP256K1 implementation against state-of-the-art solutions. The results demonstrate that the algorithm performs more effectively on the Zynq-US board compared to the Artix-7. This performance advantage is due to the advanced technology and superior architectural features of the Zynq-US board.

The number of \glspl{lut} utilized by the proposed design on the Zynq-US is generally lower than most works in the literature, except \cite{asif2018fully}. Specifically, the proposed implementation uses 12.5\% more \glspl{lut} than \cite{asif2018fully}, but this is offset by its use of 1\,036 fewer \glspl{dsp}. On average, the implementation achieves a 48\% reduction in \glspl{lut} compared to the other referenced works when targeting the Zynq-US platform.

For the Artix-7 platform, the results differ slightly. References \cite{asif2018fully}, \cite{wang2023fpga}, \cite{yang2020optimized}, and \cite{javeed2017high} show lower \glspl{lut} usage compared to the proposed design. However, these works often rely on other resources, such as \glspl{dsp}, or do not provide a complete breakdown of their resource usage, complicating direct comparisons.

The proposed SECP256K1 implementation stands out by not utilizing any \glspl{dsp}, a feature shared with \cite{romel2023fpga} and \cite{roy2012implementation}. However, both of these works require a higher number of \glspl{lut}. In contrast, other referenced designs either rely on \glspl{dsp} or do not report their usage. Similarly, our implementation does not utilize \gls{ram} blocks, whereas other designs either use these resources or omit such details. Additionally, our design is efficient in register usage, employing 10\% fewer registers than \cite{romel2023fpga}, which itself has the second-lowest register count among the compared implementations.

Overall, the findings suggest that our implementation occupies a smaller area compared to analogous designs in the literature, making it a resource-efficient solution for resource-constrained, low-power applications like \gls{cryp} wallets.

 Our implementation achieves the highest frequency on the Zynq-US. However, the maximum frequency is about three times slower on the Artix-7 platform, showcasing the contribution of superior technology. Furthermore, the proposed implementation exhibits reduced throughput and increased latency compared to some works in the literature. This performance trade-off is largely attributed to the design's focus on minimizing hardware resource usage, prioritizing efficiency over speed in resource-constrained environments.

\subsubsection{\gls{sca} Attack Analysis} As highlighted in \autoref{sec:intro}, SECP256K1 is a critical algorithm used in Ethereum wallets and is often targeted by attackers attempting to extract private keys. To evaluate the resistance of the proposed architecture against \gls{spa} and timing attacks, the architecture was deployed on the Zynq-US board, and an \gls{sca} experiment was conducted.

\begin{figure}[t]
    \centering
    \includegraphics[width=0.8\linewidth]{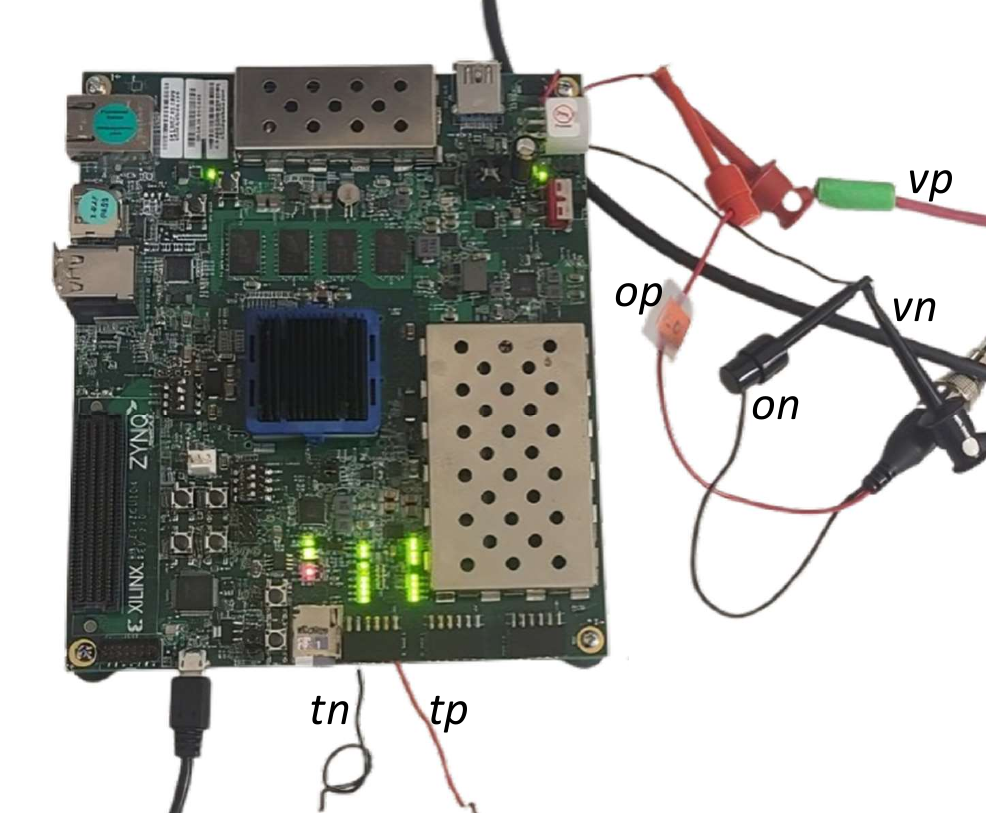}
    \caption{Setup for performing an \gls{sca} on the proposed SECP256K1 architecture deployed on a Zynq-US \gls{fpga}.}
    \label{fig:enter-label}
\end{figure}

\begin{figure*}[t]
    \centering
    \begin{subfigure}[b]{0.45\textwidth}
    \resizebox{\textwidth}{!}{
        \begin{tikzpicture}
        \begin{axis}[
            xlabel={Time (s)},
            ylabel={Power (W)},
            grid=both, % Adds a grid
            width=\textwidth, % Adjust width
            height=0.5\textwidth, % Adjust height
            scaled ticks=false, % Disable scientific notation
            ticklabel style={/pgf/number format/fixed}, % Ensure fixed-point format
        ]
        % Import data from the CSV file and plot it
        \addplot[blue, thick] 
        table[
            col sep=comma, % Set the delimiter (comma)
            x=x, % Use column 'x' for X-axis
            y=y  % Use column 'y' for Y-axis
        ] {csvFile1.CSV};
        \end{axis}
    \end{tikzpicture}
    }
    \caption{Multiple power traces}
    \label{fig:powerTraceA}
    \end{subfigure}
    \hspace{1cm}
    \begin{subfigure}[b]{0.45\textwidth}
        \centering
        \resizebox{\textwidth}{!}{
            \begin{tikzpicture}
        \begin{axis}[
            xlabel={Time (s)},
            ylabel={Power (W)},
            ylabel style={yshift=-12pt}, % Move the label closer to the y-axis
            grid=both, % Adds a grid
            xmin=0.013, xmax=0.032,  % Crop to x-range [1, 4]
            ymin=-0.1, ymax=0.2,  % Crop to y-range [2, 5]
            width=\textwidth, % Adjust width
            height=0.5\textwidth, % Adjust height
            scaled ticks=false, % Disable scientific notation
            ticklabel style={
            /pgf/number format/fixed,
            /pgf/number format/precision=3
            }, % Ensure fixed-point format
        ]
        % Import data from the CSV file and plot it
        \addplot[blue, thick] 
        table[
            col sep=comma, % Set the delimiter (comma)
            x=x, % Use column 'x' for X-axis
            y=y  % Use column 'y' for Y-axis
        ] {csvFile0.csv};

        % Draw vertical lines
        %\addplot[green, dashed, thick] coordinates {(0.013, -0.05) (0.032, -0.05)};
        
        % Horizontal
        \addplot[red, dashed, thick] coordinates {(0.015, -0.1) (0.015, 0.2)};
        \addplot[red, dashed, thick] coordinates {(0.030, -0.1) (0.030, 0.2)};
        \end{axis}
    \end{tikzpicture}
        }
        \caption{Single power trace}
        \label{fig:powerTraceB}
    \end{subfigure}
    \caption{Power trace of the proposed SECP256K1 algorithm with temporary registers deployed on the Zynq-US \gls{fpga}.}
\end{figure*}
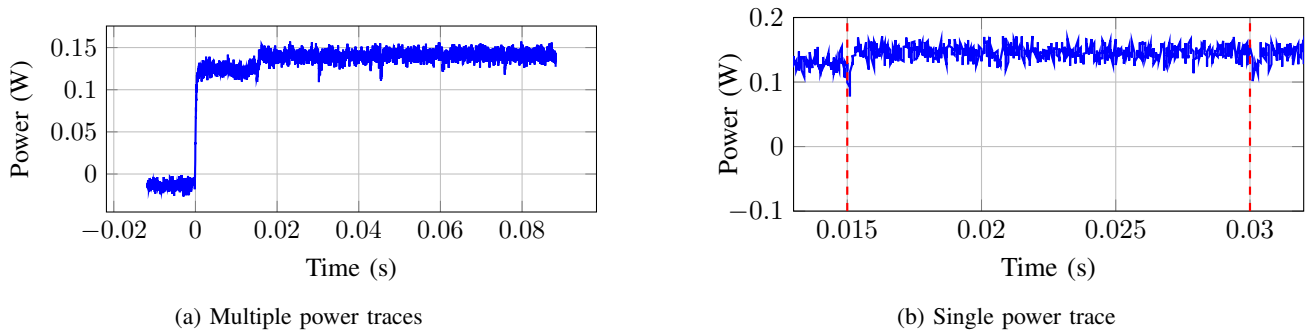

\begin{figure}[t]
    \centering
    \resizebox{0.45\textwidth}{!}{
        \begin{tikzpicture}
        \begin{axis}[
            xlabel={Time (s)},
            ylabel={Power (W)},
            grid=both, % Adds a grid
            width=\textwidth, % Adjust width
            height=0.5\textwidth, % Adjust height
            legend style={at={(0.99,0.01)}, anchor=south east}, % Adjust legend position
            scaled ticks=false, % Disable scientific notation
            ticklabel style={/pgf/number format/fixed}, % Ensure fixed-point
        ]
        % Import data from the CSV file and plot it
        \addplot[blue, thick] 
        table[
            col sep=comma, % Set the delimiter (comma)
            x=x, % Use column 'x' for X-axis
            y=y  % Use column 'y' for Y-axis
        ] {csvFile2.csv};
        \addlegendentry{Private key 1} % Legend entry for the first signal

        \addplot[red, thick] 
        table[
            col sep=comma, % Set the delimiter (comma)
            x=x, % Use column 'x' for X-axis
            y=y  % Use column 'y' for Y-axis
        ] {csvFile3.csv};
        \addlegendentry{Private key 2} % Legend entry for the second signal
        
        \end{axis}
        \node(msc)[] at (8,0.8){MSE = 0.001840};
    \end{tikzpicture}
    }
    \caption{The power traces observed when SECP256K1 processes two different private keys (MSE = 0.001840).}
    \label{fig:powerTcorr}
\end{figure}
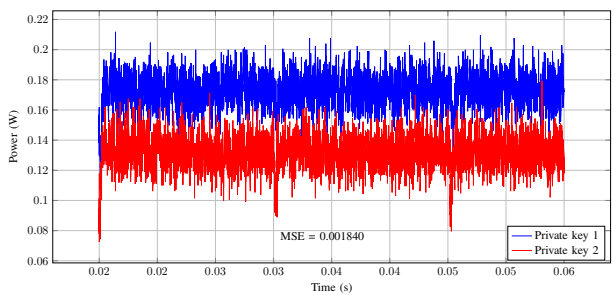

The algorithm was configured to execute continuously in a run-reset loop, operating with a 125\,MHz internal clock. The experimental setup is illustrated in \autoref{fig:enter-label}. A 12\,V DC voltage source with a current capacity of 2\,A was connected to the \gls{fpga} through probes $vp$ and $vn$. To measure the current, a current probe connected to channel one of an oscilloscope was clamped around $vp$. Moreover, channel two of the oscilloscope measured voltage through probes $op$ and $on$. Additionally, the algorithm's start input was configured to output a trigger signal via the PMOD GPIO connectors of the \gls{fpga}, which were then connected to the oscilloscope's trigger input through probes $tp$ and $tn$. The oscilloscope's math operation function was used to compute the power trace by multiplying the current (channel one) by the voltage (channel two).

Before starting the algorithm, the current probe was degaussed to eliminate the current drawn by the idle \gls{fpga}. The power trace captured after starting the algorithm is shown in \autoref{fig:powerTraceA}. A noticeable increase in the power trace occurs when the trigger signal is detected, indicating the start of the algorithm execution. Additionally, the periodic dips seen in the power trace, occurring every 15\,ms, correspond to the algorithm's reset instances. 

\autoref{fig:powerTraceB} presents the power trace from the beginning to the end of the algorithm’s execution. Based on the 125\,MHz frequency and the number of \glspl{cc} reported in \autoref{table:secp}, the measured duration of $15\,{\mu}s$ in the figure is consistent with expectations. Attackers often analyze the spikes observed in the trace during this period to distinguish between the processing of binary values (1s and 0s) \cite{trezone}. This information, if exploited, can potentially reveal the private key, emphasizing the importance of analyzing and mitigating such vulnerabilities in cryptographic implementations.

\begin{table*}[t]
    \centering
    \caption{Comparison of implementation results of \gls{hmac}\gls{sha}-512.}
    \begin{threeparttable}
       \begin{tabular}{l l l l l l l l l l l}
       \hline \hline
        Work & Platform & \multicolumn{4}{c}{Area} & Frequency & \multicolumn{2}{c}{Latency} & Throughput\tnote{a} \\
        & & kLUT & DSP & RAM (kbits) & Registers & (MHz) & (\gls{cc}) & ($\mu$s) & (bits/\gls{cc})\\
        \hline\hline
         \textbf{This work} & \textbf{Zynq-US} & \textbf{4.90} & \textbf{0} & \textbf{36} & \textbf{2\,592} & \textbf{200.00} & \textbf{335} & \textbf{1.66} & \textbf{1.53}\\
         \textbf{This work} & \textbf{Artix-7} & \textbf{4.90} & \textbf{0} & \textbf{36} & \textbf{2\,592} & \textbf{90.00} & \textbf{335} & \textbf{3.72} & \textbf{1.53}\\

         Marcio et al.\cite{juliato2011fpga} & Stratix-3 &  4.60 & N/A & 5.12 & 4\,116 & 116.04 & 81 & N/A & 6.32\\
         Nguyen et al.\cite{9842174} & Virtex-7 &  4.28 & 0 & N/A & 1\,310 & 168.56 &  N/A & 2.01 & N/A\\
         %Mercora\cite{HMAC_512} & Spartan 6 &  2.07 & N/A & 0 & N/A & 133 & N/A& 67 & 1\,016\\
    \hline \hline
    \end{tabular}
    \begin{tablenotes}
            \item[a] Throughput is estimated by authors as: 512 $\div$ \gls{cc}.
        \end{tablenotes}
    \end{threeparttable}
    \label{table:hmac}
\end{table*}

\autoref{fig:powerTcorr} presents the power traces of the proposed SECP256K1 architecture processing two distinct private keys. An offset was added to observe and compare the two traces. Attackers often exploit variations in such power traces to infer private keys by statistical analysis \cite{donjonTREZ,ghosside}. To evaluate the uniformity of the proposed algorithm, we calculate the \gls{mse} between the two traces shown in \autoref{fig:powerTcorr}. The \gls{mse} is determined as:

\begin{equation}
    MSE = \frac{1}{n} \sum_{i=1}^{n}(t_1 - t_2)^2,
    \label{eq:mse}
\end{equation}
 where $n$ is the total number of samples in each trace, $t_1$ is the first power trace, and $t_2$ is the second power trace.  

The calculated \gls{mse} for the traces is 0.001840, which is relatively small compared to the magnitudes of the traces. This low \gls{mse} value indicates a strong correlation between the two traces, demonstrating that the proposed SECP256K1 architecture ensures significant uniformity in power consumption when processing different keys. Consequently, this uniformity enhances resistance to \gls{sca} attacks, effectively reducing the system's vulnerability. 

The following part discusses the implementation results of the proposed \gls{hmac}\gls{sha}-512 algorithm.

\subsection{\gls{hmac}\gls{sha}-512}
\label{sub:hmac}

\autoref{table:hmac} presents the implementation results of the proposed \gls{hmac}\gls{sha}-512 architecture on Zynq-US and Artix-7 \gls{fpga} boards, as illustrated in \autoref{fig:eth_hmac}. The table also includes a comparison with similar implementations from the literature. 

The proposed architecture requires the same amount of resources (\glspl{lut}, \glspl{dsp}, registers, and \glspl{ram} blocks) when implemented on both the Zynq-US and Artix-7 \gls{fpga} boards. However, the maximum frequency attained is higher on the Zynq-US board due to its emphasis on high performance. 
%As a result, the Zynq-US implementation achieves better latency and throughput measurements compared to the Artix-7 implementation.

Resource utilization comparisons reveal that the proposed architecture requires approximately 1.07$\times$ more \glspl{lut} than the design in \cite{juliato2011fpga} and 1.14$\times$ more than the design in \cite{9842174}. Additionally, our design uses 1\,282 more registers than \cite{9842174} but 1\,524 fewer than \cite{juliato2011fpga}. Moreover, the proposed implementation utilizes 7$\times$ more \gls{ram} than that used by \cite{juliato2011fpga}. Although the proposed design slightly exceeds the resource requirements of prior works, this trade-off supports the reuse of \gls{sha}-512 and accommodates additional input capabilities.

Furthermore, the Zynq-US implementation attains the highest maximum operating frequency among the compared designs in \cite{juliato2011fpga} and \cite{9842174}, while achieving a 1.2$\times$ reduction in time latency relative to \cite{juliato2011fpga}. Conversely, the Artix-7 implementation exhibits a 1.9$\times$ increase in time latency compared to \cite{9842174}. In addition, both the \gls{cc} latency and throughput of our implementations on the two boards are lower than those reported in \cite{9842174}. This is attributed to the relatively higher number of \glspl{cc}, which is likely a consequence of the resource-conservation strategy adopted in our design.

The following section discusses the implementation results of the \gls{bip}-39 protocol and the \gls{ckd} function.

\begin{table}[t]
    \centering
    \caption{Implementation results of \gls{bip}-39 and the \gls{ckd} function using a frequency of 167\,MHz on the Zynq UltraScale. }
  %  \resizebox{\columnwidth}{!}{%
  \resizebox{5cm}{!}{
    \begin{tabular}{l c c }
    \hline \hline
        Metrics & \gls{bip}-39 & \gls{ckd}\\ 
    \hline \hline
        Area \\
        ~~~k\glspl{lut} & 17.08 & 25.43 \\
        ~~~Registers & 6\,392 & 17\,792\\
        ~~~DSP & 0 & 0  \\
        ~~~RAM (kbits) & 0 & 36  \\
        \hline
        Latency\\
            ~~~(k\gls{cc}) & 692.83 & 1\,887.86\\
            ~~~(ms) & 4.149 & 11.305 \\
    \hline \hline
    \end{tabular}
    }
    \label{table:ckdf}
\end{table}

\begin{table*}[t]
    \centering
    \caption{Comparison of hardware area and latency across individual modules and the complete EthVault system.}
    %\resizebox{5cm}{!}{
    \begin{threeparttable}
    \begin{tabular}{lcccccccccc}
    \hline \hline
        Metrics                     &RAM & KECCAK256 & SHA256 & HMACSHA512   & BIP39     & SECP256K1  & CKDF  & ECDSA & \multicolumn{2}{c}{\textbf{EthVault}}\\ 
    \hline \hline
        Area \\
        ~~~\glspl{lut}  & 84    & 2\,484  & 1\,043 & 4\,365   & 17\,083   & 21\,067  & 25\,430  &10\,043 & \multicolumn{2}{c}{\textbf{62\,209(27\%)}} \\
        ~~~Registers    & 0     & 1\,607  & 915    & 2\,079   & 6\,414    & 13\,872   & 17\,792  & 4\,609&\multicolumn{2}{c}{\textbf{31\,180(7\%)}}  \\
        ~~~RAM (kbits)  & 648     & 0       & 0      &36         & 0         & 0     & 36     & 0 & \multicolumn{2}{c}{\textbf{684(6\%)}}\\
        \hline                                                                                                                                       
        Latency\tnote{a}\\                                                                                                             
            ~~~(\gls{cc})   & 1     & 25      & 73     & 335      & 692\,827  & 1\,887\,520  & 1\,887\,855\tnote{e} & 1\,888\,550\tnote{d} &\textbf{6\,356\,729}\tnote{b} &  \textbf{3\,775\,064}\tnote{c} \\
            ~~~($\mu$s)     & 0.006 & 0.150    & 0.437   & 2.000        & 4\,149    & 11\,303    & 11\,305 & 11\,309 & \textbf{38\,064} &  \textbf{22\,605} \\
        %Frequency (MHz)             & & & & & & & & 166.67 \\
        %Throughput\tnote{a} (Mbps)  & & & & & & & & 22.52 \\
    \hline \hline
    \end{tabular}
    \begin{tablenotes}
            \item[a] Frequency of 167\,MHz is used.
            \item[b] First private-public key pair.
            \item[c] Subsequent private-public key pairs.
            \item[d] Signing data.
            \item[e] Time to create normal keys.
        \end{tablenotes}
    \end{threeparttable}
    %}
    \label{table:eth_wallet}
\end{table*}

\subsection{\gls{bip}-39 and the \gls{ckd} Function}
\label{subsec:bipCkdf}

\autoref{table:ckdf} presents the implementation results of the proposed \gls{bip}-39 architecture, shown in \autoref{fig:bip39RTL}, alongside those of the \gls{ckd} function, depicted in \autoref{fig:ckdf_fig}.  
While \gls{bip}-39 relies on the \gls{hmac}\gls{sha}-512 algorithm, its contribution to the area is not included in the calculation of the \gls{bip}-39 architecture's area. This is because the \gls{hmac}\gls{sha}-512 module is part of the \gls{ckd} function. However, \autoref{table:ckdf} demonstrates that the area of the \gls{bip}-39 architecture is very close to that of the \gls{ckd} function.  

The significant size of the \gls{bip}-39 architecture primarily stems from the MNG module, shown in \autoref{fig:bip39RTL}, which stores the English mnemonic word list \cite{bip39wordlist}. This module accounts for 71.4\% of the total area utilized by \gls{bip}-39, highlighting it as a major contributor to the architecture's resource usage. Alternatively, an external memory could be utilized to store the words, effectively reducing on-chip resource requirements while maintaining functionality.

Notably, a comparative analysis was not performed as no hardware implementations of the \gls{ckd} function were found in the literature. 

The next section presents the implementation results of EthVault, along with the resource utilization and timing analysis of its core modules.

\subsection{Hardware Resource Utilization and Latency of EthVault} %how many keys can be stored in this wallet
\label{sub:eth_imp}

\autoref{table:eth_wallet} presents the resource utilization and latency of the individual modules that comprise EthVault. Specifically, it reports the area and timing characteristics of RAM, KECCAK256, SHA256, HMACSHA512, BIP39, SECP256K1, CKDF, and ECDSA. The table shows that CKDF, SECP256K1, and ECDSA modules dominate the total latency, contributing the most to the overall system delay. This is expected, as these are computationally intensive cryptographic operations. Similarly, the same modules utilize the highest number of \glspl{lut} and registers. However, the RAM module uses the highest number of \gls{ram} blocks, as it stores the generated child keys and addresses. The table also includes the overall resource consumption and latency of the implemented EthVault architecture.

%\autoref{table:eth_wallet} outlines the resource utilization of the proposed Ethereum \gls{hd} cold wallet, illustrated in \autoref{fig:ethWalArch}. 
The results indicate that EthVault utilizes only 27\% of the available \glspl{lut}, 7\% of the registers, and 6\% of the \gls{ram} blocks on the Zynq UltraScale+ \gls{fpga}, with no usage of \gls{dsp} blocks. The design operates at a maximum frequency of 167\,MHz. These results demonstrate efficient resource utilization while upholding high-security standards, making the design ideal for secure hardware wallet applications.

The latency of EthVault is measured as the time required to generate the first private–public key pair along with its corresponding address, and subsequently the second key pair and address. As discussed in \autoref{sec: hw_arch}, generating the second key pair requires less time. Notably, it takes 3\,775\, 06\,\glspl{cc} which is equivalent to 22.61\,ms. In comparison, calculating the first key pair takes 6\,356\,729\,\glspl{cc}. This significant reduction in \glspl{cc} for the second key pair improves the overall throughput of the wallet when generating subsequent keys. In particular, we estimate the throughput of the wallet as:

\begin{equation}
    \text{Throughput} = \frac{\text{Frequency (MHz)}}{\text{Latency (CC)}} \times \text{Key Size (bits)},
\end{equation}

where ``Key Size" corresponds to the total number of output bits (private key, public key, and address). For Ethereum, this value is 856 bits. Hence, the throughput of generating child keys and addresses in EthVault 37.87\,kbps, where frequency is 167\,MHz and latency is 3\,775\, 06\,\glspl{cc}.

Moreover, \autoref{table:eth_wallet} shows that the latency of EthVault for signing contracts is about 1\,888\,550\,\glspl{cc}, corresponding to the execution time of the \gls{ecdsa} algorithm. Given the 512-bit size of the signature ($r$, $s$), the resulting throughput for signing transaction data in  EthVault is 45.27\,kbps.

The following sections compare the throughput of EthVault with that of Trezor One and the Ethereum blockchain.

\subsection{Comparing EthVault, Trezor One, and Ethereum Blockchain}
\label{subsec:EthvaultVsTrez}

\autoref{fig:bip32} shows that the \gls{ckd} function computes keys using a key-derivation path. The latency used by Trazor One physical wallet to execute the \gls{ckd} function and generate the master public key is 386.59\,ms. Moreover, the \gls{ckd} function takes 94.30\,ms to execute each element in the given path \cite{trezone}. Hence, assuming that Trezor One employs the same technique as EthVault, storing the partial \gls{ckd} path as discussed in \autoref{subsec:ETHwallet}, the latency for deriving subsequent keys is 188.6\,ms. Therefore, the latency of generating the second key in EthVault is about 8$\times$ less than that of Trezor One. Moreover, we can estimate the child key generation throughput of the Trezor One wallet as 4.5\,kbps (Throughput = $\frac{1}{\text{latency}(s)}$ $\times$ 856). This suggests that the child key generation throughput of EthVault is about 8$\times$ higher.

Also, the current Ethereum blockchain network has a transaction rate of 15 to 20 \gls{tps} \cite{FOBS}. This suggests that a cold wallet signing transaction may have a minimum throughput of 10.24\,kbps (Authors estimate Ethereum's throughput = \gls{tps} $\times$ (size of \gls{ecdsa} signature)). Therefore, the 45.27\,kbps transaction data signing throughput of EthVault is sufficient to support user transactions in the current Ethereum blockchain.

The following section presents a detailed power evaluation of the EthVault and Trezor One wallets.

\begin{table*}[t]
    \centering
    \caption{Performance and efficiency comparison of EthVault and Trezor One wallets in terms of platform, security features, and key hardware metrics.}
  %  \resizebox{\columnwidth}{!}{%
  \resizebox{18.1cm}{!}{
  \begin{threeparttable}
    \begin{tabular}{cclccccccccc }
    \hline \hline
       Wallet & Platform & Security Features & Power & Frequency & Latency & Throughput\tnote{a} & Area & TPA\tnote{b}   & EPO\tnote{c} & PD\tnote{d}\\ 
        &&  &(mW)  & (MHz) & ($\mu$s) & (kbps) &  & (bps/Area) & ($\mu$J/b) & ($\mu$W/Area) \\
    \hline \hline
        EthVault & FPGA & -Firmware proof & 140 & 167 & 22\,605 & 37.87 & 62\,209\,\glspl{lut} & 0.61 & 3.70 & 2.25\\
        && -Physical isolation &&&&&&&&\\
        && -PIN \& passphrase entry &&&&&&&&\\
        &&  -\gls{sca}-resistant SECP256K1&&&&&&&&\\
        Trezor One & CPU & -MCU-level protections & 140 & 120 & 188\,600 & 4.50 & 100\,$\text{mm}^2$ & 45 & 31.11 & 1\,400\\
        &&-PIN \& passphrase entry &&&&&&&&\\
        &&-Public security and design &&&&&&&&\\
    \hline \hline
    \end{tabular}
    \begin{tablenotes}
  \item[]
  % First row: a, b, c
  \begin{minipage}[t]{0.32\linewidth}
    \textsuperscript{a} Throughput = $\frac{\text{1}}{\text{Latency}} \times 856$
  \end{minipage}%
  \begin{minipage}[t]{0.23\linewidth}
    \textsuperscript{b} TPA = $\frac{\text{Throughput}}{\text{Area}}$
  \end{minipage}%
  \begin{minipage}[t]{0.23\linewidth}
    \textsuperscript{c} EPO = $\frac{\text{Power}}{\text{Throughput}}$
  \end{minipage}
  \begin{minipage}[t]{0.20\linewidth}
    \textsuperscript{d} PD = $\frac{\text{Power}}{\text{Area}}$
  \end{minipage}

  %\vspace{0.3em} % Optional spacing between rows

  % Second row: d,
  %\begin{minipage}[t]{0.48\linewidth} 
  %\end{minipage}
\end{tablenotes}

    \end{threeparttable}
    }
    \label{table:power}
\end{table*}

\subsection{Power Evaluation}
\label{subsec:powEv}

Post-implementation power estimation of the EthVault architecture was performed using Vivado. The total on-chip power consumption on the Zynq UltraScale+ \gls{fpga} board was 2\,204\,mW, with dynamic power accounting for 72\% (1\,581\,mW) and static power contributing 28\% (623\,mW). The primary sources of dynamic power were signal activity, contributing 42\%, and logic operations, contributing 33\%. This reflects the intensive cryptographic computations inherent in Ethereum wallet functionalities. The estimated junction temperature is 27.2\,$^\circ\mathrm{C}$. While the analysis was based on vectorless estimation, future work will incorporate switching activity from post-implementation to enhance accuracy and guide low-power optimization.

Additionally, power measurements for EthVault were performed using a Tektronix TDS 3012 Oscilloscope. The EthVault algorithm was loaded onto the ZCU104 \gls{fpga} board, where a voltage probe was connected to the FPGA’s power supply to measure voltage, and a current probe was clamped onto the active power cable to measure current. The oscilloscope’s built-in math function was then used to compute power as P = V $\times$ I. Prior to running the algorithm (from mnemonic generation to signing, where a random entropy, $e$, was provided), the current probe was degaussed (zeroed) to ensure that only the current drawn during EthVault execution was captured. A trigger signal was connected to the start input of EthVault to capture power traces precisely at the beginning of computation on the Oscilloscope. Furthermore, Vivado’s clocking wizard was used to configure a phase-locked loop to generate the 167\,MHz, driving EthVault from ZCU104’s 300\,MHz clock.

For comparison, the power consumption of a Trezor One wallet was also measured. The \gls{usb} cable connecting the wallet to a computer was stripped to expose the positive and negative supply lines. Voltage and current probes were connected to channels one and two of the oscilloscope, respectively, and the math function was again used to compute instantaneous power consumption during wallet operation. The wallet was not password-protected, and funds were transferred from the wallet using the Trezor Suite desktop application \cite{trezor_suite}.

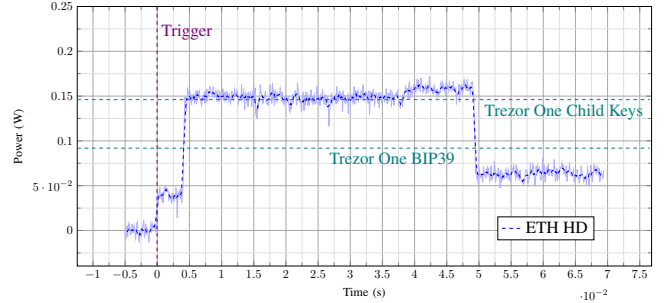
\begin{figure}[t]
    \centering
    \resizebox{0.48\textwidth}{!}{ 
        \begin{tikzpicture}
  \begin{axis}[
    width=\textwidth, % Adjust width
    height=0.5\textwidth, % Adjust height
    xlabel={Time (s)},
    ylabel={Power (W)},
    ymax=0.25, % set y-axis maximum to 3
    legend style={
        at={(0.9,0.1)},
        anchor=south east,
        font=\Large % Increase legend font size
    },
    grid=both,
    minor tick num=1,         % Number of subdivisions between major ticks
    major grid style={line width=0.2pt,draw=gray!70},
    minor grid style={line width=0.1pt,draw=gray!25},
  ]

     % Plot ETH HD Orig
    \addplot [
      color=blue!30,
      thick,
      forget plot,
    ] table [x=x, y=y, col sep=comma] {p11.csv};
    %\addlegendentry{ETH HD Origl}

    % Plot ETH ND MA
    \addplot [
      color=blue,
      dashed,
      thick,
    ] table [x=x, y=y_ma, col sep=comma] {p11.csv};
    \addlegendentry{ETH HD}

    %\addplot[red, dashed, thick] coordinates {(0, -0.09) (0, 0.29)};
  \end{axis}

  \draw[thick, dashed, violet] (2.3,0) -- node[right, yshift=3cm,  font=\Large]{Trigger}(2.3,7.5);
  \draw[teal, dashed] (0,3.4) -- node[pos=0.55, below, font=\Large]{Trezor One BIP39}(16.5,3.4);
  \draw[teal, dashed] (0,4.8) -- node[pos=0.85, below, font=\Large]{Trezor One Child Keys}(16.5,4.8);
\end{tikzpicture}
    }
    \caption{
    Measured power utilization of HardVault during Ethereum HD key generation compared with the Trezor One. The curve shows two phases: a low-power \gls{bip}-39 stage ($\approx$40\,mW) and a higher-power child key derivation stage ($\approx$\,140 mW).
    }
    \label{fig:powerEthVTrez}
\end{figure}

\autoref{fig:powerEthVTrez} compares the measured power consumption of EthVault and the Trezor One wallet. EthVault requires approximately 140\,mW, which is about 16$\times$ lower than the power estimated by Vivado. We attribute this gap primarily to the conservative default assumptions in Vivado's Power Estimator, which tend toward worst-case switching activity and operating conditions. In contrast, the Trezor One wallet also utilizes around 140\,mW. Additionally, the figure shows that EthVault requires little under 40\,mW during the first 4\,ms, corresponding to the execution of the \gls{bip}-39 algorithm for mnemonic generation. This measurement is consistent with the runtime characteristics of the BIP39 module reported in \autoref{table:eth_wallet}.

Similarly, the complete execution cycle lasts approximately 50\,ms, which corresponds to the combined duration of generating the first private–public key pair (38.06\,ms) and signing a transaction (11.31\,ms), as reported in \autoref{table:eth_wallet}.

To provide context for EthVault relative to existing market solutions, \autoref{table:power} presents a comparison with Trezor One. The table highlights selected security features, as well as performance and efficiency metrics, namely \gls{tpa}, \gls{epo}, and \gls{pod}. Due to differences in the target platforms, a direct comparison of area and area-related metrics is not meaningful. Nevertheless, these values are included to enable reference in future related works.

The area of Trezor One is estimated based on the physical size of the STM32F205RET6 \gls{mcu} it uses, which is approximately 100\,${\text{mm}^2}$ \cite{stm32f205_ds}. In contrast, the area of EthVault is expressed in terms of the number of utilized \glspl{lut}.

As shown in \autoref{table:power}, EthVault requires about 8$\times$ less \gls{epo} than Trezor One, demonstrating improved energy efficiency. Furthermore, EthVault achieves a higher maximum frequency, lower power consumption, and greater throughput. However, \gls{tpa} and \gls{pd} cannot be directly compared due to the platform-dependent area disparities noted earlier.

The next section presents the system integration of EthVault and its real-world deployment.

% trezor - https://trezor.io/learn/a/trezor-hardware-built-in-security 
\subsection{System Integration and Real-World Deployment}
\label{subsec:sysInteg}
As illustrated in \autoref{fig:lit_wallet}, a hardware cold wallet interacts with a hot wallet to sign and authorize transactions without exposing sensitive credentials. In this section, we provide insight into how EthVault enables secure interaction with both a hot wallet and the end user.

\autoref{fig:ethWalArch} illustrates the internal structure of the wallet. Generated child keys, private keys, and their associated Ethereum addresses are stored in a dedicated \gls{ram} module. The private key is used exclusively for signing transaction data using the \gls{ecdsa} algorithm. To maintain a high level of security, only the signature, public key, and derived Ethereum address are accessible externally, ensuring that the private key never leaves the secure hardware boundary. EthVault outputs can be accessed through controlled interfaces, such as USB or JTAG ports.

The \gls{sha}-256 digest of the transaction data can be transmitted via the USB interface. Once received, the \gls{fpga} internally processes it using \gls{ecdsa} to generate the signature, which is then sent to the hot wallet. This approach ensures that the private key remains fully protected at all times.

Moreover, a display is used to show the 24-word mnemonics through the $mcs$ output during wallet initialization. For security reasons, these mnemonics are not stored within the wallet at any point. Instead, users are instructed to write them down and store them safely. When a key recovery is needed, the user must manually re-enter the mnemonic phrase via the $mcsIn$ module, using an attached input interface such as a keyboard or touchscreen. This approach ensures that sensitive recovery information never resides permanently within the device, reducing the risk of extraction in the event of physical compromise.

The following section outlines potential \gls{sca} attacks that could still affect EthVault, assessing their likelihood and possible mitigation strategies.

\section{Residual \gls{sca} Threats and Mitigation}
\label{sec:Threats}

While many possible attacks exist, this section focuses on \gls{dpa}, \gls{fia}, and \gls{ema} attacks. \gls{dpa} is a statistical technique used to extract secret information by analyzing data-dependent correlations in measured signals. The method involves recording multiple traces of a signal, partitioning them into subsets, computing the average of each subset, and then evaluating the differences between these averages. By examining these differences, sensitive information can be extracted \cite{kocher2011introduction, quisquater2025electromagnetic}. Although EthVault’s use of temporary registers may provide protection against \gls{spa} attacks (as discussed in \autoref{subsec:arch_secp}), \gls{dpa} attacks could still target power traces in the BIP39, HMAC-SHA512, or SECP256K1 modules to recover the master key. To mitigate \gls{dpa}, EthVault can introduce countermeasures such as amplitude masking or noise injection. The former can be achieved by adding circuits that draw variable power, while the latter can be realized by inserting variations in timing or execution order \cite{kocher2001using}. These techniques help obscure the power consumption patterns during key generation, thereby reducing vulnerability to \gls{dpa}.

In \gls{fia}, an attacker deliberately introduces faults into a computing system to disrupt normal operation and extract sensitive information. Such faults can be induced by exposing the target device to high heat, injecting irregularities into the clock, or radiating \gls{em} pulses \cite{barenghi2012fault}. EthVault could be vulnerable to \gls{fia}, particularly during data signing, where the \gls{sha}-256 digest of the transaction data $z$ is received from the software wallet. Faults introduced into $z$ could disrupt normal execution and compromise the signing process. To mitigate this risk, EthVault can employ redundant encryption of transaction data and compare the resulting hashes before signing. This approach assumes that faults are transient and unlikely to affect both executions simultaneously. Additionally, EthVault could be encased in a tamper-resistant enclosure equipped with sensors to detect physical tampering attempts \cite{barenghi2012fault}.

An \gls{ema} attack targets a device’s \gls{em} emissions during operation to extract secret data. \gls{ema} can take the form of \gls{sema} or \gls{dema}. In the former, an attacker relies on a single \gls{em} measurement to directly recover part or all of the secret data, while in the latter, multiple \gls{em} measurements are collected to reduce noise, and statistical methods are applied to extract the secret information \cite{de2005electromagnetic}. Since EthVault generates keys in stages, with only certain parts of the architecture active at a time, it may radiate unique \gls{em} signatures that could be exploited to extract private keys. To mitigate \gls{ema}, EthVault can employ \gls{em} shielding to prevent emissions from escaping the device. Additionally, injecting artificial noise can reduce the \gls{snr}, thereby lowering the probability of successful information extraction \cite{rohatgi2009electromagnetic}.

The following section outlines the limitations of this work and potential directions for future research.

\section{Limitations and Future Work}
\label{sec:Limitations}

EthVault currently only supports the Ethereum blockchain with a \gls{hd} wallet structure. Future work will extend support to additional \gls{cryp}, starting with Bitcoin, and introduce an \gls{nd} mode to offer users a choice between \gls{hd} and \gls{nd} key generation methods.

EthVault implements countermeasures against \gls{spa} and timing attacks. Nevertheless, residual vulnerabilities remain, as discussed in the previous section. As future work, we plan to transition to an \gls{asic} implementation and to incorporate additional protections against advanced adversaries, including \gls{dpa}, \gls{fia}, and \gls{em} side-channel attacks. We also intend to perform a comprehensive security analysis of the implemented countermeasures.

At present, critical secrets (e.g., seed values and private keys) reside in \gls{ram}. To ensure survivability without exposure in the event of a reset or power loss, we will adopt secure retention mechanisms. A potential approach would be to use an encrypted battery-backed \gls{ram} with integrity protection, so that data can be recovered on reboot only after successful verification. Furthermore, to ensure reliable key storage during operation, a hardware-based error detection and correction mechanism, such as a parity bit or Hamming code \cite{singh2016error}, may be incorporated to identify and, if possible, correct memory bit errors.

Moreover, the current version of EthVault does not verify whether the mnemonic words entered by the user are valid. As a result, users may input words that are not part of the official mnemonic list. Future versions of the wallet will include a validation mechanism within drivers that interact with the wallet. In the meantime, users are strongly encouraged to carefully double-check their mnemonic words during key recovery.

The current sources of entropy for $e$, used during the \gls{bip}-39 phase, and $k$, used in the signature generation phase, are implemented as constant random values. In a practical deployment, these parameters must be generated using a cryptographically secure entropy source to ensure adequate security. Future versions of the wallet will incorporate a \gls{qrng} module designed to provide true randomness for all entropy-dependent operations within EthVault \cite{ma2016quantum}.

\section{Conclusion}
\label{sec:Conclusion}

In this work, we present a hardware architecture for an Ethereum \gls{hd} cold wallet. In doing so, we propose a hardware architecture for the \gls{ckd} function. Additionally, we propose a SECP256K1 architecture designed to enhance security against \gls{spa} and timing attacks. This architecture leverages complete point addition equations, temporary registers, and parallel processing to achieve robust protection. 

Our implementation results demonstrate that the building blocks of the proposed design are more compact compared to analogous implementations in the existing literature, suggesting a smaller overall size for the wallet. Furthermore, EthVault complies with \gls{bip}-32, \gls{bip}-39, and \gls{bip}-44 standards, which blockchain users highly value.

\section*{Acknowledgement}

This work was supported in part by an FRQNT Établissement de la relève professorale grant \#309366 and in part by MITACS through project IT28459.

\balance
\bibliographystyle{IEEEtran}
\bibliography{IEEEabrv, ConfAbrv, references}
\end{document}